\documentclass[]{svmult}

\usepackage{mathptmx}       
\usepackage{helvet}         
\usepackage{courier}        
\usepackage{type1cm}        
\usepackage{amsmath}
\usepackage{color}
\usepackage{makeidx}         
\usepackage{graphicx}        
\usepackage{multicol}        
\usepackage[bottom]{footmisc}

\def\apj{\rm ApJ~}
\def\KK{\rm ~K}

\def\EE#1{\times 10^{#1}}

\def\ccm{\rm ~cm^{-3}}
\def\kms{\rm ~km~s^{-1}}
\def\ergs{\rm ~erg~s^{-1}}

\def\La{${\rm Ly}\alpha$}
\def\Ha{${\rm H}\alpha$}

\def\Msun{M_\odot}
\def\Rsun{R_\odot}
\def\Msunyr{M_\odot~\rm yr^{-1}}
\def\Mu5{\dot{\cal{M}}}
\def\Mdot{\dot M}
\DeclareTextCommandDefault{\textcopyright}{\textcircled{c}}
\newcommand{\lsim}{\raise0.3ex\hbox{$<$}\kern-0.75em{\lower0.65ex\hbox{$\sim$}}}
\newcommand{\gsim}{\raise0.3ex\hbox{$>$}\kern-0.75em{\lower0.65ex\hbox{$\sim$}}}
\newcommand\apss{\rm{Ap\&SS}~}
\newcommand\mnras{\rm{MNRAS}~}
\newcommand\aap{\rm{A\&A}~}
\newcommand\apjl{\rm{ApJL}~}
\newcommand\araa{\rm{ARA\&A}~}
\newcommand\nat{\rm{Nature}~}
\newcommand\ssr{\rm{SSRv}}
\newcommand\aj{\rm{AJ}}
\newcommand\apjs{\rm{ApJS}}
\newcommand\prd{\rm{PhRvD}}
\newcommand\memsai{\rm{MmSAI}}
\makeindex             

\begin{document}
\tableofcontents

\title*{Thermal and non-thermal emission from circumstellar interaction}
\author{Roger A. Chevalier and Claes Fransson}
\institute{Roger A. Chevalier \at Dept. of Astronomy, University of Virginia, P.O. Box 400325 Charlottesville, VA 22904, USA, \email{rac5x@virginia.edu}
\and Claes Fransson \at Stockholm University,
Department of Astronomy and Oskar Klein Centre,
AlbaNova,
S - 106 91 Stockholm,
Sweden,  \email{claes@astro.su.se}}
%
%
\maketitle

\abstract*{
Each chapter should be preceded by an abstract (10--15 lines long) that summarizes the content. The abstract will appear \textit{online} at \url{www.SpringerLink.com} and be available with unrestricted access. This allows unregistered users to read the abstract as a teaser for the complete chapter. As a general rule the abstracts will not appear in the printed version of your book unless it is the style of your particular book or that of the series to which your book belongs.
Please use the 'starred' version of the new Springer \texttt{abstract} command for typesetting the text of the online abstracts (cf. source file of this chapter template \texttt{abstract}) and include them with the source files of your manuscript. Use the plain \texttt{abstract} command if the abstract is also to appear in the printed version of the book.}

\abstract{It has become clear during the last decades that the interaction between the supernova ejecta and the circumstellar medium is playing a major role both for the observational properties of the supernova and for understanding the evolution of the progenitor star leading up to the explosion. In addition, it provides an opportunity to understand  the shock physics connected to both thermal and non-thermal processes,  including relativistic particle acceleration, radiation processes and the hydrodynamics of shock waves. This chapter has an emphasis on the information we can get from radio and X-ray observations, but also their connection to observations in the optical and ultraviolet.  We first review the different physical processes involved in circumstellar interaction, including hydrodynamics, thermal X-ray emission, acceleration of relativistic particles and non-emission processes in the radio and X-ray ranges. Finally, we discuss applications of these to different types of supernovae.}

\section{Introduction}
\label{sec:1}

This review is partly an update of our previous review from 2003 on circumstellar interaction \cite{Chevalier2003}, mainly adding new developments during the last decade. In this review we are mainly concentrating on the physical processes, and only discuss individual objects of different supernova types as examples of these. While the focus is on radio and X-rays there is an especially close connection between the X-rays and optical observations, and where relevant we also discuss observations in this wavelength band. 

Massive stars are known to have strong winds during their lives.
At the time of the supernova (SN hereafter) explosion, the rapidly expanding gas plows into the surrounding medium, creating shock waves.
The shock waves heat the gas, giving rise to X-ray emission. A fraction of this may be reprocessed into optical and ultraviolet (UV) radiation, which in extreme cases may dominate the emission from the SN ejecta. 
The shock waves also accelerate particles to relativistic speeds.
Relativistic electrons radiate synchrotron emission in the magnetic fields that are present;
the magnetic field may be amplified by turbulence near the shock and/or the downstream region.

\section{Initial conditions}
\label{sec:2}

For the most part, the structure of the supernova ejecta can be separated from that of the ambient medium and then
their interaction can be discussed.
This is not the case around the time of shock breakout, when the diffusion time for photons in the shocked region first becomes
comparable to the age of the supernova.
Radiative acceleration\index{radiative acceleration} of the pre-shock gas leads to the dissolution of the radiation dominated shock wave, which is followed
by the formation of a viscous shock wave in the surrounding medium.
Radiative pre-acceleration of surrounding gas gives a velocity $v\propto r^{-2}$ due to the flux divergence and is typically
only important at early times.

\subsection{Ejecta structure}

\index{ejecta structure} After the shock wave has passed through the progenitor star, the gas evolves to free expansion, $v=r/t$ where $t$ is the age
of the explosion.
In free expansion, the density of an element of gas drops as $t^{-3}$ so that the density profile is described by a
function $\propto t^{-3}f(r/t)$.

The profile $f(r/t)$ can typically be described by a function that is a steep power law at high velocities and flat in the central region.
The outer steep power law region is especially important for circumstellar interaction because it is the region that typically
gives rise to the observed emission.
The outer edge of a star has a density profile that can be approximated by $\rho = a (r_*-r) ^{\delta}$, where $a$ is a constant
and $r_*$ is the outer edge of the star.
For a radiative envelope, $\delta = 3$ and for a convective envelope, $\delta \approx 1.5$.  

The supernova shock wave accelerates through the outer layers of the star due to the cumulation effect of energy going into a
vanishingly small amount of matter.
The acceleration of the shock wave through the outer layers of the star and the subsequent evolution to free expansion can
be described by a self-similar solution  \cite{Sakurai1960}.
The power law exponent in this case is not determined by dimensional analysis, but by the passing of the solution through a critical point;
it is a self-similar solution of the second kind.
The solution applies to a planar shock breakout, i.e. the breakout occurs over a distance that is small compared to stellar radius.
In that case the power law profile is steepened by two powers of $r$ in going to spherical expansion.
The result of the self-similar solution is that $n=10.2$ for $\delta =3$ and $n=12$ for $\delta=1.5$.
There is also a self-similar solution for an accelerating shock in an atmosphere with an exponential density profile.
The resulting value of $n$ is 8.67, which is also the value obtained in the limit $\delta \rightarrow \infty $.

The overall result of these considerations is that the outer part
of a core collapse supernova can be approximated by a steep power
law density profile, or $\rho_{\rm ej}\propto r^{-n}$ where $n$
is a constant. After the first few days the outer parts of the
ejecta expand with constant velocity, $V(m)
\propto r$ for each mass element, $m$, so that $r(m) = V(m) t$ and $\rho(m) =
\rho_o(m) (t_o/t)^3$. Therefore
\begin{equation}
\rho_{\rm ej} = \rho_o
\left(\frac{t}{t_0}\right)^{-3} \left(\frac{V_0 t} {r}\right)^{n}.  
\label{eq1a}
\end{equation}
This expression takes into account the free expansion of the gas.

The inner density distribution cannot be analytically calculated in a straightforward way, but physical arguments imply
a roughly $r^{-1}$ density profile \cite{Chevalier1989,Matzner1999}.
Although the outer power law density index is found from the self-similar solution, the coefficient of the density must be determined
by numerical simulation of the whole explosion.
The outer power law is only approached asymptotically so the power law index of the density profile applicable to a given situation is
smaller than the asymptotic value.

The self-similar solution assumes that the flow is adiabatic, which is not the case once radiation can diffuse from the shocked layer.
For a radiation dominated shock, the shock thickness corresponds to an optical depth of approximately $\tau_s =c/v$, where $v$ is
the shock velocity and $c$ is the speed of light.
Once the shock wave reaches an optical depth of $\tau_s$ from the surface, radiation can escape and the shock wave acceleration
through the layers of decreasing density ends.

\subsection{Ambient medium}

Because of their large luminosity massive stars lose mass in all evolutionary stages. What is most important for at least the early interaction between the supernova and ambient medium is the mass loss\index{radiative acceleration} immediately before the explosion. Red supergiants have slow winds with velocities $10-50 \kms$. The mass loss rates of most red supergiants are in the range  $10^{-6}-10^{-5} \Msunyr$, but there are also a number of stars undergoing a super-wind phase with very high mass loss rate,  $10^{-4}-10^{-3} \Msunyr$. This includes stars like VY CMa, NML Cyg and IRC10420. It is obvious both from the mass loss rates and their small fraction of the total number of red supergiants that this is a short-lived phase.  Compact progenitors, like WR-stars and blue supergiants, have similar mass loss rates $10^{-6}-10^{-4} \Msunyr$ but wind velocities $1000-3000 \kms$. See \cite{Smith2014b} for an extended discussion.

There are supernovae that show evidence for larger mass loss\index{mass loss} rates occurring somewhat before the supernova explosion, in particular Type IIn SNe (see Sect. \ref{sec_iin} and \cite{Smith2014}).
The driving mechanism for the mass loss is not understood, nor is the reason why it should occur close to the time
of the explosion.
Such strong mass loss rates are observed during the outbursts of Luminous Blue Variables (LBVs) and they are
frequently mentioned in this context.
Other possibilities are turbulence in the late phases of stellar evolution and/or binary interaction \cite{Chevalier2012,Shiode2014,Smith2014}.

If the mass loss parameters stay approximately constant 
 leading up to the explosion, the circumstellar density
is given by
\begin{equation}
\rho_{\rm cs}=\frac{\dot M}{4\pi u_{\rm w} r^2}.
\label{eq1b}
\end{equation}
In most cases it is the CSM density which is most observationally relevant  and can be measured. It is therefore convenient to scale this by the mass loss rate and wind velocity to that typical of a red supergiant. We therefore introduce a mass loss rate parameter $C_*$ defined by
\begin{equation} 
{\dot M \over u_{\rm w}} =  6.303\times 10^{14} \left({\dot M \over 10^{-5} \Msunyr}\right) \left({u_{\rm w} \over 10 \kms}\right)^{-1} \equiv 6.303\times 10^{14}  C_* \ \rm g \ cm^{-1}
\label{eq107}
\end{equation}
The CS density is then 
\begin{equation}
\rho_{\rm cs}=  5.016\times 10^{-17} C_* \left({r \over 10^{15} \ {\rm cm}}\right)^{-2} \ \rm g~ cm^{-3}.
\label{eq1bb}
\end{equation}
 Note that $C_*$ varies by a large factor for different types of progenitors. For a red supergiant with $\Mdot =10^{-5} \  \Msunyr$ and $u_{\rm w} \approx 10 \kms$ $C_* \sim 1$, while for a fast wind, like that from a WR star or blue supergiant with $u_{\rm w} \approx 1000 \kms$,  $C_* \sim 0.01$ for the same mass loss rate. An LBV with $\Mdot = 0.1 \  \Msunyr$ and $u_{\rm w} \approx 100 \kms$ has $C_* \sim 10^3$.
 
 Most of our discussion in the following  will be based on the assumption of a spherical geometry. There is, however, strong evidence that both the ejecta and the CSM may be very complex. An example of a both anisotropic and clumpy CSM medium is that of the red supergiant VY CMa \cite{Smith2009}, while the famous Eta Car nebula has a more regular, bipolar structure, with a dense shell containing $\sim 10 \ \Msun$ of mainly molecular gas \cite{Smith2006}. Both anisotropies and clumpiness can have strong observational consequences for the CSM interaction with the SN ejecta.  

\section{Shock interaction}
\label{sec:3}
\index{shock structure}
\subsection{Shock structure and evolution}

When the radiation dominated
shock front in a supernova nears the stellar surface,
a radiative precursor to the shock forms when the radiative
diffusion time is comparable to the propagation 
time.   There is radiative
acceleration of the gas and
the shock disappears when optical depth $\sim c/v$ is reached \cite{Ensman1992}.
The fact that the velocity decreases with radius implies that
the shock will re-form as a viscous shock in the circumstellar
wind.
This occurs when the supernova has approximately doubled in radius.

The interaction of the ejecta, expanding with velocity $\gsim 10^4 \kms$, 
and the
nearly stationary circumstellar medium results in a reverse shock\index{reverse shock} wave  
propagating inwards (in mass),
and an outgoing circumstellar shock. The
density in the circumstellar gas is given by Eq. (\ref{eq1b}).
 As discussed above, hydrodynamical calculations show that to a good 
 approximation
the  ejecta density can be described by Eq. (\ref{eq1a}).
A useful self-similar solution \index{similarity solution} for the interaction
can then be found \cite{Chevalier1982a,Chevalier1982b,Nadezhin1985}. 
Here we  sketch a simple derivation. More details can be found in
these papers, as well as in the review \cite{Chevalier1990}. 

Assume that the shocked
gas can be treated as a thin shell with mass $M_{\rm s}$, 
velocity $V_{\rm s}$, and  radius $R_{\rm s}$. 
Balancing the ram pressure from the circumstellar   gas and the
impacting ejecta, the momentum equation for the shocked shell of 
circumstellar gas and
ejecta is
\begin{equation}   
M_{\rm s}{dV_{\rm s} \over dt}
= 4 \pi R_{\rm s}^2[\rho_{\rm ej} (V_{\rm ej}  - V_{\rm s})^2 -
 \rho_{\rm cs}  V_{\rm s}^2] .
\label{eq2}
\end{equation}  
Here, $M_{\rm s}$ is the sum of the mass of the shocked ejecta and circumstellar   gas. The swept up
mass behind the circumstellar shock is $M_{\rm cs}=\Mdot R_{\rm s}/u_{\rm w}$, and that behind the
reverse  shock $M_{\rm rev}=4 \pi \int_{R_{\rm s}}^\infty \rho(r) r^2 dr = 4 \pi ~t_o^3 V_o^n (t/R_{\rm s})^{n-3}/(n-3)$, assuming that $R_{\rm s} >>
R_{\rm p}$, the radius of the progenitor. With $V_{\rm ej}  =R_{\rm s}/t$ we 
obtain  
\begin{eqnarray}
&&\left[{\Mdot \over u_{\rm w}} R_{\rm s} 
+ {4 \pi ~\rho_o~t_o^3 ~V_o^n~
t^{n-3} \over (n-3) ~R_{\rm s}^{n-3}} \right] {d^2 R_{\rm s} \over dt^2} = \nonumber \\
&&4 \pi R_{\rm s}^2 
\left[ 
{\rho_o ~t_o^3 ~V_o^n~t^{n-3} \over R_{\rm s}^n} 
\left({R_{\rm s} \over t} - {dR_{\rm s} \over dt}\right)^2 
- {\Mdot \over 4 \pi~u_{\rm w} R_{\rm s}^2}  \left({dR_{\rm s} \over
dt}\right)^2 \right].  
\label{eq3}
\end{eqnarray}
This equation has the power law solution
\begin{equation}
R_{\rm s}(t) = \left[{8 \pi \rho_o~t_o^3 ~V_o^n~u_{\rm w} \over (n-4) (n-3)
 ~\Mdot}\right]^{1/(n-2)} ~t^{(n-3)/(n-2)} .
\label{eq4}
\end{equation}
The form of this similarity solution can be written down directly by
dimensional analysis from
the only two independent quantities available, $\rho_o~t_o^3 ~V_o^n$ and
$\Mdot/u_{\rm w}$.  The solution applies after a few expansion times, when
the initial radius has been `forgotten.'  
The requirement of a finite energy in the flow implies $n > 5$.
More accurate
similarity solutions, taking  the structure within the shell into
account, are given in \cite{Chevalier1982a,Nadezhin1985}. In general,
these solutions differ by less than $\sim 30 \%$ from the thin shell
approximation. 
 
 \subsection{Postshock conditions and radiative cooling}
 \label{sec_pscond}
 
\begin{figure}[!t]
\begin{center}
\includegraphics[width=11cm]{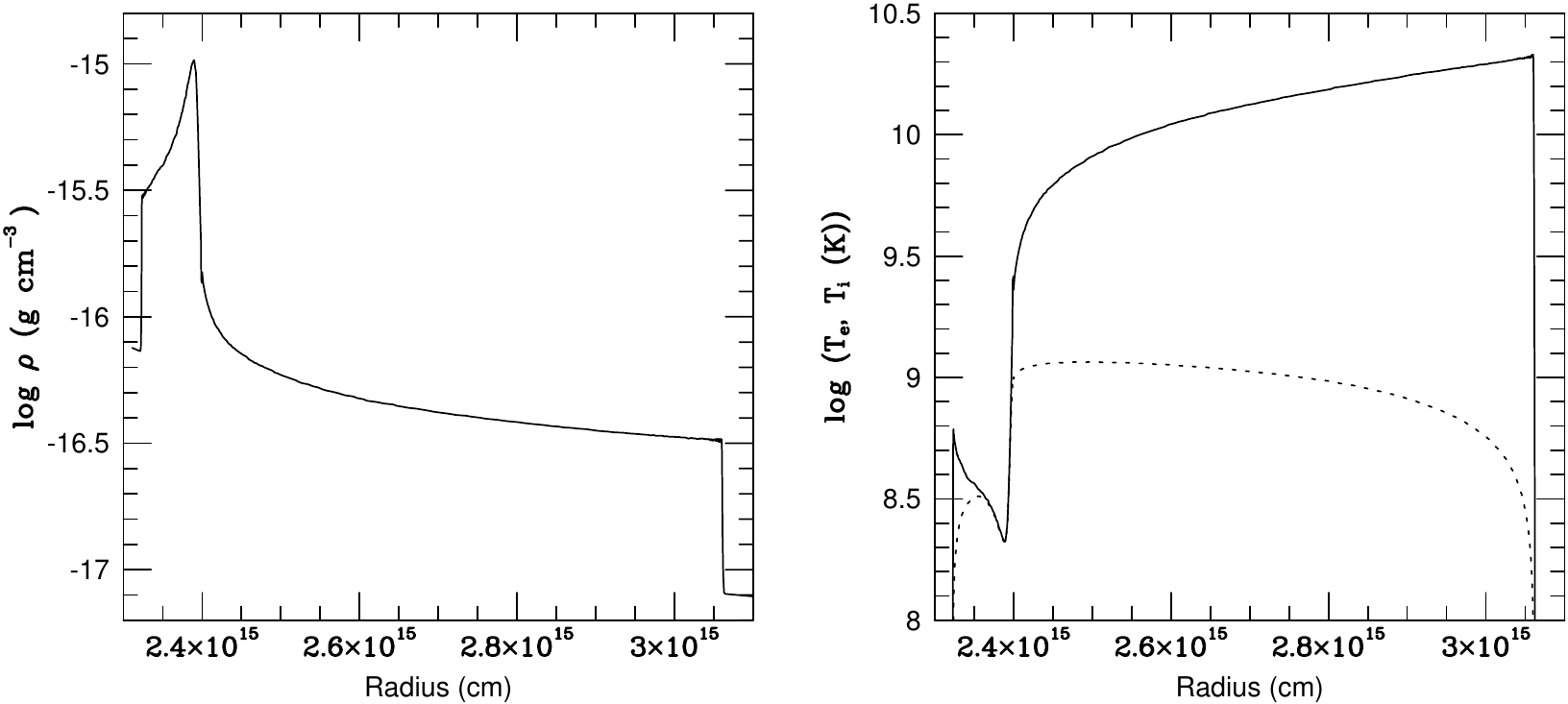}
\end{center}
\caption[]{Density and temperature structure of the reverse and circumstellar shocks for
$n=7$ and a velocity of $2.5\EE4 \kms$ at 10 days. Both shocks are assumed 
to be adiabatic. Because of the 
slow Coulomb equipartition\index{electron ion equipartition} the 
electron temperature (dotted line) is much lower than the ion temperature
 (solid line)
 behind the circumstellar shock.}  
\label{fig1}
\end{figure}

The maximum ejecta velocity close to the reverse shock depends on
time as $V_{\rm ej} = R_{\rm s}/t \propto t^{-1/(n-2)}$. 
 The velocity of the circumstellar shock is  
 \begin{equation}
V_{\rm s}(t) = {dR_{\rm s}(t) \over dt}= {(n-3) \over (n-2)} {R_{\rm s}(t) \over t}  = {(n-3)\over (n-2) } V_{\rm ej}  \propto t^{-1/(n-2)} \ .
\label{eq4b}
\end{equation}
with $R_{\rm s}(t)$ given by Eq. (\ref{eq4}), while the reverse shock velocity is 
\begin{equation}
V_{\rm rev}=  V_{\rm ej} - V_{\rm s} = {V_{\rm ej} \over (n-2)} \ .
\label{eq4c}
\end{equation} 
Assuming cosmic abundances and equipartition between ions and electrons,
the  temperature of the shocked circumstellar gas is
\begin{equation}
T_{\rm cs} =
1.36\EE9 ~Ê\left({n-3\over n-2}\right)^2 \left(V_{\rm ej}\over 10^4 \kms \right)^2 
 ~\rm K
\label{eq5}
\end{equation}
and at the reverse shock\index{reverse shock}
\begin{equation}
T_{\rm rev}= {T_{\rm cs}\over (n-3)^2}.
\label{eq6}
\end{equation}
\index{electron ion equipartition} The time scale for
equipartition between electrons and ions\index{electron ion equipartition} is 
\begin{equation}
t_{\rm eq} \approx 2.5\EE7 ~ \left({T_{\rm e} \over 10^9
\rm ~ K}\right)^{1.5}~  \left({n_{\rm e} \over 10^7 {\rm~cm}^{-3}}\right)^{-1} \rm~ s.
\label{eq6b}
\end{equation}
One finds that the
reverse shock is  marginally in equipartition, unless the temperature is $\gsim 5\EE8$
K. The ion temperature behind the
circumstellar shock is $\gsim 6 \EE{9}$ K for $V_4 \gsim 1.5$, and the
density a factor $\gsim 4$ lower than behind the reverse shock. 
Ion-electron collisions are therefore ineffective, and $T_{\rm e} <<
T_{\rm ion}$, unless efficient plasma instabilities heat the
electrons collisionlessly (Fig. \ref{fig1}).

For typical parameters,
the electron temperatures of the two shocks are
very different,  $\sim (1-3)\EE9 \KK$ for the circumstellar shock and
$10^7 - 5\EE8 \KK$ for the reverse shock, depending on $n$. 
The radiation from 
the reverse shock is 
mainly below $\sim 20$ keV, while that from the circumstellar shock 
is above $\sim 50
\rm~keV$.  

With $\rho_{\rm rev} / \rho_{\rm cs} = \rho_{\rm ej} / \rho_{\rm wind} = 4 \pi ~u_{\rm w} \rho_o  t_o^{3} V_o^{n} t^{n-3}  / (\Mdot R^{n-2})$ one gets for the  density behind the
reverse shock\index{reverse shock} 
\begin{equation}
\rho_{\rm rev}= 
{(n-4)(n-3)\over 2} 
 \rho_{\rm cs}  \ .
\label{eq7}
\end{equation}
Therefore the density behind the reverse shock is   much higher than behind the
circumstellar shock for $n \gsim 7$. 
There is a drop in density across the contact discontinuity, moving from
the shocked ejecta to the circumstellar medium (see Fig. \ref{fig1}).

\begin{figure}
\begin{center}
\includegraphics[width=9cm]{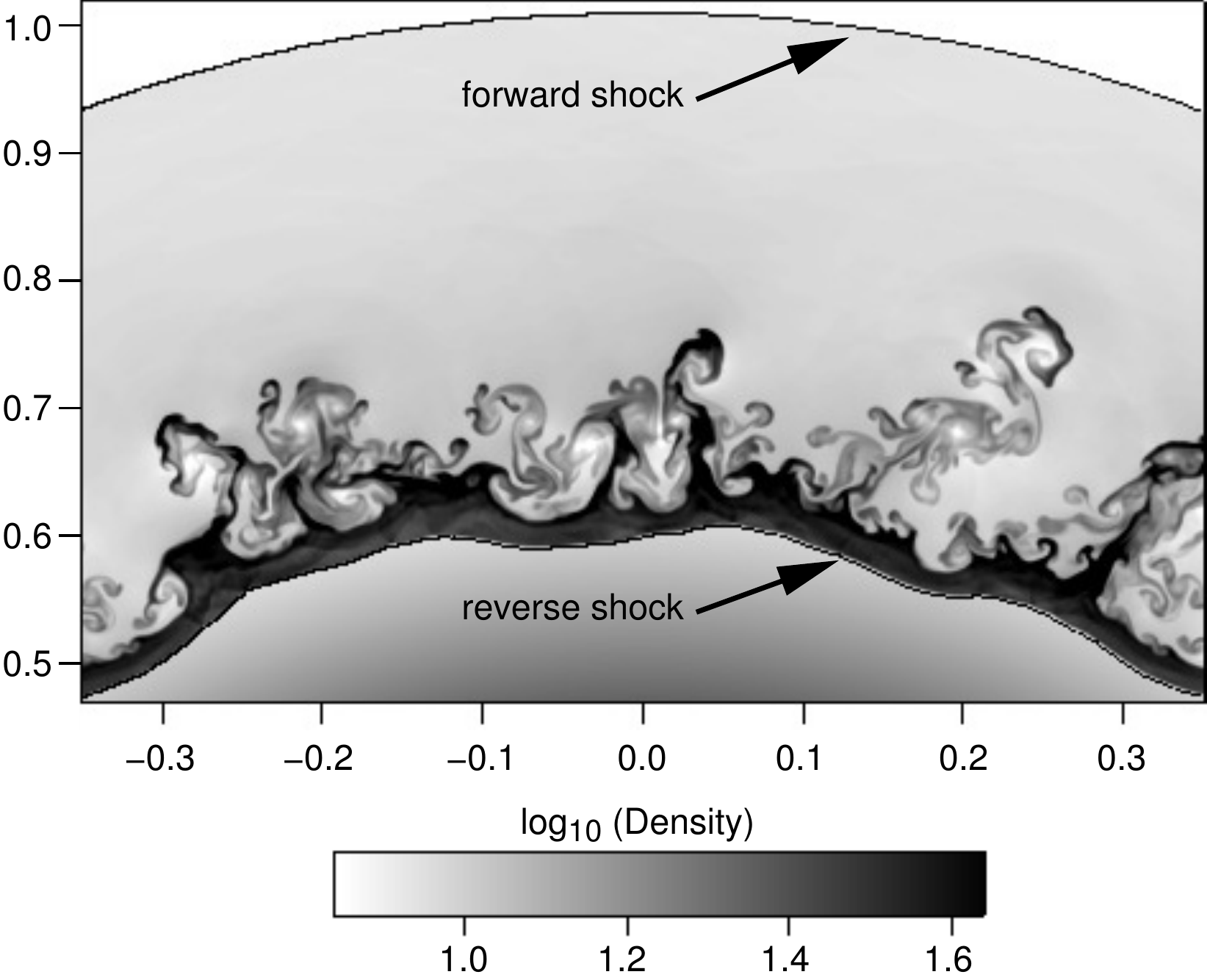}
\end{center}
\caption[]{Two-dimensional calculation of the shock structure for a
supernova with $n=6$ in a stellar wind. Cooling is not included here. (Courtesy John Blondin).}
\label{fig2}
\end{figure}

\index{Rayleigh-Taylor instability}The fact that low density gas is decelerating higher density gas
leads to a Rayleigh-Taylor instability.
Chevalier, Blondin \& Emmering \cite{Chevalier1992} have calculated the
structure using a two-dimensional PPM (piecewise parabolic method)
hydrodynamic code. They indeed find that
instabilities develop, with dense, shocked ejecta gas penetrating into
the hotter, low density shocked circumstellar gas
(Fig. \ref{fig2}). The instability mainly distorts the contact surface,
and does not seriously affect the general dynamics.  The calculation
assumed that cooling is not important. If the gas at the
reverse shock cools efficiently,
the extent of the instability is similar, although the 
Rayleigh-Taylor fingers are narrower \cite{Chevalier1995}.

\index{cooling shock} The above expressions for the density and temperature apply to the gas immediately behind the reverse shock. However,  radiative cooling  may be important in many cases  resulting in lower temperature, higher densities and higher luminosity of the gas.  
For  $T_{\rm e} \gsim 2.6\EE7 \KK$ free-free cooling dominates with a cooling rate $\Lambda \approx 1.0\EE{-23}~(T_{\rm e}/10^7 \ {\rm K})^{0.5} \ergs cm^{3}$. Assuming isobaric cooling behind the shock, one obtains for the cooling time $t_{\rm cool} =5 kT_{\rm e}/n_{\rm H}\Lambda(T_{\rm e})$
 \begin{equation}
t_{\rm cool} =  {13.5 \over (n-2) (n-3) (n-4)}~C_*^{-1}~\left({V_{\rm ej} \over 10^4 \kms}\right)^{3}  \left({t\over {\rm days}}\right)^2~~{\rm days},
\label{eq11a}
\end{equation}
At $T_{\rm e} \lsim 2.6\EE7 \KK$, line emission increases the cooling rate and
$\Lambda(T_{\rm e}) \approx 2.4\EE{-23}~(T_{\rm e}/10^7 \ {\rm K})^{-0.48} \ \ergs cm^{3}$ for solar abundances. The cooling time then becomes
\begin{equation}
t_{\rm cool} =  {1.0 \times 10^3 \over  (n-3) (n-4) (n -
2)^{3.34}}~C_*^{-1}~\left({V_{\rm ej} \over 10^4 \kms}\right)^{5.34} ~ \left({t\over {\rm days}}\right)^2~~{\rm days},
\label{eq11ab}
\end{equation}
again assuming solar abundances \cite{Fransson1996}. From these expressions it is clear
that the cooling time is very sensitive to the density gradient, as
well as the shock velocity and mass loss rate. 
 If the
temperature of the reverse shock\index{reverse shock} falls below $\sim 2\EE7$ K, a thermal
instability may occur and the gas cools to $\lsim 10^4 \KK$, where
photoelectric heating from the shocks balances the cooling \cite{Fransson1984,Chevalier1985}. Because of its low temperature and high density this is usually referred as  the cool, dense shell\index{cool, dense shell}, and can have important observational consequences. In case the reverse shock has propagated into the processed material, as may occur comparatively early in Type IIb and Ib/c SNe, the cooling may be further enhanced due to the higher metallicity. The above expressions should then be modified as discussed in \cite{Nymark2006}, resulting in a prolonged radiative phase for the reverse shock.

SNe with high mass loss
rates, $C_* \gsim 5$, generally have radiative reverse shocks
for $\gsim 100$ days, 
while SNe with lower mass loss rates, like the Type IIP SNe, have adiabatic shocks from early times. 

 \index{cooling shock} If cooling, the total energy emitted from the reverse shock is \cite{Fransson1984}
\begin{eqnarray} 
L_{\rm rev} &=& 4 \pi R_{\rm s}^2 ~{1 \over 2}~ \rho_{\rm ej} V_{\rm rev}^3 = {(n-3)(n-4)\over 4
(n-2)^3}~Ê{\Mdot  V^3\over u_{\rm w} } \cr
&=& 1.6\EE{41} ~ {(n-3)(n-4)\over
(n-2)^3} ~C_* V_4^3 ~ \ergs.
\label{eq12}
\end{eqnarray}
Most of this is emitted as soft X-rays. Because of the large column density of the cool, dense shell most of this radiation will, however, be thermalized into UV and optical emission. For high $\Mdot/u_{\rm w}$ the
luminosity from the reverse shock may therefore contribute appreciably, or even dominate, the
bolometric luminosity. 

Because $V \propto t^{-1/(n-2)}$, $L_{\rm rev} \propto t^{-3/(n-2)}$ in
the cooling case. 
However, one of the most important effects of the cooling is that the cool gas may
absorb most of the emission from the reverse shock. Therefore, in
spite of the higher intrinsic luminosity of the reverse shock, little
of this will be directly observable. The column density of the cool gas
is given by $N_{\rm cool} = M_{\rm rev}/(4 \pi R_{\rm s}^2 m_{\rm p})$, or
\begin{equation} 
N_{\rm cool} 
\approx 1.0\EE{21} (n-4)  ~~C_*~\left(V_{\rm ej}\over 10^4 \kms \right)^{-1} \left({t \over  
100 \rm ~days}\right)^{-1}~\rm cm^{-2}.
\label{eq11b}
\end{equation}
\index{photoelectric absorption} Because the threshold energy due to photoelectric absorption is
related to $N_{\rm cool}$ by $E(\tau=1)=1.2 (N_{\rm cool}/10^{22} ~\rm
cm^{-2})^{3/8}$ keV, it is clear that the emission from the reverse
shock\index{reverse shock} is strongly affected by the cool shell, and a transition from
optically thick to optically thin is expected during the first months,
or year. Examples of this are discussed in connection to SN 1993J in Sect. \ref{sec_iib} and SN 2010jl in Sect. \ref{sec_iin}.

Because of the high density in combination with low temperature the cool, dense shell may also be a site of dust formation. This, however, requires a temperature of $\lsim 2000$ K. To balance the heating by the X-rays from the reverse shock cooling has to be very efficient and may require enrichment by heavy elements. Observationally there is evidence for dust formation\index{dust} in the cool, dense shell\index{cool, dense shell} from \index{SN 1998S} SN 1998S, where an infrared (IR) excess as well as a fading of the red side of the line profiles developed after about one year \cite{Pozzo2004}. This became stronger with time at the same time as the dust temperature decreased. Dust condensation was also consistent with the different line profiles of the Ly$\alpha$ and H$\alpha$ lines, where the former developed a fading at earlier epochs compared to the latter \cite{Fransson2005}. Also the Type Ibn SN 2006jc had a similar evolution \cite{Smith2008,Mattila2008} and is a good case for dust formation. In this case the reverse shock may already be in the metal enriched regions, promoting the dust formation.

The considerations so far apply to stars with mass loss rates $C_* \lsim 10$, which includes most stars, including red supergiants with dense winds.
At higher mass loss rates, like the ones relevant to LBV progenitors and Type IIn SNe, the observed X-ray emission from the forward shock\index{forward shock}  comes to dominate
the emission from the reverse shock.
One reason is the absorption effect of the dense cool shell described above; another is that 
once the reverse shock becomes radiative, the luminosity rises proportional to density while the luminosity
from the forward shock continues to grow as density squared.
The emission from the forward shock is expected to be primarily free-free emission and the cooling time is
\begin{equation}
t_{\rm cool} =  {3.5   (n-3) \over (n-2) }~C_*^{-1}~\left({V_{\rm ej} \over 10^4 \kms}\right)^{3} ~ \left({t\over {\rm days}}\right)^2~~{\rm days}. 
\label{eqfscool}
\end{equation}
The  cooling time is therefore long except in cases of high mass loss density and slow shock wave; these properties are expected to go together.

At early epochs  \index{Compton cooling} Compton cooling may also be important with a cooling rate 
\begin{equation} 
{dE\over dt}= { 4 k T_{\rm e}  \sigma_{\rm T} \over m_{\rm e} c^2} {L(t) \over 4 \pi r^2} 
\label{eq112}
\end{equation}
where $L(t)$ is the luminosity of the SN. The cooling time then becomes 
\begin{equation} 
t_{\rm cool}= 1.67  \left({V_{\rm ej} \over 10^4 \kms}\right)^2  \left({t\over {\rm days}}\right)^2 \left({L\over 10^{42} \ergs}\right)^{-1}  \ \rm days.
\label{eq112b}
\end{equation}
which may be short for slow shocks and when the SN luminosity is high.

If the \index{cooling shock} forward shock is cooling, the power emitted by the forward shock is
\begin{eqnarray} 
L_{\rm for} &=& 4 \pi R_{\rm s}^2 ~{1 \over 2}~ \rho_{\rm cs}{ V_{\rm s}^3 }= {(n-3)^3 \over 2
(n-2)^3}~Ê{\Mdot  V_{\rm ej}^3\over u_{\rm w} } \cr
&=& 3.2\EE{41} ~ {(n-3)^3\over
(n-2)^3} ~ C_* \left({V_{\rm ej} \over 10^4 \kms}\right)^3 ~ \ergs.
\label{eqforlum}
\end{eqnarray}
The postshock gas temperature for the forward shock case is higher than for the reverse shock, giving an indication of the
type of shock that is dominating the emission.
These considerations can be applied to the case of SN 2010jl\index{SN 2010jl} where the inference of $C_* \approx 10^3$
came from the bolometric light curve, which was dominated by optical radiation  \cite{Fransson2014}.
Application of Eq. (\ref{eqfscool}) for $n=8$ gives $t_{\rm cool}=0.35  V_{\rm ej, 4}^{-3}    C_* $ days, so with $C_* \approx 10^3$ and $V_{\rm s}=4000\kms$ gives  $\sim 5400$ days for
the duration of the radiative forward shock phase.
The evolution of the luminosity in Eq. (\ref{eqforlum}) is determined by the value of $n$, which relates to the deceleration of the interaction shell;
$n=7.6$, a reasonable value for the ejecta profile, is indicated for the first 300 days of evolution \cite{Fransson2014}

To relate the shock properties to the supernova model, we make the approximation that the supernova is described by an outer
region with $n=7$ and an inner region with a constant density; the two regions are separated at a transition velocity, $V_t$.
The inner density profile is $\rho= A t^{-3}$ and the outer profile is $\rho=A t^{-3}(V/V_t)^{-7}$  where $A$ and $V_t$ are constants
to be determined from the mass and energy: $V_t=2.9\EE{3}E_{51}^{1/2}M_1^{-1/2}\kms$ and $A=1.13\EE{8}E_{51}^{-3/2}M_1^{5/2}$ g s$^3$ cm$^{-3}$,
where $E_{51}$ is in units of $10^{51}$ ergs and $M_1$ in units of $10~\Msun$.
Substituting the outer density profile parameters in Eq. (\ref{eq4}) gives
\begin{equation}
V=\frac {R_{\rm s}}{t}=1.48 \times 10^{4} E_{51}^{2/5}M_1^{-1/5}  C_*^{-1/5}t_{100}^{-1.5}\kms, 
\end{equation}
where $t_{100}$ is in units of 100 days.
For $C_*=10^3$, as inferred for SN 2010jl, the velocity coefficient is $3730\kms$.
A somewhat higher energy or lower mass would bring the velocity into agreement with the value of $V_{\rm s} =4000\kms$ from X-ray observations
\cite{Ofek2014,Chandra2015}.
With these parameters, the interaction observed in SN 2010jl\index{SN 2010jl} for the first 300 days occurs in the outer, steep density power law region of the supernova; after 300 days, the more rapid decline can be attributed to a more rapidly decreasing density profile.
Different parameters could give a situation in which the reverse shock wave has moved in to the flat part of the ejecta density distribution,
which would also give a more rapid decline \cite{Ofek2014}.

\subsection{Asymmetries, shells and clumping}
\label{sec_asymm}

\index{asymmetries} \index{circumstellar clumps}In view of the evidence for dense equatorial winds from red supergiant
stars, Blondin, Lundqvist, \& Chevalier \cite{Blondin1996} simulated the interaction
of a supernova with such a wind.
They found that for relatively small values of the angular density
gradient, the asymmetry in the interaction shell is greater than,
but close to, that expected from purely radial motion.
If there is an especially low density close to the pole, the flow
qualitatively changes and a protrusion emerges along the axis,
extending to $2-4$ times the radius of the main shell.
Protrusions have been observed in the probable supernova remnant
41.9+58 in M82, although the nature of the explosion is not
clear in this case \cite{McDonald2001}.

In addition to asymmetric winds, there is evidence for supernova
shock waves interacting with clumps of gas in the wind, as have
been observed in some red supergiant winds.
In some cases the clumps can be observed by their very narrow
lines in supernova spectra, as in Type IIn supernovae.
The velocity of a shock wave driven into a clump, $V_{\rm c}$, can be estimated
by approximate pressure balance 
\begin{equation}
V_{\rm c} \approx V_{\rm s} \left({\rho_{\rm s} \over \rho_{\rm c}} \right)^{1/2} \ ,
\label{eq_vc}
\end{equation}
where $V_{\rm s}$ is the shock velocity in the smooth wind with density
$\rho_{\rm s}$ and $\rho_{\rm c}$ is the clump\index{circumstellar clumps} density.
The lower shock velocity and higher density can lead to radiative
cooling of the clump shock although the main shock is non-radiative.
Optical line emission of intermediate velocity observed in Type IIn
(narrow line) supernovae like SN 1978K, SN 1988Z, and SN 1995N \index{SN 1978K} \index{SN 1995N}
can be explained in this way \cite{Chugai1994,Chugai1995,Fransson2002}.  \index{SN 1988Z}

The presence of many clumps can affect the hydrodynamics of the interaction.
Jun, Jones, \& Norman \cite{Jun1996} found that propagation in a region
with clumps gives rise to widespread turbulence in the shocked region
between the forward shock and the reverse shock, whereas the
turbulence is confined to a region near the reverse shock for the
non-clump case (Fig. \ref{fig2}).
Their simulations are for interaction with a constant density medium,
but the same probably holds true for interaction with a
circumstellar wind.

The interaction between the winds\index{interacting winds} at the different evolutionary stages of the progenitor may result in dense shells around the SN. This may in particular be the case for a fast wind sweeping up a slower one. This may occur for a fast Wolf-Rayet wind sweeping up the slow red supergiant wind, or the wind from a blue supergiant sweeping up a slow wind, as may have been the case for SN 1987A \cite{Blondin1993}.

The interaction of the SN ejecta with a dense shell depends on the ratio of the densities of the shell and the SN ejecta in the same way as was discussed for the clump interaction above (Eq. \ref{eq_vc}). The shells resulting from the interaction may in fact break up into clumps because of instabilities, as may have produced the clumps we now see as hotspots in SN 1987A.

For a high shell density a slow shock will be transmitted into the shell, while a reflected shock will propagate back into the SN ejecta and set up a reverse shock. The dynamics of this has been discussed in \cite{Chevalier1989b} where it is found that the ratio of the radius of the reverse shock\index{reverse shock} to the contact discontinuity between shocked shell and shocked ejecta is given by 
\begin{equation}
{R_{\rm rev}\over R_{\rm c}}=\left({4n-20\over 4n-15}\right)^{1/3}
\end{equation}
For $n=9$ one gets $R_{\rm rev}/ R_{\rm c}=0.91$. 

\section{Thermal emission from hot gas}
\subsection{Optically thin emission}
The optical depth to
electron scattering\index{electron scattering} behind the circumstellar shock is  
\begin{equation}  
\tau_{\rm e} = 0.18 ~ C_* ~V_4^{-1}~Êt_{\rm days}^{-1}.
\label{eq8}
\end{equation}  
For mass loss rates $\lsim 10^{-4} \Msunyr$ and shock velocities $\gsim 10^4 \kms$ the CSM will be optically thin and the radiation from the shocks will be only mildly affected. We therefore treat this case first. In Sect. \ref{sec_abs_scatt} we discuss the optically thick case. 
 
If the temperature in the post-shock gas is $\gsim 2\times10^7$ K free-free emission dominates the cooling and one can estimate the luminosity from the circumstellar and
reverse shocks from
\begin{equation} 
L_{\rm i} = 4 \pi
\int \Lambda_{\rm ff}(T_{\rm e}) n_{\rm e}^2 r^2 dr \approx \Lambda_{\rm ff}
(T_{\rm i}) {M_{\rm i}
\rho_{\rm i}\over (\mu_{\rm e} m_H)^2 }.
\label{eq10}
\end{equation} 
where the index $i$ refers to quantities connected  either with the reverse shock\index{reverse shock} or circumstellar shock\index{forward shock}.  The density behind the circumstellar shock is $\rho_{\rm cs} = 
  4 ~\rho_0 = \Mdot /
(\pi u_{\rm w} R_{\rm s}^2) $, while that behind the reverse shock\index{reverse shock} is given by Eq. (\ref{eq7}).  The swept up mass behind the circumstellar 
shock
is $M_{\rm cs}=\Mdot R_{\rm s}/u_{\rm w}$ and that behind the reverse 
shock $M_{\rm rev}= (n-4) M_{\rm cs}/2$. With
$\Lambda_{\rm ff}= 2.4\EE{-27} \bar g_{\rm ff}~ T_{\rm e}^{0.5}$, we get 
\begin{equation} 
L_{\rm i} 
\approx 3.0\EE{39} ~\bar g_{\rm ff}~ C_{\rm n}  ~C_*^2 \left({t \over  
10 \rm ~days}\right)^{-1}~\ergs ,
\label{eq11}
\end{equation}  
where $\bar g_{\rm ff}$ is the free-free Gaunt factor, including relativistic
effects. For the reverse shock
$C_{\rm n} = (n-3) (n-4)^2 / 4 (n-2) $, and for the circumstellar shock $C_{\rm n}=1$.
This assumes electron-ion equipartition\index{electron ion equipartition}, which is highly questionable for
the circumstellar shock (see Fig. \ref{fig1}).  
Because of occultation by the optically thick ejecta only half of the
above luminosity escapes outward.  

\begin{figure}[t]
\begin{center}
\includegraphics[scale=0.6,angle=0,origin=c]{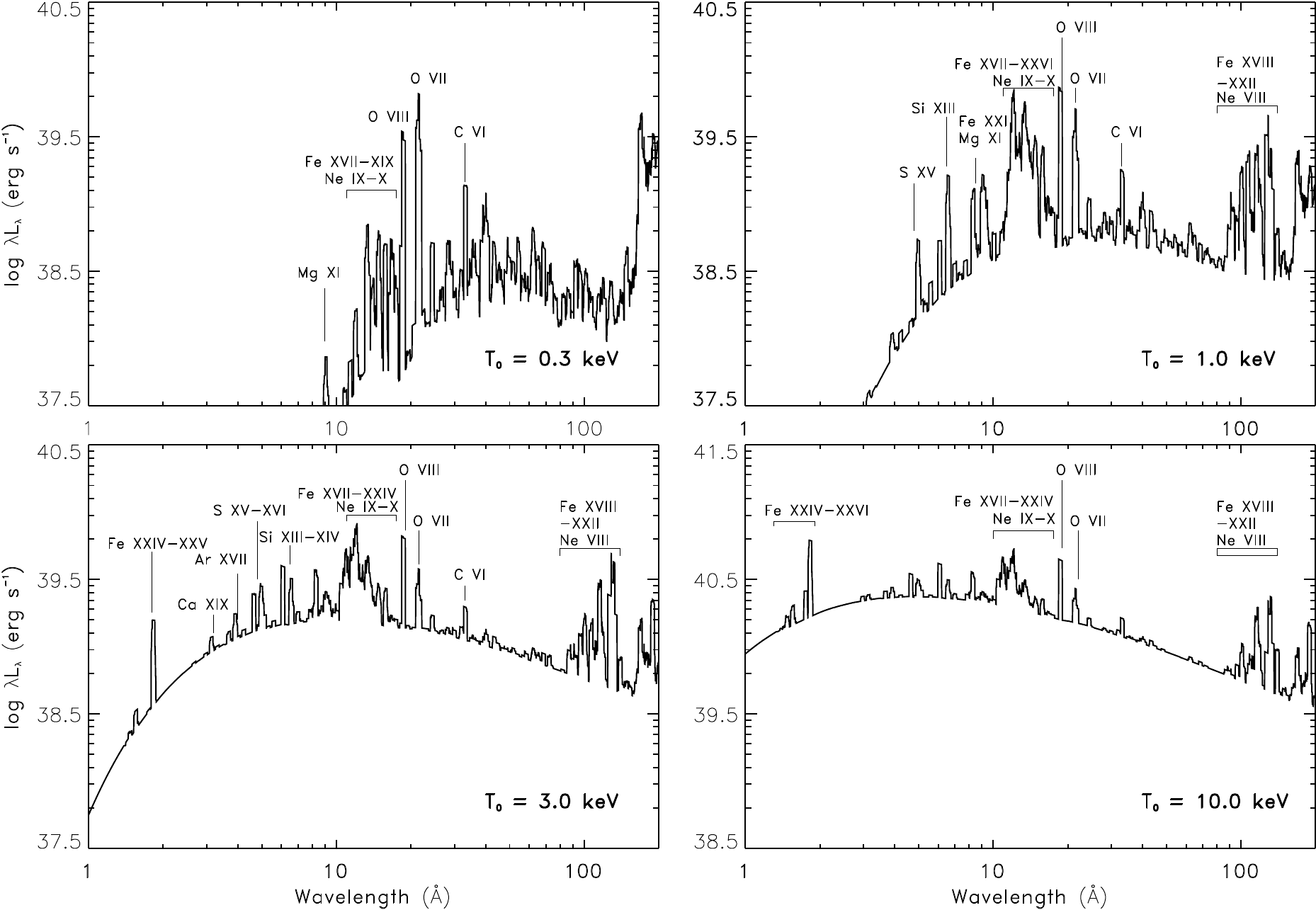}
\end{center}
\caption[]{Spectra, $\lambda L_{\lambda}$, of radiative shocks with different reverse shock temperatures,  $T_{0}=0.3,\ 1.0,\ 3.0,\ {\rm and}\ 10.0$ keV. The spectra are a convolution of spectra with decreasing temperature of the post-shock gas. All for solar composition. Note the increasing dominance of line emission for with decreasing temperature. From \cite{Nymark2006}. Reproduced with permission from Astronomy \& Astrophysics, \textcopyright  ~ESO.}
\label{fig3b}
\end{figure}

 \index{cooling shock} For temperatures $\lsim 2\times 10^7$ K  behind the reverse shock the spectrum is
dominated by line emission from metals (Fig. \ref{fig3b}). An important
point is that for a radiative shock the observed spectrum is formed in gas with widely
different temperatures, varying from the reverse shock temperature down to
$\sim 10^4$ K. A spectral analysis based on a one temperature model can be
therefore  misleading  \cite{Nymark2006}.

Another way of observing the hot gas is through the emission from
collisionally heated \index{dust} dust grains in the gas \cite{Dwek1987}.
Dust formation in the rapidly expanding ejecta is unlikely,
so the forward shock front must be considered.
Evaporation by the supernova radiation creates a dust-free zone
around the supernova.
The time for the supernova shock wave to reach the dust depends
on the supernova luminosity and the shock velocity; it is probably
at least several years.
The infrared luminosity\index{infrared luminosity} from dust can be up to $\sim 100$ times
the X-ray luminosity of the hot gas for typical parameters,
and the dust temperature is a measure of the density of the gas \cite{Dwek1987}.
If the X-ray emission from a supernova like
 SN 1986J is from circumstellar clumps
that are out in the region where dust is present, there is the
possibility of a large infrared luminosity. \index{SN 1986J}

\subsection{Optically thick emission}
\label{sec_abs_scatt}

\index{Comptonization} Equation (\ref{eq8}) shows that electron scattering \index{electron scattering} can be a factor if the circumstellar density is large,
e.g., the case of  $C_* =10^3$ for SN 2010jl.
The lower shock velocity expected at high density also brings up the optical depth.
The effect of significant Comptonization is a decrease in the energies of the most energetic photons.
At each scattering, the photon wavelength is decreased by about $h/m_{\rm e} c$, 
where $h$ is Planck's constant and $m_{\rm e}$ is the electron mass.
In each scattering the photon loses an energy $\Delta E \approx E^2/m_{\rm e} c^2$. Approximating this as a continuous process, this energy will  be lost in a time $\sim \lambda_{\rm mfp}/c$, where $\lambda_{\rm mfp}$ is the mean free path. Therefore a  photon with initial energy $E_0$ will after a time $t$ have an energy given by 
\begin{equation}
\int_0^t {dt' \over \lambda_{\rm mfp}/c} = \int_{E_0}^E {m_{\rm e} c^2 \over E^2} dE 
\end{equation}
or
\begin{equation}
 {c t \over \lambda_{\rm mfp}} =  m_{\rm e} c^2({1 \over E} - {1 \over E_0}) \approx {m_{\rm e} c^2 \over E}  \  \rm .
\end{equation}
In the time $t$ the number of scatterings, $N$,  is given by $t = N \lambda_{\rm mfp}/c \approx \tau_{\rm e}^2 \lambda_{\rm mfp}/c$. Therefore, 
\begin{equation}
 E \approx    {m_{\rm e} c^2 \over \tau_{\rm e}^2 }=\frac {511}{\tau_{\rm e}^2}  ~ \rm keV.
\end{equation}

A collisionless shock can form with $\tau_{\rm e}$ as high as $c/V_{\rm s}$, so there is the possibility of setting a strong
upper limit on the X-ray spectrum. 

The electron scattering opacity  for a medium with cosmic abundances and ionized H and He is 0.34 cm$^2$ g$^{-1}$,
while the photoabsorption\index{photoelectric absorption} opacity is 106 $E_{keV}^{-8/3}$ $(Z/Z_{\odot})$  cm$^2$ g$^{-1}$, where $Z$ is the metallicity.
The two opacity sources are comparable for $E\approx 10$ keV.
If $\tau_{\rm e}\approx 1$, photons near 10 keV can escape, but lower energy photons are absorbed.
This was the case in the early observations of SN 2010jl \cite{Chandra2015,Ofek2014} (see Section 6.4).

Photoabsorption does not occur if the surrounding circumstellar medium is completely ionized by the supernova emission.
If the supernova is in the cooling regime, a $10,000\kms$ shock wave is capable of completely ionizing the surrounding medium,
but a $5000\kms$  shock wave is not \cite{Chevalier2012b}. 
This is consistent with the absorption observed in SN 2010jl.
At lower densities, complete ionization is more easily achieved, which is consistent with the early soft X-ray emission
observed from SN 1993J \cite{Fransson1996}.

\subsection{Reprocessing of X-rays}

\subsubsection{Ionization of the circumstellar gas}
\label{sec_ioni}

\index{shock break-out.} The earliest form of circumstellar interaction occurs at shock
break-out. As the shock approaches the surface, radiation leaks out on
a time scale from seconds to hours, depending on the size of the progenitor. 
Matzner \& McKee \cite{Matzner1999} have found approximate solutions for the effective temperature and energy of the shock break-out for both convective and radiative envelopes, approximated as polytropes. For the convective case (applicable to red supergiants) they find 
\begin{equation}
T_{\rm max} \approx 5.6 \times 10^5 \left({R_* \over 500 \ \Rsun}\right)^{-0.54} \ \rm K 
\end{equation}
\begin{equation}
E \approx 1.7 \times 10^{48} \left({R_* \over 500 \ \Rsun}\right)^{1.74} \  \rm ergs, 
\end{equation}
while for the radiative case (blue supergiants, WR stars) they find 
\begin{equation}
T_{\rm max} \approx 1.6 \times 10^6 \left({R_* \over 50 \ \Rsun}\right)^{-0.48} \ \rm K 
\end{equation}
\begin{equation}
E \approx 7.6 \times 10^{46} \left({R_* \over 50 \ \Rsun}\right)^{1.68} \  \rm ergs. 
\end{equation} 
The color temperature, which determines the ionization, may differ from $T_{\rm max}$ because of scattering and other effects, but these expressions show the strong dependence on the radius of the progenitor, $R_*$, for both the temperature and total energy of the shock break-out. This is also the case for the duration of the burst, which is usually set by the light travel time over the progenitor, $\sim R_*/c$, which varies from seconds to hours for different types of progenitors \cite{Matzner1999}.  

This burst of EUV (extreme ultraviolet)
 and soft X-rays ionizes and heats the
circumstellar medium on a time scale of a few hours. In addition, the
momentum of the radiation may accelerate the circumstellar gas to a
high velocity.  Most of the emission at energies $\gsim 100$ eV is
emitted during the first few hours, and after 24 hours little ionizing
energy from the shocked ejecta remains.

\index{X-ray ionization} In addition to the initial burst, the X-ray emission from the shocks  ionizes and heats both the
circumstellar medium and the SN ejecta as it propagates through the CSM. Observationally, these
components are easily distinguished  by the different velocities. The
circumstellar component is expected to have velocities typical of the
progenitor winds, i.e., $\lsim 1000 \kms$, while  the ejecta
have considerably higher velocities. The density is likely to be of the
order of the wind density $10^5-10^7 \ccm$, or higher if clumping is
important.  The ionizing X-ray flux depends strongly on how much of
the flux from the reverse shock  can penetrate the cool shell, or in the case of extreme mass loss rates the CSM gas.  
The state of ionization in the circumstellar
gas is characterized by the value of the ionization parameter,
$\zeta=L_{\rm cs}/(r^2 n) = 10^2 (L_{\rm cs}/10^{40} {\ergs}) (r/10^{16} \rm~
cm)^{-2} (n / 10^6 \ccm)^{-1}$ \cite{Kallman1982}. The
comparatively high value of $\zeta \approx 10-10^3$ explains the
presence of narrow coronal lines\index{coronal lines} of [Fe V-XIV] seen in objects like the Type IIn SNe 1995N \index{SN 1995N}
\cite{Fransson2002}, 1998S \cite{Fassia2001}, 2006jd \cite{Stritzinger2012} and 2010jl\index{SN 2010jl} \cite{Fransson2014}.

The radiative effects of the soft X-ray burst were clearly seen from the
ring of SN 1987A\index{SN 1987A}, where a number of narrow emission lines from highly
ionized species, like C III-IV, N III-N V, and O III-IV were first seen in the UV
\cite{Fransson1989}. Later, a forest of lines came to dominate also the
optical spectrum (see e.g., \cite{Groningsson2008}). Imaging with HST (e.g., \cite{Jakobsen1991,Burrows1995}) showed
that the lines originated in the now famous circumstellar ring of SN
1987A at a distance of $\sim 200$ light days from the SN. The presence
of highly ionized gas implied that the gas must have been ionized and
heated by the radiation at shock break-out. 
Because of the finite light travel time across the ring, the
observed total emission from the ring is a convolution of the emission
at different epochs from the various part of the ring. Detailed
modeling \cite{Lundqvist1996} shows that while the ionization of the ring
occurs on the time scale of the soft X-ray burst, the gas recombines
and cools on a time scale of years, explaining the persistence of the
emission decades after the explosion. The observed line emission
provides sensitive diagnostics of both the properties of the soft
X-ray burst, and the density, temperature and abundances of the gas in
the ring. In particular, the radiation temperature must have reached
$\sim 10^6$ K, in agreement with the 
modeling of the shock break-out. The strong N III]  relative to the C III] and O III] lines indicated a strong N enrichment from CNO processing, as well as an enhanced He abundance. 

\index{radiative acceleration} The soft X-ray burst may also pre-accelerate the gas in front of the
shock. In the conservative case that Thompson scattering dominates,
the gas immediately in front of the shock will be accelerated to
\begin{equation}
V_{\rm acc} = 1.4\EE3~ \left({E \over 10^{48}~ {\rm
ergs}}\right)~
\left({V_{\rm s} \over 1\EE4 \kms}\right)^{-2} ~
 \left({t \over {\rm days }}\right)^{-2}~ \kms, ~
\label{eq16b}
\end{equation}
where
$E$ is the total radiative energy in the burst. 
If the gas is pre-accelerated, the line
widths are  expected to decrease with time. After about one
expansion time ($\sim R_{\rm p}/V$) the reverse and circumstellar shocks are
fully developed, and the radiation from these will dominate the
properties of the circumstellar gas. 

The increasing blue-shifts with time of the H$\alpha$  line in SN 2010jl\index{SN 2010jl}  has been proposed to arise as a result of radiative acceleration by the extremely large radiated energy ($\gsim 6 \times 10^{50}$ ergs) \cite{Fransson2014}, although part of this may also result from the dense shell \cite{Dessart2015}.

\subsubsection{Emission from the SN ejecta and cool, dense shell}
\label{sec_cds}

\begin{figure}[t]
\begin{minipage}[c]{0.5\textwidth}
\centering 
\includegraphics[scale=0.3]{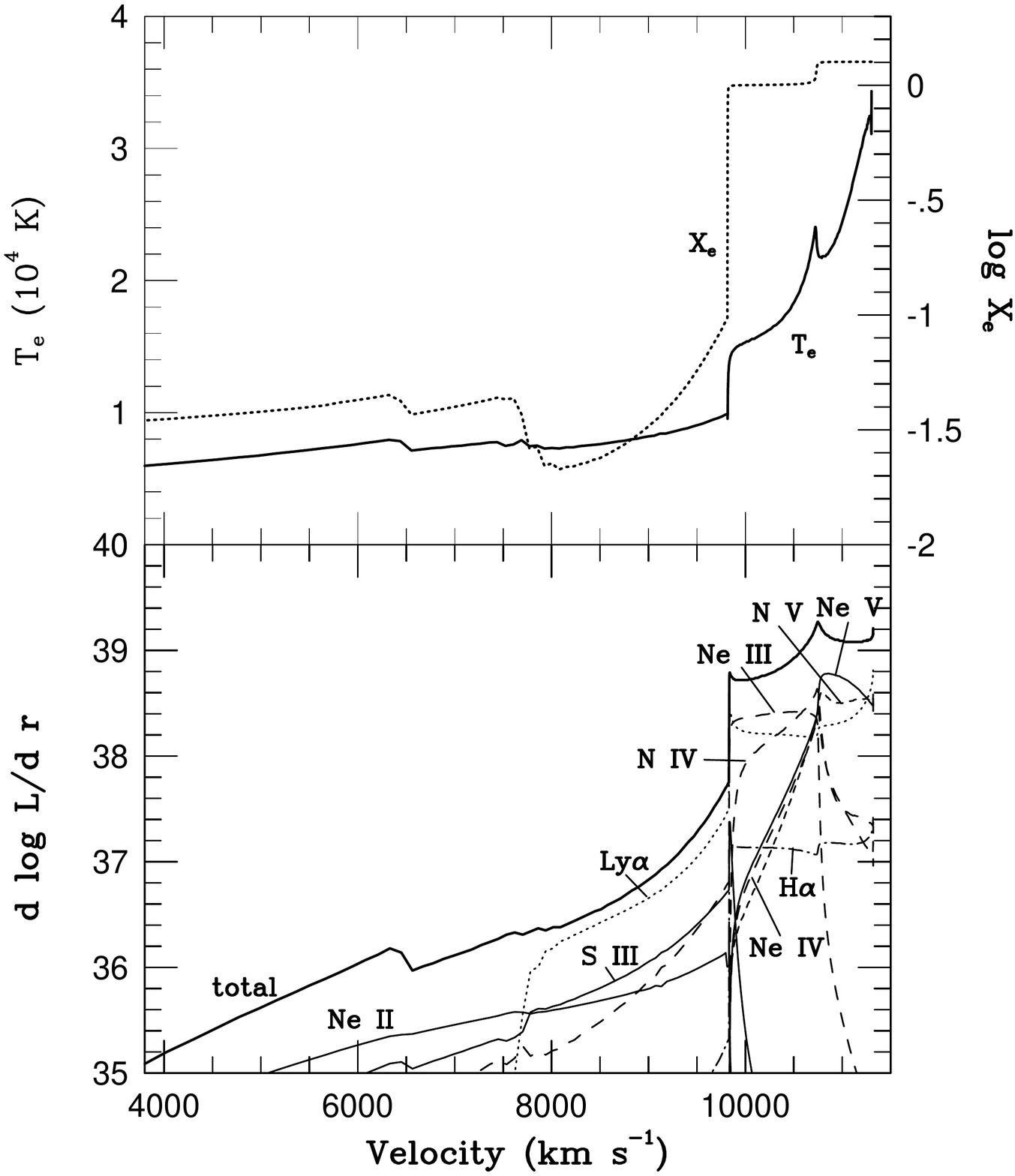}
\end{minipage}
\begin{minipage}[c]{0.5\textwidth}
\centering 
\includegraphics[scale=0.3]{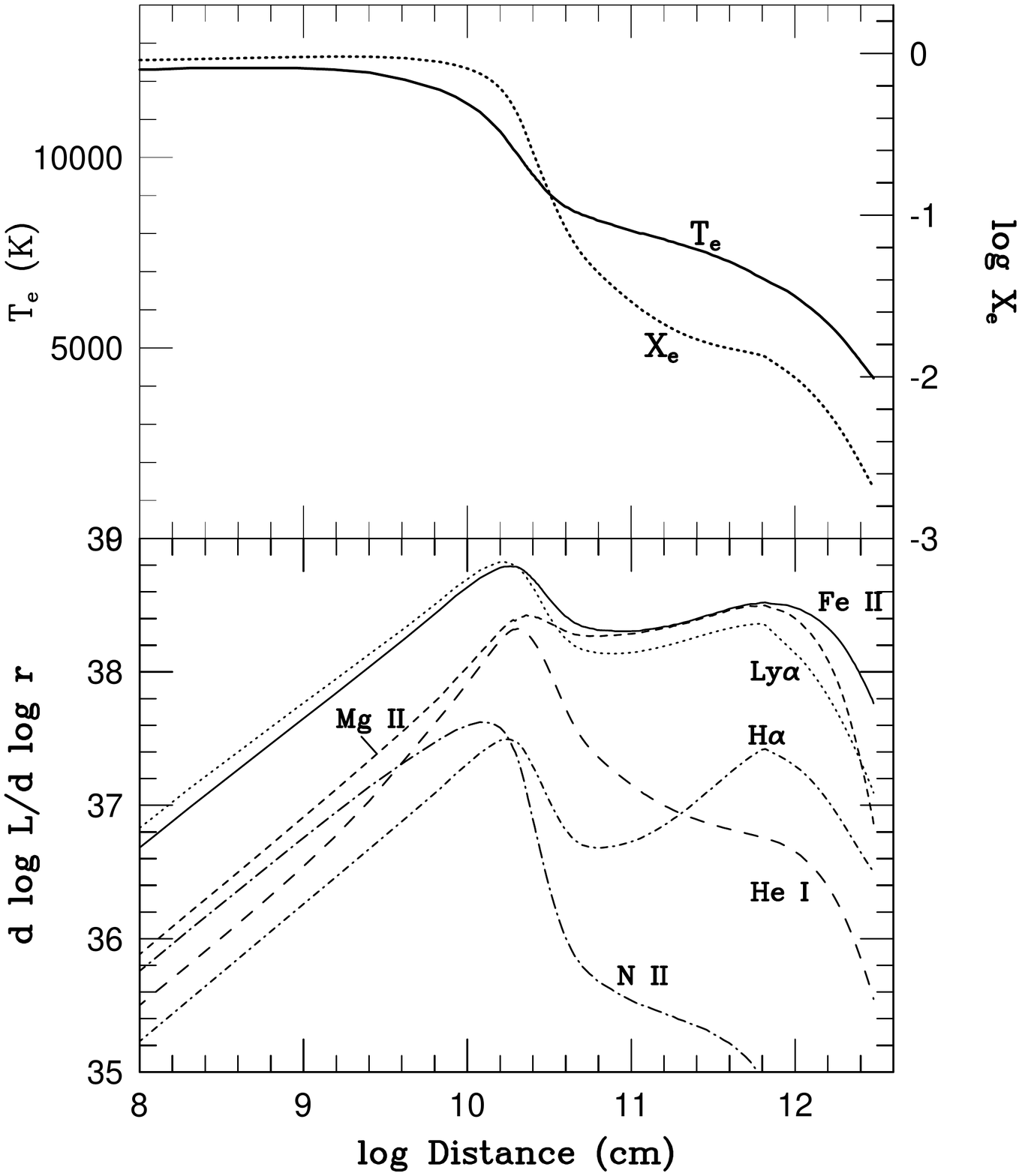}
\end{minipage}
\caption[]{Structure of the ejecta (left panels) and cool shell (right panels) ionized by the
reverse shock at 500 days for parameters appropriate to SN 1993J ($\Mdot = 5\EE{-5} \Msunyr$ for $u = 10 \kms$). Upper
panels show the temperature and ionization of the ejecta 
and the cool shell, while the lower panels show the
corresponding luminosities per unit distance. Note the different
length scales in the two panels. The ejecta region has low density,
high temperature and ionization, while the cool shell has a high
density, is extremely thin, has a low temperature, and is only
partially ionized.}
\label{fig5}
\end{figure}

 \index{cool, dense shell} The ingoing X-ray flux from the reverse shock  ionizes the outer
parts of the ejecta \cite{Fransson1982,Fransson1984}. The state of highest ionization  therefore is
close to the shock, with a gradually lower degree of ionization
inwards. Unless clumping in the ejecta is important, the ejecta
density is $\sim 10^6-10^8 \ccm$ during the first year. In the left panel of Fig.
\ref{fig5} we show temperature and ionization structure of the ejecta for a reverse shock velocity of $1300 \kms$,
as well as the emissivity of the most important lines. The ejecta velocity was $1.3\times 10^4 \kms$ and the ejecta gradient $n=12$. The temperature of the outer ejecta
 close to the shock is $\sim 3\EE4$ K, decreasing inwards in. Most of
the emission here is  emitted as UV lines of highly ionized ions, like
\La, C III-IV, N III-V, and O III-VI. Inside the ionized shell
there is an extended partially ionized zone, similar to that
present in the broad emission line regions of AGN's. Most of the
emission here comes from Balmer lines.

As we have already discussed, the outgoing flux from the reverse shock
is to a large extent absorbed by the cool shell between reverse shock
and the contact discontinuity if radiative cooling has
been important. The whole region behind the reverse
shock is in approximate pressure balance, and the density of this gas
is therefore a factor $\sim 4 T_{\rm rev}/T_{\rm cool} \approx 10^3-10^4$
higher than that of the ejecta. Because of the high density, the gas
is only partially ionized and the temperature only $(5-8)\EE3$ K. 
Most of the emission comes out as Balmer lines, Mg II and Fe II
lines (Fig. \ref{fig5}, right panel). The thickness of the emitting
region is also very small, $\sim 3\times 10^{12}$ cm. In one
dimensional models, the velocity is marginally smaller than the
highest ejecta velocities. Instabilities in the shock are, however,
likely to erase this difference.

An important diagnostic of the emission from the cool shell and the
ejecta is the \Ha~ line \cite{Chevalier1985}. This line arises as a result of recombination
and collisional excitation. In \cite{Chevalier1994}, it is estimated that $\sim 1 \%$ of
the reverse shock luminosity is emitted as \Ha, fairly independent of
density and other parameters. Observations of this line 
permit us to follow the total luminosity from the reverse shock,
complementary to the X-ray observations. 
In SN 1993J\index{SN 1993J}, the \Ha~ line had the box-like shape that is expected
for shocked, cooled ejecta \cite{Matheson2000a,Matheson2000b}.
The top of the line showed structure that varied with time; this
could be related to hydrodynamic instabilities of the reverse
shocked gas.

\section{Non-thermal emission from relativistic particles}
\label{sec:4}

\subsection{Particle acceleration and magnetic field amplification in shocks}
\label{sec:_acc}

Although the basic principles of 
particle acceleration in shocks, based on first order Fermi acceleration across a shock discontinuity ('diffusive shock acceleration'\index{diffusive shock acceleration}), were proposed in the 1970s, e.g., \cite{Longair2011}, there are many aspects that are not well understood. This includes in particular the efficiency of the acceleration mechanism for ions and electrons and the strength of the magnetic field. The energy density of these are usually parameterised in terms of the shock energy density, $\rho_{\rm cs}V_{\rm s}^2$, as $\epsilon_B$, $\epsilon_{\rm e}$ and $\epsilon_{\rm ion}$. Based on equipartition arguments these are often assumed to be of the order of $\sim 0.1$, but without strong observational support. For interpreting  radio and X-ray observations, these parameters are, however, crucial.
 
Except for the general spectrum, which mainly depends on the kinematics of the shock, in particular, the compression ratio, the microphysics determining the injection efficiency and magnetic field amplification are determined by a complex interplay between different plasma instabilities and their implications for  the relativistic particles. For quantitative results advanced numerical simulations are therefore needed. A main complication here is the different physical scales which are important, including  the electron skin depth, the gyro radii of the electrons and ions and the scale of the acceleration region. These differ by many orders of magnitude and different approximations are needed. Here we only mention some recent important results. For a general review of the physics of collisionless shocks see e.g., \cite{Caprioli2015,Marcowith2016}.

One approach, which is well suited for studying the ion component, is known as hybrid-PIC simulations. In these the ions are treated kinetically and electrons as a neutralizing fluid. This allows an accurate treatment of the acceleration of the ions (i.e., cosmic rays) but not the electrons.  The advantage is that the simulation box can be a factor $(m_{\rm p}/m_{\rm e})^{1/2}=43$ larger compared to the case when  electrons are also included kinetically. 

Using this approach Caprioli \& Spitkovsky \cite{Caprioli2014a,Caprioli2014b,Caprioli2014c} have made an extensive study of the acceleration efficiency and magnetic field amplification of shocks with different Mach numbers, $M=5 - 50$,  and obliquity. These are the most advanced simulations to date and we discuss some of the main results. 

The particles are injected with a thermal spectrum,  but soon develop a non-thermal tail  with a power law slope $dN(p)/dp \propto p^{-2}$, where $p$ is the momentum of the particles, as expected from diffusive shock acceleration. The temperature was $\sim 20 \%$ of that without any cosmic ray acceleration, with the rest of the energy residing in the cosmic rays. The escaping ions show a clear separation between the thermal and non-thermal particles at an energy $\sim (4-5)  m_{\rm p} V_{\rm s}^2/2$. 

Independent of the Mach number, Caprioli \& Spitkovsky  find a significant drop in the
acceleration efficiency for shocks with obliquity $\theta < 45^\circ$, and it is close to zero for nearly perpendicular shocks ($\theta > 60^\circ$). The efficiency in general increases with the Mach number, but is in the range $\epsilon_{\rm ion}=5-15  \ \%$.  

The \index{magnetic field amplification} magnetic field amplification is induced by the cosmic ray streaming \cite{Bell2004,Bell2001} and  therefore depends on both the Mach number and obliquity, as discussed above.  The largest amplifications therefore occur for quasi-parallel shocks with high Mach numbers. For $\theta = 90^\circ$ and $M=50$ the amplification is $\sim 10$ above the background field, and is approximately proportional to $M$ for the range studied. 

An efficient acceleration of cosmic rays may have consequences for the shock structure, with a weak sub-shock from the cosmic ray precursor ahead of the main shock \cite{Drury1983,Jones1991}. This will in turn affect the spectra of the accelerated electrons. Low energy electrons will then only scatter across part of the shock, with a smaller compression, while the high energy particles would scatter across the full shock and therefore feel the full compression. This would therefore result in a steep low energy spectrum and a flatter high energy spectrum, similar to that without cosmic rays. In the hybrid simulations discussed above there is a steeper section of the particle spectrum close to the shock, joining the thermal and non-thermal regions. It is not clear if this would result in a steeper integrated synchrotron spectrum. The total compression ratio in the quasi-parallel case is $\sim 4.3$, somewhat larger than in the test-particle case.  

\begin{figure}[t]
\begin{center}
\includegraphics[scale=0.4,angle=0,origin=c]{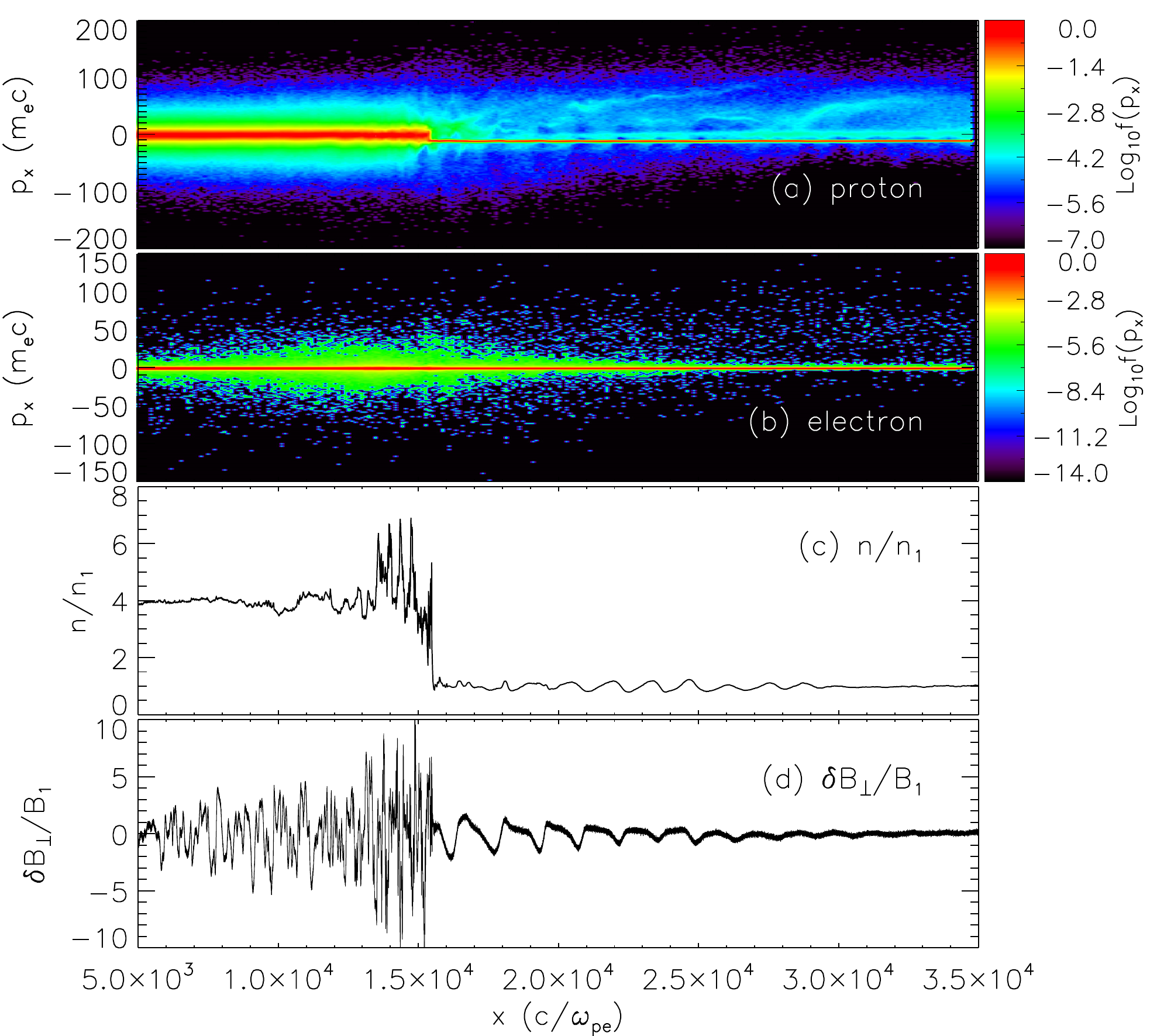}
\includegraphics[scale=0.4,angle=0,origin=c]{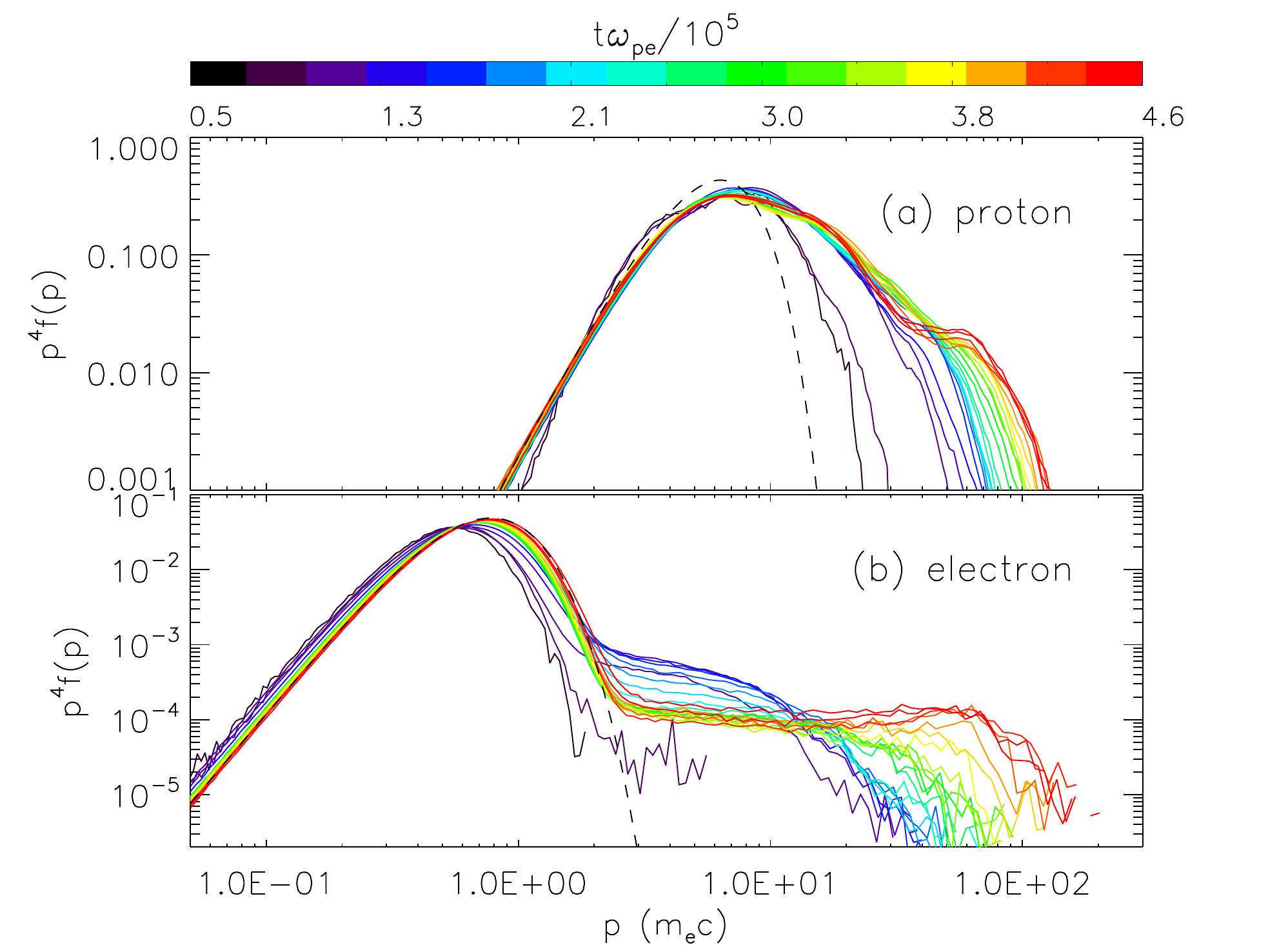}
\end{center}
\caption[]{ Upper figure: Upper panels: Momentum distributions of electrons and protons along the x-axis in the pre- and post-shock regions. Middle panel: Density. Lower panel: The transverse magnetic field along the same axis. Note the increased B-field in the pre-shock region due to the Bell-instability and the continued amplification  behind the shock. Lower figure: Evolution in time of the electron and proton distributions at a point behind the shock. Note that $p^4 f(p) = p^2 dN(p)/dp$ is plotted. The flat electron spectrum above $p \sim 3 m_{\rm e} c$ has  $ dN(p)/dp \propto p^{-2}$ in agreement with diffusive shock acceleration.  With permission from \cite{Park2015}.}
\label{fig5a}
\end{figure}

Recently, full \index{particle-in-cell simulations} PIC (particle-in-cell) simulations have allowed a study of the simultaneous acceleration of both electrons and protons in a collisionless shock \cite{Park2015} (Fig. \ref{fig5a}). Although they are  only one-dimensional (1D) in space for computational reasons, they include all components of velocity and magnetic field. To limit the effect of the different gyro radii of the electrons and protons the proton to electron mass was set to $m_{\rm p}/m_{\rm e}=100$ . 

As above, the streaming of the protons excites magnetic turbulence in the upstream plasma by the non-resonant Bell instability \cite{Bell2004}, which leads to an amplification by a factor of about two in the upstream plasma. This turbulence is then amplified further at the shock transition. 

The acceleration of the electrons initially occurs by shock drift acceleration across the shock and scattering by the magnetic turbulence. As the momentum increases this transitions into a standard diffusive shock acceleration process. 
In the simulations both the electrons and ions relax to the same power law spectra, with a slope close to $dN(p)/dp \propto p^{-2}$. 
At non-relativistic energies this results in an energy spectrum $dN/dE\propto E^{-3/2}$ and in the relativistic regime $dN/dE\propto E^{-2}$. The similar spectra result in an energy independent ratio of $N_{\rm e}/N_{\rm p} \approx 10^{-3}-10^{-2}$ between the electrons and protons, with the ratio increasing with shock velocity.

The general result  that efficient acceleration only seems to occur for quasi-parallel shocks, i.e., where the B-field is approximately aligned with the direction of the shock velocity, may have consequences for shocks in a stellar wind from a rotating progenitor star, where the magnetic field may have the topology of a Parker spiral, causing the shock to be closer to  perpendicular with respect to the magnetic field direction. Small scale turbulence and irregularities may, however, result in a situation intermediate between the two cases.

Although there has been impressive recent advances,  especially connected to these kinetic simulations, many questions remain. This includes the dependence on the shock velocity, the efficiency for perpendicular shocks, the level of the magnetisation, the dependence on the assumed proton to electron mass ratio, non-linear effects from the cosmic ray pressure, etc.

\subsection{Optically thin synchrotron emission}
\label{sec_synchro}
The most important emission mechanism of the non-thermal electrons is usually synchrotron emission. This mechanism is discussed in many textbooks e.g., \cite{Longair2011,Rybicki1986} and here only a few important results are given. 

A relativistic electron with a Lorentz factor $\gamma$ in a magnetic field $B$ will radiate in a narrow frequency range around the critical frequency 
\begin{equation}
\nu_c =  {eB \over m_{\rm e} c} \gamma^2= 1.76\times 10^7 B \gamma^2  ~ \rm Hz, 
\label{eq102}
\end{equation}
where an average over pitch angles has been taken. Here $e$ and $m_{\rm e}$ are the electron charge and mass, respectively, and $c$ the velocity of light. The total energy lost by the electron is
\begin{equation} 
{dP \over dt}= {4 \over 3} \sigma_{\rm T} c \gamma^2 {B^2 \over 8 \pi} \ .
\label{eq101}
\end{equation} 
where $\sigma_{\rm T} $ is the Thompson cross section. We assume that the electrons are created with a power law distribution of relativistic electrons, $dN/dE = N_0 E^{-p}$, where $E=\gamma  m_{\rm e} c^2$. Note that $p$ is here and from now on the spectral index of the electron distribution, not momentum as in the previous section. Approximating the spectrum of an individual electron with a delta function and using Eqs (\ref{eq102}) and (\ref{eq101}),  the  emissivity per frequency at frequency $\nu$ becomes
\begin{equation} 
j(\nu) d\nu \approx {1 \over 4 \pi}  {dN(E)\over dE} {dP \over dt} dE =  {\rm const.} ~ N_0 ~ B^{(p+1)/2} ~\nu^{-(p-1)/2} d\nu \ .
\label{eq103}
\end{equation}
It is  a power law with spectral index $\alpha=(p-1)/2$ determined by the power law index of the non-thermal electrons. 

The \index{synchrotron cooling} synchrotron cooling time of an electron is 
\begin{equation} 
t_{\rm synchro}={\gamma m_{\rm e} c^2 \over dP/ dt}= 6 \pi  {m_{\rm e} c \over \sigma_{\rm T} }  B^{-2} \gamma^{-1}=7.7\times10^8  B^{-2} \gamma^{-1} \rm s
\label{eq104}
\end{equation}
or in terms of the emitted frequency
\begin{equation} 
t_{\rm synchro}= 6 \pi  {(e m_{\rm e} c)^{1/2} \over \sigma_{\rm T} } B^{-3/2}  \nu^{-1/2}=3.2\times 10^7 B^{-3/2}  \nu_{10}^{-1/2} \ \rm s.
\label{eq105}
\end{equation}
where $\nu_{10}=\nu/10$ GHz.

It is common to relate the magnetic field to the mass loss rate parameters by scaling the magnetic energy density to the shock energy density, $\rho_{\rm cs} V_{\rm s}^2$. With the scaling parameter $\epsilon_{\rm B}$ we get
\begin{equation} 
{B^2 \over 8 \pi}= \epsilon_{\rm B}  \rho_{\rm cs}  V_{\rm s}^2 = \epsilon_{\rm B} {\dot M \over 4 \pi u_{\rm w} r^2} V_{\rm s}^2 = \epsilon_{\rm B} {\dot M \over 4 \pi u_{\rm w}  t^2} \ .
\label{eq106}
\end{equation} 
We then obtain
\begin{equation} 
B= 4.1 \epsilon_{\rm B}^{1/2}  C_*^{1/2} \left({t \over 10 \ \rm days}\right)^{-1} \ \rm G
 \label{eq108}
\end{equation}
and
\begin{equation} 
{t_{\rm synchro} \over t}= 4.4 \epsilon_{\rm B}^{-3/4} C_*^{-3/4} \left({t \over 10 \ \rm days}\right)^{1/2}  \nu_{10}^{-1/2}  \ .
\label{eq108b}
\end{equation}

If relativistic electrons are created by the acceleration process with a spectrum $N_0 E^{-p}$ according to Eq. (\ref{eq104}) they will  have a lifetime proportional to $E^{-1}$. For the electrons for which the cooling time is short compared to the time since they were created, the integrated electron spectrum of the source is proportional to $N_0 E^{-p-1}$. From Eq. (\ref{eq103})  the source  then has a synchrotron spectrum $j(\nu) \propto ~\nu^{-p/2}$, i.e., steeper by 1/2 in the spectral index.

\subsection{Inverse Compton X-ray emission}

\index{inverse Compton emission} Besides synchrotron emission, inverse Compton emission may  be important for the UV and X-ray emission. The most important sources of photons are the ones from the photosphere and from any optical/UV emission from the shock region. The cooling time scale of the same non-thermal electrons emitting the synchrotron emission is identical to Eq. (\ref {eq101}), with $B^2/8 \pi$ replaced by the radiation density, $U=L/4 \pi r^2 c$, so

\begin{equation} 
{t_{\rm Compton} \over t }= 0.85 \  \epsilon_{\rm B}^{1/4} C_*^{1/4}  \left({L \over 10^{42} \ergs}\right)^{-1}  \left({ V_{\rm s} \over 10^4 \ \kms}\right)^{2} \left({t \over 10 \ \rm days}\right)^{1/2}  \nu_{10}^{-1/2}  \ .
\label{eq109}
\end{equation}
The ratio between the synchrotron and Compton cooling time scales is  
\begin{equation} 
{t_{\rm synchro} \over t_{\rm Compton} }=  {2 L  \over c B^2 r^2 } = 5.2 \times 10^{-2}\ \epsilon_{\rm B}^{-1} C_*^{-1} \left({L \over 10^{42} \ergs}\right) \left({ V_{\rm s} \over 10^4 \ \kms}\right)^{-2} \ .
\label{eq110}
\end{equation}

Inverse Compton cooling is therefore mainly important for the cooling at early phases when the luminosity is high, for low mass loss rates and for slow shocks. Because the cooling time scale has the same $\gamma^{-1}$ dependence as the synchrotron cooling, the effect on the electron distribution and emitted spectrum will be the same for energies and frequencies where the cooling time is short, i.e., $F_\nu \propto  \nu^{-p/2}$. 

For a Compton dominated shock the radio spectrum and light curve is given by
 \begin{equation} 
F_\nu(t) \propto  \epsilon_{\rm B}^{-1} C_*^{-1} L(t)^{-1}  t^{m-p/2} \nu^{-p/2} \ .
\label{eq110b}
\end{equation}
\cite{Fransson1998,Nymark2006}. 
The radio light curve is therefore inversely correlated with the optical light curve, which may result in a flattening of the radio light curve at the time of optical peak for high frequencies \cite{Chevalier2006}. 

The energy lost by the relativistic electrons is transferred to the soft photons, which will be scattered to an energy $E_x \sim \gamma^2 E_0$, where $E_0$ is the initial energy of the photon. For a SN the main soft photon flux is in the optical. The Compton scattered photons are then typically in the X-rays. The total energy loss of the electron is  $dP/dt = 4/3 \sigma_{\rm T} c U_{\rm rad}\gamma^2$, same as in Eq. (\ref{eq101}) with $B^2/8 \pi$ replaced by $U_{\rm rad}$. With a non-thermal electron spectrum $dN/dE = N_0 E^{-p}$, we get for the inverse Compton spectrum at energy $E_x$
\begin{equation*}
{dF\over dE_x} dE_x \approx 4 \pi r^2 {dP\over dt} {dN \over dE}  dE = {16 \pi r^2 \over 3} \sigma_{\rm T} c U_{\rm rad}\gamma^2 {dN\over dE}  dE \ .
\end{equation*}
With a non-thermal electron spectrum $dN/dE = N_0 E^{-p}$ we get 
\begin{equation*}
{dF\over dE_x} dE_x = {16 \pi r^2\over 3} \sigma_{\rm T} c U_{\rm rad} N_0 (m_{\rm e} c^2)^{-2} E^{-p+2}  dE \ .
\end{equation*}
Using $E_x\approx  \gamma^2 E_0= (E/m_{\rm e} c^2)^2 E_0$ one finds for the inverse Compton spectrum at energy $E_x$
\begin{equation*}
{dF\over dE_x} dE_x = {8 \pi r^2\over 3} \sigma_{\rm T} c {U_{\rm rad}\over E_0} N_0 (m_{\rm e} c^2)^{-2} \left({E_x \over E_0}\right)^{-(p-1)/2}  (m_{\rm e} c^2)^{-p+1}  dE_x \  \rm  .
\end{equation*}
Averaging over a blackbody spectrum $E_0 \approx 3.6 kT_{\rm bb}$ \cite{Felten1966}. Far from the source of the soft photons $U_{\rm rad}=L/4 \pi r^2 c$, so finally
\begin{equation}
{dF(E_x)\over dE_x} = {2\over 3} \sigma_{\rm T} \  L(t) \ N_0  (m_{\rm e} c^2)^{-p-1} (3.6 kT_{\rm bb})^{(p-3)/2}   E_x^{-(p-1)/2}   \  \rm .
\end{equation}
The spectrum is therefore the same as the corresponding synchrotron spectrum from the same electron distribution and the flux is directly proportional to the luminosity of the soft photon source. Also note the insensitivity to the temperature of the blackbody radiation for $p\approx 3$.

 Compton scattering is also  important as a
 cooling process of the {\it thermal} electrons of the circumstellar shock. In contrast to free-free and line cooling this does not depend on the density and only on the radiation density of the photons from the photosphere. Heating by the hard radiation from the shocks may be important for both the CSM and the outer layers of the SN ejecta. 
 
 The average energy transfer rate of a photon with energy $h \nu$ interacting with thermal electrons with temperature $T_{\rm e}$ is
 \begin{equation} 
\Delta E= {h \nu\over m_{\rm e} c^2} (4 kT_{\rm e} - h \nu)
\label{eq111}
\end{equation}
Depending on the energy of the photon the radiation relative to the temperature of the electrons can therefore act as either a heating or cooling process.

The photospheric photons
have energies  $\sim 3 ~kT_{eff} \approx 1 - 10$ eV. 
A fraction $\tau_{\rm e}^N$ of the photospheric photons will scatter $N$ times
in the hot gas. In each scattering the photon increases its energy by a  factor
$\Delta\nu/\nu\approx {4 ~k T_{\rm e}/ m_{\rm e} c^2} \gsim 1$.
The multiple scattering creates a power law continuum
that may reach as far up in energy as  the X-ray regime.
Typically, $1 \lsim \alpha \lsim 3$. 
This type of emission may have been observed in the ultraviolet
emission from SN 1979C \cite{Fransson1982} \index{SN 1979C}.
 For $T_{\rm e} \gsim 10^9 \KK$ relativistic effects become important and
considerably increase the cooling \cite{Lundqvist1988}.

\subsection{The radio spectrum including absorption processes}
\label{sec_abs}

Unambiguous evidence for the presence of relativistic electrons comes from radio
observations of SNe. A characteristic is the wavelength-dependent
turn-on of the radio emission \cite{Sramek2003}, first
seen at short wavelengths, and later at longer wavelengths. 
This behaviour is interpreted  as a result of 
decreasing absorption due to the expanding emitting region \cite{Chevalier1982b}.

Depending on the magnetic field and the density of the circumstellar
medium, the absorption may be produced either by free-free absorption\index{free-free absorption}
in the surrounding thermal gas, or by synchrotron self-absorption\index{synchrotron self-absorption} by
the same electrons that are responsible for the emission. The
relativistic electrons are believed to be produced close to the
interaction region, which provides an ideal environment for the
acceleration of relativistic particles. As discussed above, the details of the
acceleration and injection efficiency are still not well understood. 
 Here we just parameterize the
injection spectrum with the power law index $p_{\rm i}$ and an
efficiency, $\eta$, in terms of the  post-shock energy density. Without
radiation or collisional losses the spectral index of the synchrotron
emission will then be $F_\nu \propto \nu^{-\alpha}$ with $\alpha=(p_{\rm i}-1)/2$.
 As discussed earlier, diffusive acceleration
predicts that $p_{\rm i} = 2$ in the test particle limit, but can be steeper
if the particle acceleration is very efficient and nonlinear effects
are important. 

For \index{free-free absorption} free-free absorption, the optical depth $\tau_{\rm ff} =
\int_{R_{\rm s}}^{\infty} \kappa_{\rm ff} n_{\rm e} n_{\rm i} dr$
 from the radio emitting
region close to the shock through the circumstellar medium decreases as
the shock wave expands,
explaining the radio turn-on.  Assuming a fully ionized and a temperature $T_{\rm e}$ wind with
constant mass loss rate and velocity, so that Eq. (\ref{eq1b})
applies, the free-free optical depth at wavelength $\lambda$ is
\begin{eqnarray} 
\tau_{\rm ff}(\lambda) &\approx& 6.0\times 10^{2} \left[1+{n({\rm He}) \over n({\rm H})}\right]
\lambda^{2.118}~C_*^2 ~ \nonumber \\
&& \left({ T_{\rm e} \over 10^5 \ \rm K}\right)^{-1.323}
\left({ V_{\rm s} \over 10^4 \ \kms}\right)^{.-3} \left({t \over \rm days}\right)^{-3} \  ,
\label{eq13}
\end{eqnarray}
including the slight variation of the Gaunt-factor with temperature and wavelength.  
One should also note that especially for stars that have undergone strong mass loss the helium to  hydrogen ratio may have been substantially enhanced compared to the solar value. 

From
the radio light curve, or spectrum, the epoch of $\tau_{\rm ff} = 1$ can
be estimated for a given wavelength, and from the line widths in the
optical spectrum the maximum expansion velocity, $V$, can be
obtained.  From $t[\tau(\lambda)_{\rm ff}=1]$ the mass loss parameter $C_*$ and therefore the
ratio  $\Mdot/u_{\rm w}$ can be calculated. Because $\Mdot/u_{\rm w} \propto
T_{\rm e}^{0.66} x_{\rm e}^{-1}$, errors in $T_{\rm e}$ and $x_{\rm e}$ may 
lead to large
errors in $\Mdot$. 
Both $T_{\rm e}$ and $x_{\rm e}$  are set by the radiation from the SN and have
to be estimated from models of the ionization and heating of the circumstellar medium.

Calculations show
that the radiation from the shock break-out (Sect. \ref{sec_ioni}) heats the gas to $T_{\rm e} 
\approx 10^5 \KK$ \cite{Lundqvist1988}.
 $T_{\rm e}$ then decreases with time, and after a year $T_{\rm e} \approx
(1.5-3)\EE4 \KK$. In addition, the medium may recombine, which further
decreases the free-free absorption.
If the shock breakout radiation is soft and the X-ray flux from the continuing circumstellar interaction is weak,  heating resulting from the free-free absorption may  determine the temperature, giving a temperature $\lsim 10^5$ K \cite{Bjornsson2014}. 
 We also note that because $\tau_{\rm ff} \propto   n_{\rm e} n_{\rm i} $  Eq. (\ref{eq13})
 may lead to an
overestimate of $\Mdot/u_{\rm w}$ if the medium is clumpy\index{circumstellar clumps}.

Under many circumstances (see below), synchrotron self-absorption\index{synchrotron self-absorption}
(SSA) by
the same relativistic electrons emitting the synchrotron radiation is important \cite{Slysh1990,Chevalier1998,Fransson1998}. In Eq. (\ref{eq103}) we gave an approximate expression for the synchrotron emissivity. 
Here we give the exact expressions for the emission and for the absorption effects, assuming a constant power law, $p$, for the electrons and a randomly oriented magnetic field. For a detailed account of synchrotron processes see e.g., \cite{Chevalier1998,Pacholczyk1970,Rybicki1986}.  

The emissivity is given by 
\begin{equation}
j(\nu) =   c_{\rm em}(p)  ~ B^{(p+1)/2} ~ N_0  ~\left({\nu \over \nu_0}\right)^{-(p-1)/2} 
\label{eqss3b}
\end{equation}
where $\nu_0 = 1.253 \times 10^{19} \ {\rm Hz}$ and
\begin{eqnarray}
c_{\rm em}(p) &=& 4.133 \times 10^{-24} \left({p+7/3\over p+1}\right) \Gamma\left({3p-1\over 12}\right) \Gamma\left({3p+7\over 12}\right) \times \nonumber \\ &&\Gamma\left({p+5\over 4}\right) \Gamma\left({p+7\over4}\right)^{-1}.
\end{eqnarray}
The self absorption opacity, $\kappa(\nu)$, can be obtained from the emissivity and detailed balance (see e.g., \cite{Pacholczyk1970,Rybicki1986}).  Under the same assumptions as above for the emissivity one obtains
\begin{equation}
\kappa(\nu)=  c_{\rm abs}(p) ~ N_0 ~ B^{(p+2)/2} \left({\nu \over \nu_0}\right)^{-(p+4)/2}  \ \rm ,
\label{eqss3}
\end{equation}
where 
\begin{eqnarray}
c_{\rm abs}(p) &=& 1.183 \times 10^{-41} \left({p+10/3}\right) \Gamma\left({3p+2\over 12}\right) \Gamma\left({3p+10\over 12}\right)  \times \nonumber \\ &&\Gamma\left({p+6\over4}\right) \Gamma\left({p+8\over4}\right)^{-1}  \ .
\label{eq_c6}
\end{eqnarray}
\begin{table}[t!]
\caption{Synchrotron constants in c.g.s. units for values of the energy power law index $p$ for  emission and absorption (Eqs. \ref{eqss3b} and \ref{eqss3}). }
\label{tab_p}       
\begin{tabular}{p{2cm}p{1.5cm}p{1.5cm}p{1.5cm}p{1.5cm}p{1.5cm}p{1.5cm}}
\hline\noalign{\smallskip}
$p$ & 1.0& 1.5 & 2.0 & 2.5 & 3.0 & 3.5 \\
\noalign{\smallskip}\svhline\noalign{\smallskip}
$c_{\rm em}/10^{-23}$& 3.8359&1.6954&0.9874&0.6688&0.5013&0.4050 \\
$c_{\rm abs}/10^{-41}$&8.4809&6.7024&5.7410&5.2179&4.9697&4.9198\\
\noalign{\smallskip}\hline\noalign{\smallskip}
\end{tabular}
\end{table}
Values for $c_{\rm em}$ and $c_{\rm abs}$ for selected values of $p$ are given in Table \ref{tab_p}.

Taking the self-absorption of the synchrotron emission  \index{synchrotron self-absorption} into account, the observed flux is
\begin{equation}
F_\nu={\pi R_{\rm s}^2 \over D^2}  S(\nu)[1 - \exp(-\tau(\nu))] \ ,
\label{eqss1}
\end{equation}
where 
\begin{equation}
\tau(\nu) = \int_0^s \kappa(\nu) ~ds \approx \kappa(\nu) ~ s
\label{eqtau}
\end{equation}
is the optical depth of a slab with thickness $s$, and $S(\nu)$ is the source function
\begin{equation}
S(\nu)= {j(\nu)\over \kappa(\nu)} = {c_{\rm em}(p) \over c_{\rm abs}(p)} ~B^{-1/2}\left({\nu \over \nu_0}\right)^{5/2} \ .
\label{eqss2}
\end{equation}
For a spherical source the effective thickness $s$ can be taken to be  equal to the ratio of the emitting volume, $4 \pi R_{\rm s}^3 f /3$ and the projected area $\pi R_{\rm s}^2$, where $f$ is the volume filling factor of the synchrotron emitting volume.

In the
optically thick limit we  therefore have 
\begin{equation}
F_\nu(\nu) \approx 
{\pi R_{\rm s}^2 \over  D^2}~{c_{\rm em} \over  c_{\rm abs}} \left({\nu \over \nu_0}\right)^{5/2} B^{-1/2} \ ,
\label{eqss5}
\end{equation} 
independent of $N_{\rm 0}$. A
fit of this part of the spectrum therefore gives the quantity $R_{\rm s}^2~
B^{-1/2}$. 

In the optically thin limit $F(\nu) \propto j(\nu) \propto R_{\rm s}^2
B^{(p+1)/2} ~ N_0  ~\nu^{-(p-1)/2}$. The spectrum will therefore have a break at the frequency where $\nu(\tau  \approx1)$.
The condition $\tau(\nu) = \kappa(\nu) s = 1$  with  $\kappa(\nu)$ from Eq. (\ref{eqss3})
therefore gives a second condition on $B^{(p + 2)/2}~N_{0} ~ R_{\rm s}^3$. 
Now, if $R_{\rm s}(t)$ is known in some independent way, one can
 determine both the magnetic field and the column density of
relativistic electrons, independent of assumptions about equipartition,
etc.  

In some cases, most notably for
SN 1993J\index{SN 1993J}, the shock radius, $R_{\rm s}$, can be determined directly from very long baseline interferometry (VLBI)
observations. If this is not possible, an alternative is from
observations of the maximum ejecta velocity seen in, e.g., the \Ha
~line, which should reflect the velocity of the gas close to the
shock.  Because the SN expands homologously, $R_{\rm s} = V_{\rm max} t$. A fit
of the spectrum at a given epoch can therefore yield both $B$ and
$N_{\rm rel}$ independently. From observations at several epochs the
evolution of these quantities can then be determined \cite{Fransson1998}.

Even if $R_{\rm s}$, and therefore $V_{\rm s}$, is not known a  useful estimate for these quantities can be derived from the observation of the flux and time of the peak of the radio light curves for a given frequency \cite{Chevalier1998}.  With $s=4 f  R_{\rm s}/3$ and  $\tau(\nu)\approx \kappa(\nu) s  \approx 1$, with $\kappa(\nu)$ from Eq. (\ref{eqss3}),  this gives a relation between $N_0$, $B_{\rm peak}$ and $R_{\rm s~  peak}$. If we assume an electron spectrum with lower energy limit $E_{\rm min}$, which we can take to be equal to $\sim m_{\rm e} c^2$, one gets for the total energy density $U_{\rm e} = N_0 E_{\rm min}^{-p+2}/(p-2)$. Finally, we assume a constant ratio between this and the magnetic energy density equal to $\zeta$, we get $N_0 = \zeta (p-2) B^2  E_{\rm min}^{p-2}/8 \pi$. From Eq. (\ref{eqss3}) we then get 
\begin{equation}
B =  \left[{6 \pi \over f R_{\rm s}c_{\rm abs} (p-2) \zeta E_{\rm min}^{p-2}}\right]^{2/(6+p)}  \left({\nu \over \nu_0}\right)^{(p+4)/(p+6)}   \ .
\label{eqss5b}
\end{equation}
Estimating the flux at the peak with the optically thick expression, Eq. (\ref{eqss5}), and using the above expression for $B$ and solving for $R_{\rm s}$ one gets
\begin{eqnarray}
R_{\rm s}(t_{\rm peak}) = \left({c_{\rm abs}  \over \pi}\right)^{(5+p)/(13+2p)} && \left[{6\over f (p-2) \zeta E_{\rm l}^{p-2}}\right]^{1/(13+2p)}   \times \nonumber \\
&&\left[{F_\nu(t_{\rm peak}) D^2 \over c_{\rm em}}\right]^{{(6+p)}/(13+2p)}  \left({\nu \over \nu_0}\right)^{-1} \ .
\label{eqss6}
\end{eqnarray}
The velocity of the ejecta at the shock is then given by $V_{\rm shock}(t_{\rm peak})=(n-3)/(n-2) ~R_{\rm s}(t_{\rm peak})/t_{\rm peak}$. This velocity depends on the uncertain parameters $f$ and especially $\zeta$. However, this dependence is very weak. With $p=3$ one gets $R_{\rm s}(t_{\rm peak})  \propto (f \zeta)^{-1/19}$. For $\zeta=10^{-3}$ the radius and velocity are only $\sim 30 \%$ larger than for $\zeta=1$. 

\begin{figure}[t]
\centering
\includegraphics[viewport=  0 10 600 600,scale=0.40,angle=0,origin=c,clip]
{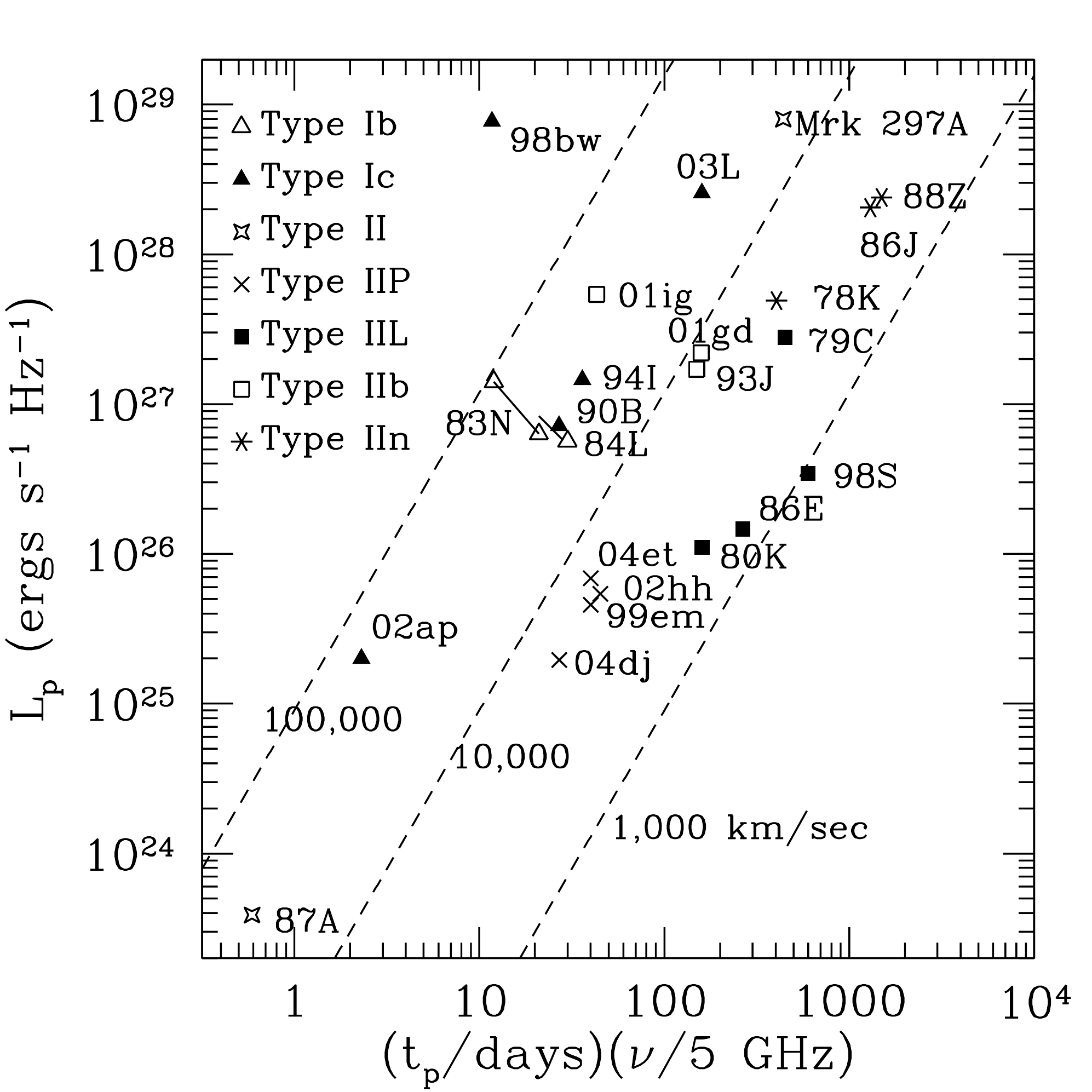}
\caption[]{Peak luminosity and corresponding epoch for 
well-observed radio SNe. The dashed lines give curves of constant
expansion velocity, {\it assuming} SSA. With permission from \cite{Chevalier2006}). 
}
\label{fig4d}
\end{figure}

Using Eq. (\ref{eqss6}), we can plot lines of constant shock
velocity into a diagram with peak radio luminosity versus time of peak
flux, {\it assuming} that SSA dominates (Fig. \ref{fig4d}).
Each SN
can now be placed in this diagram to give a predicted shock
velocity. If this is lower than the observed value (as measured by
VLBI or from line profiles) SSA gives a too low flux and should
therefore be relatively unimportant and free-free absorption instead
dominate. In Fig. \ref{fig4d} we show an updated version of the
figure in \cite{Chevalier1998}. The most interesting point is that most Type
Ib/Ic like SNe 1983N, 1994I, 1998bw and 2002ap, fall into the high
velocity category, while Type IIP, like SNe 1999em, 2004et and 2004dj, as well as the Type IIL SNe 1979C\index{SN 1979C}  and 1980K, and
the Type IIn SNe 1978K\index{SN 1978K}, 1988Z\index{SN 1988Z}, 1998S\index{SN 1998S} fall in the
free-free group \index{SN 1978K}. Strong, direct observational evidence for free-free absorption, however, only exist for SN 1979C\index{SN 1979C} and SN 1980K\index{SN 1980K} \cite{Weiler1986}. SN 1987A\index{SN 1987A}  is clearly special with its low mass loss
rate, but is most likely dominated by SSA \cite{Storey1987,Slysh1990,Chevalier1995b}.

Although the injected electron spectrum from the shock is likely to be
a power law with $p_{\rm i} \approx 2$ ($\alpha \approx 0.5$), the
integrated electron spectrum is affected by  synchrotron, Compton and Coulomb cooling. 
The column density of electrons with a Lorentz factor $\gamma$ is
\begin{equation}
N(\gamma)  s = {U_{\rm rel} (p-2)  \over E_{\rm min}^2}   \ s(\gamma)
\ \left({\gamma \over \gamma_{\rm min}}\right) ^{-p_{\rm i}}
\label{eqnrel}
\end{equation}
where $s(\gamma)$ is the effective thickness of the region of relativistic electrons, which depends on their energy. For energies where cooling is not important $s\approx V_{\rm s} t/4$. For electrons with short cooling times, i.e.,  $ t_{\rm cool} \ll t$, one has instead $s(\gamma)\approx V_{\rm s}\  t_{\rm cool}(\gamma)/4$. Including all loss processes one can then write 
\begin{equation}
N(\gamma)  s \approx {U_{\rm rel} (p-2)  \over E_{\rm min}^2}  {V_{\rm s} t \over 4}
\ \left({\gamma \over \gamma_{\rm min}}\right) ^{-p_{\rm i}}
\left[1 + {t \over t_{\rm Coul}(\gamma)} + 
{t \over t_{\rm Comp}(\gamma)} + {t \over t_{\rm synch}(\gamma)}\right]^{-1}.
\label{eqnrel2}
\end{equation}
Here $ t_{\rm Coul}$, $t_{\rm Comp}$ and $t_{\rm synch}$ are the Coulomb, Compton and synchrotron time scales, respectively. The expressions for $t_{\rm synch}$ are given by Eq. (\ref{eq105}) and for $t_{\rm Comp}$ by Eq. (\ref{eq109}). 

The total number of relativistic electrons may either be assumed to be
proportional to the total mass, if a fixed fraction of the shocked
electrons are accelerated, or be proportional to the swept up thermal
energy. In the first case, $U_{\rm rel} \propto \Mdot /u_w$, while in the second $U_{\rm rel} \propto \Mdot V_{\rm s}^2 /u_w$, so that in general $U_{\rm rel} \propto \Mdot V_{\rm s}^{2\epsilon} /u_w$, where $\epsilon = 0$ or $1$ in
these two cases. 
If the
magnetic field is in equipartition, as above, $B^2\propto \rho V_{\rm s}^2\propto \Mdot/t^2$ and using $V_{\rm s} = (n-3)/(n-2) R_{\rm s}/t
\propto t^{-1/(n-2)}$, we find for the non-cooling case
\begin{equation} 
F_\nu(\nu) 
\propto C_*^{(5+p)/4} ~ t^{-(p-1)/2 -
(1+2\epsilon)/(n-2)} ~\nu^{(1-p)/2} \ ,
\label{eq13d}
\end{equation}
while for the synchrotron cooling case we instead get
\begin{equation} 
F_\nu(\nu) 
\propto C_*^{(2+p)/4} ~ t^{-(p-2)/2 -
(1+2\epsilon)/(n-2)} ~\nu^{-p/2}.
\label{eq13r}
\end{equation}
The cooling case therefore  steepens the spectrum by $1/2$, but flattens the light curves compared to the non-cooling case, for the same $p$. Inverse Compton losses have a similar effect. At low energy,
Coulomb losses may be important, causing the electron spectrum to
flatten. These complications are discussed in e.g.,  \cite{Fransson1998}. 
The main thing to note
is, however, that the optically thin emission is expected to be sensitive to the mass loss\index{mass loss}  parameter, $C_*$, and that the decline rate
depends on whether the number of relativistic particles scale with the
number density or the shock energy, as well as
spectral index. Observations
of the decline rate can therefore test these possibilities \cite{Fransson1998}.

The relative importance of SSA and free-free absorption depends on a
number of parameters and we refer to \cite{Chevalier1998,Fransson1998} for a more
detailed discussion. The most important of these are the mass loss
rate, $\Mdot/u_w$, the shock velocity, $V_{\rm s}$, and the
circumstellar temperature, $T_{\rm e}$. In general, a high shock
velocity and a high circumstellar temperature favor SSA, while a high
mass loss rate favors free-free absorption\index{free-free absorption}.  

\subsection{Radio and high energy signatures of cosmic ray acceleration}

\index{cosmic rays} The very dense CSM of especially the Type IIns (Sect. \ref{sec_iin}) may give rise to efficient cosmic ray acceleration, mainly protons. The interaction of the accelerated protons with energies $\gsim  280$ MeV   will give rise to pions,  $p + p \rightarrow \pi^{\pm}, \pi^0$.
The pions then decay into gamma-rays and neutrinos, $\pi^0 \rightarrow 2 \gamma$ and $\pi^{\pm} \rightarrow , \mu^{\pm} + \nu_\mu (\bar\nu_\mu)$, and finally $\mu^{\pm}   \rightarrow  e^{\pm} +\bar\nu_\mu (\nu_\mu) + \bar\nu_{\rm e} (\nu_{\rm e})$.  The end-result of this chain is therefore high energy gamma-rays and neutrinos and a population of \index{secondary electrons}  secondary electrons, in addition to the primary electrons which are injected from the thermal pool behind the shock. The secondary electrons are distinguished from the primary by being injected with a higher Lorentz factor than the primary, $\gamma \sim 68$. For proton-proton collisions\index{proton-proton collisions} to be important the effective optical depth for the high energy protons, $\sim 2.7 (C_*/10^3) (V_{\rm s}/ 5000 \kms)^{-2} (t/ 100 \  \rm days)^{-1}$ , has to be $\gsim 0.1$.  Proton - proton collisions are therefore mainly important for shocks in very high density CS media, both for high energy gamma-ray and neutrino detections, but also as a source of electrons, giving rise to synchrotron emission in radio \cite{Murase2011}. 

\index{gamma-rays, high energy} \index{neutrinos, high energy} For a sufficiently dense CSM, $C_* \gsim 10^2$, and a high proton acceleration efficiency, $\gsim 10\%$, the neutrino flux in the GeV-PeV range may be detectable with IceCube out to a distance of 10--20 Mpc. There may also be the possibility of detecting GeV  gamma-rays by Fermi in very long integrations (of the order of one year). The Cherenkov Telescope Array (CTA) will have more than an order of magnitude higher sensitivity in only 100 hours at TeV energies. Depending on the proton spectrum a strongly interacting SN may be detected out to 100 -- 200 Mpc \cite{Murase2014}.

The typical frequency of the radio emission from the secondary electrons is in the range $3 - 3000$ GHz \cite{Murase2014}, higher than the typical frequency of the primary electrons. The main problem for detecting the radio emission from these secondary electrons (and of course also the primary)  is free-free absorption by the dense CSM (Eq. \ref{eq13}). However, for high frequencies, $\gsim 100$ GHz and large shock radii, $\gsim 10^{16}$ cm, i.e., late epochs, the CSM may be transparent for the radio photons, for parameters where the optical depth to the proton-proton collisions is large enough for significant secondary electron production. Detection of this emission would be highly interesting for understanding both the structure of collisionless shocks and cosmic ray acceleration in SNe.

\section{Examples of circumstellar emission from different SN types.}
\label{sec:5}

As illustrations of the processes described above we discuss in this section a number of well observed SNe of different types. These should be seen as selected examples, but without any attempt of a complete list of references. 

\begin{figure}[t]
\centering
\begin{center}
\includegraphics[scale=0.40,angle=90,origin=c]{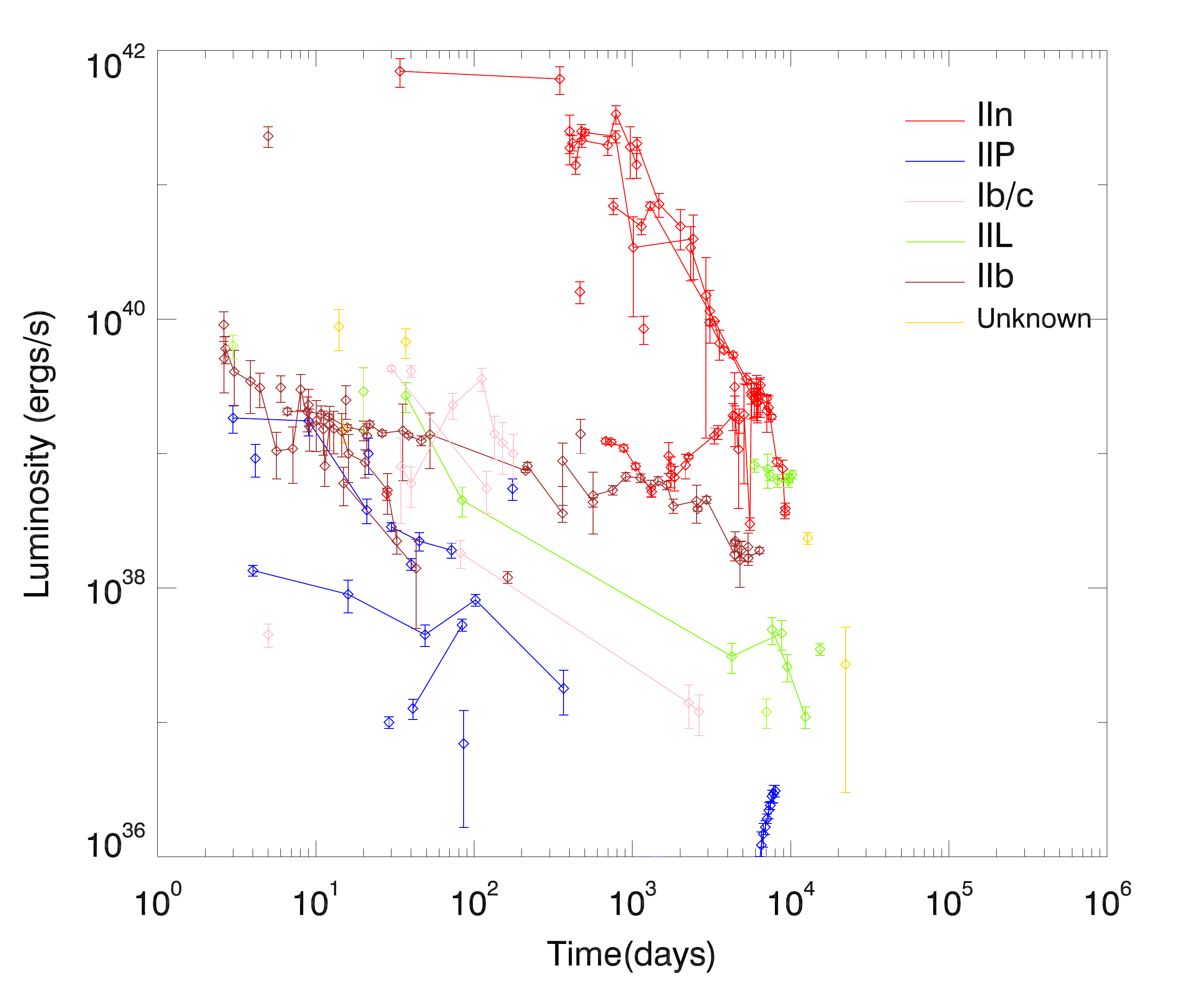}
\end{center}
\caption[]{Compilation of X-ray light curves of different types of supernovae. Note the large range, from faint objects, like SN 1987A (blue points coming up at $\sim 5000$ days), to the very bright Type IIn SNe. With permission from \cite{Dwarkadas2014}.}
\label{figxraylc}
\end{figure}

For later reference we show in Fig. \ref{figxraylc}  a compilation of light curves from a sample of SNe of different types. We note the large range, from bright Type IIns to faint objects like SN 1987A. Radio light curves\index{radio light curve} show a similar diversity.

\subsection{Type IIP}

\index{Type IIP} Type IIP SNe are the most common class of supernovae, assumed to arise from $9-17 \ \Msun$ red supergiant progenitors. In general, these have moderate  mass loss\index{mass loss} rates $\dot M \sim 10^{-6} \ \Msunyr$ and slow winds, $u_{\rm w}= 10-20 \kms$. Because of their comparatively low mass loss rates the X-ray emission is not very strong (Fig.  \ref{figxraylc}). This is also the case for the radio emission and absorption effects are only seen very early. Some of the best studied examples include SNe 1999em, 2004et and 2004dj \cite{Argo2005,Beswick2005,MartiVidal2007}.

\begin{figure}[t]
\centering
\begin{center}
\includegraphics[scale=0.35,angle=-0,origin=c]{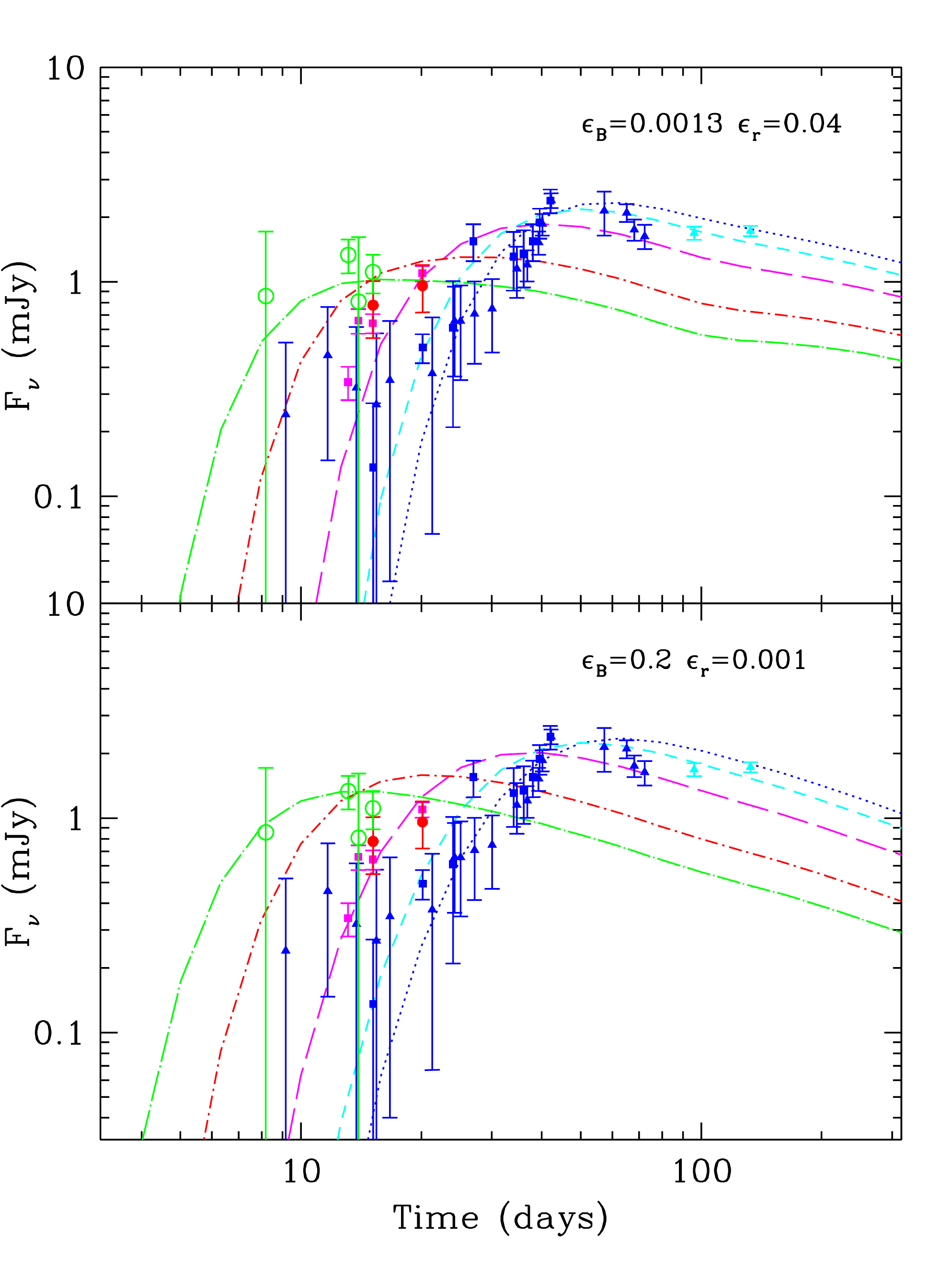}
\end{center}
\caption[]{Observed and model radio light curves (solid lines) of SN 2004et for different assumptions for the magnetic and relativistic electron energy densities, given in the figures.  With permission from \cite{Chevalier2006}.}
\label{fig4ab}
\end{figure}

In \cite{Chevalier2006}  the X-ray and radio observations were discussed for different assumptions about the efficiency for electron acceleration and magnetic field, $\epsilon_{\rm e}$ and $\epsilon_{\rm B}$, and in Fig. \ref{fig4ab} we show fits to the light curves of SN 2004et  for different assumptions. For low magnetic fields, small $\epsilon_{\rm B}$, inverse Compton cooling by the photospheric emission was important for the relativistic electrons, while for a high $\epsilon_{\rm B}$ synchrotron cooling was important. 
In both cases, the light curves were strongly affected by cooling during the first $\sim 100$ days. In the upper panel with a weak magnetic field the cooling of the relativistic electrons at early times is dominated by inverse Compton cooling by the optical radiation from the SN, resulting in a flat light curve. In the lower panel the situation is the opposite, with synchrotron cooling dominating early. The turn-on is caused by decreasing free-free absorption in both cases.

From the free-free absorption turn-on at $\sim 20$ days at 5 GHz, the mass loss rate was estimated to be $\Mdot \approx 1 \times 10^{-5} (T_{\rm e}/10^5 \ {\rm K})^{3/4} (u_{\rm w} /10 \ {\kms})\ \Msunyr$, again depending on the uncertain temperature of the CSM. Because of the extended  red supergiant progenitor the CSM temperature may have been somewhat lower than for the Type IIL SNe in \cite{Lundqvist1988}, i.e., $\lsim 10^5$ K. The mass loss rate would then be similar to those of galactic red supergiants.

No boxy optical line profiles have been observed in optical spectra for a Type IIP SN, implying that the reverse shock in these cases is adiabatic, as is expected for the low mass loss rates. 

\subsection{Type IIL}

\index{Type IIL} The first well-observed radio SNe were the optically  bright Type IIL SNe 1979C\index{SN 1979C} and 1980K\index{SN 1980K}  \cite{Weiler1986}. Although observed with the new VLA immediately after explosion, SN 1979C was first detected only $\sim 1$ year after explosion at 5 GHz, while SN 1980K was first detected $\sim 64$ days after explosion  \cite{Weiler1981,Weiler1986}. The wavelength dependent turn-on and sharp rise clearly showed that free-free absorption of a synchrotron spectrum could explain the observed light curves \cite{Chevalier1984,Weiler1986,Lundqvist1988}. 

Modeling of the radio emission, including a calculation of the CSM ionization and temperature, is discussed in \cite{Lundqvist1988}. The deduced mass loss rates of SN 1979C and 1980K were $\sim 1\times 10^{-4} \ \Msunyr$ and $\sim 3\times 10^{-5} \  \Msunyr$, respectively, (assuming a wind velocity of $10 \ \kms$). These high mass loss rates were the first indications that these SNe were not undergoing normal mass loss, characteristic of red supergiants, but more typical of a short-lived a super-wind phase. 

There is also evidence for a periodic modulation of $\sim 4.3$ years of the radio light curve of SN 1979C, which has been interpreted as either pulsations of the progenitor or binary interaction by a massive companion in an eccentric orbit \cite{Weiler1992}.  

\index{VLBI}For SN 1979C\index{SN 1979C}  VLBI observations have been possible up to 2005 \cite{Bartel2003,Marcaide2009b}. An analysis show an expansion of the radio shell that is close to free expansion.

Both SN 1979C and SN 1980k have been observed in the optical up to very recent epochs, now showing broad lines of especially [O I-III] \cite{Milisavljevic2012}. It is therefore likely that they are now evolving into oxygen rich remnants, although the reverse shock has not yet reached the oxygen rich regions. The optical lines are instead likely to arise as a result of heating of the un-shocked interior by the X-rays from the reverse shock \cite{Chevalier1994}, as discussed in Sect. \ref{sec_cds}.

\subsection{Type IIb}
\label{sec_iib}

\begin{figure}[t]
\centering
\begin{center}
\includegraphics[scale=0.5,angle=0,origin=c]{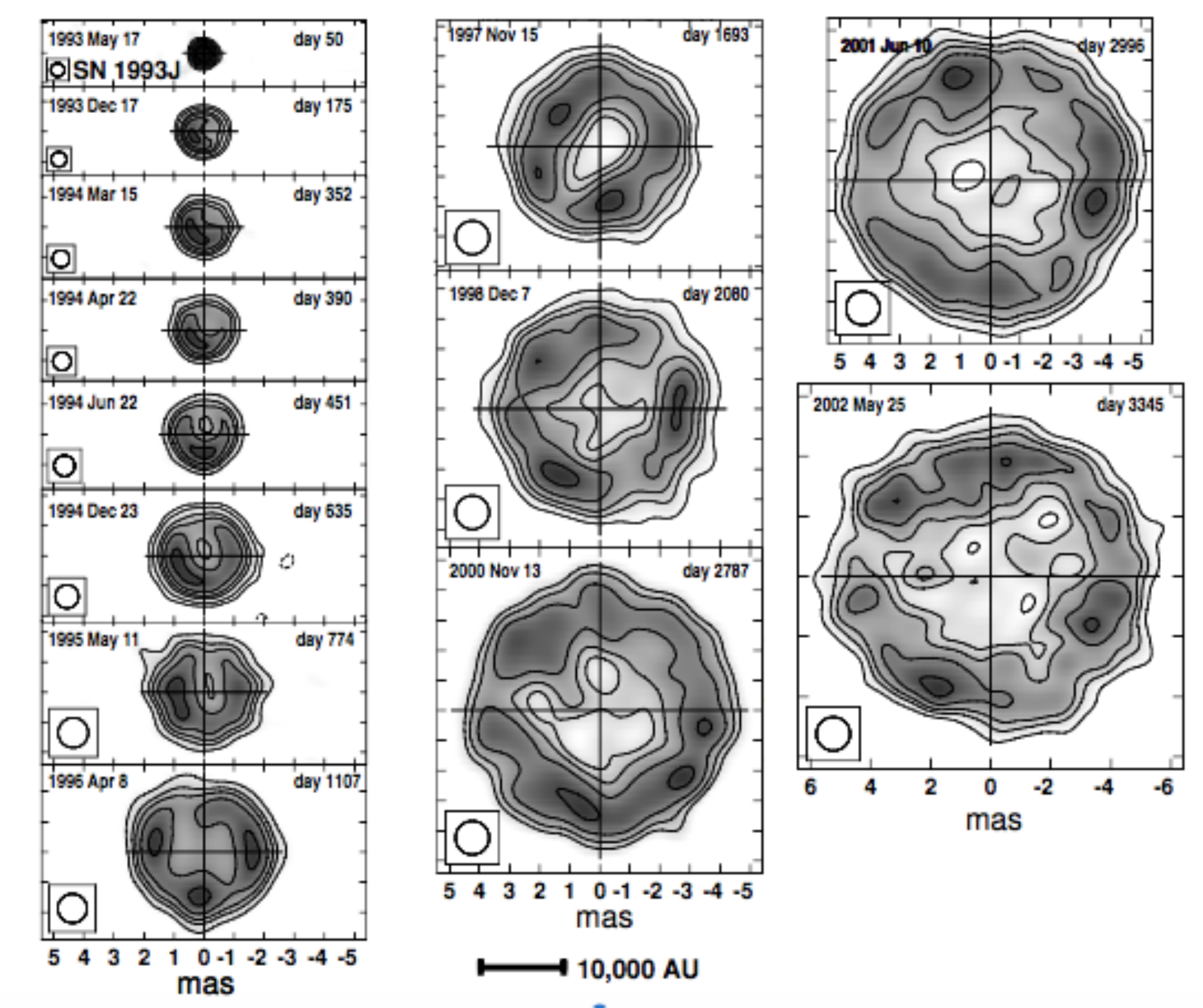}
\end{center}
\caption[]{Sequence of VLBI images of SN 1993J at 8.4 GHz between 50 to 3335 days after explosion. With permission from \cite{Bartel2009}.}
\label{figvlbi}
\end{figure}

\index{Type IIb} The best radio observations of any SN were obtained for SN
1993J\index{SN 1993J}. This SN was observed from the very beginning until more than a decade after explosion
with the VLA at wavelengths between 1.3 -- 90 cm \cite{vanDyk1994,Weiler2007}, producing a
set of beautiful light curves. In
addition, the SN was observed  with \index{VLBI} VLBI by two groups up to at least 16.9 years after explosion  at 1.7 GHz \cite[and references therein]{Bietenholz2003,Marcaide2009,Bietenholz2010},
resulting in an impressive sequence of images 
in which the radio emitting plasma could be
directly observed (Fig. \ref{figvlbi}). These images showed a remarkable degree of symmetry
and clearly resolved the shell of emitting electrons. The evolution of
the radius of the radio emitting shell could be well fitted by $R_{\rm s}
\propto t^{0.86}$, implying a deceleration of the shock front, although there has been some discussion of the exact value of the exponent.

From a fit of the observed spectra for the different epochs the
magnetic field and number of relativistic electrons could be
determined for each epoch, as described in Sect. \ref{sec_abs} and in \cite{Fransson1998}. 
Both these parameters showed a remarkably  smooth
evolution, with $B \approx 6.4 (R_{\rm
s}/10^{16} ~{\rm cm})^{-1}$ G, and $n_{\rm rel} \propto \rho V^2
\propto t^{-2}$. The
magnetic field\index{magnetic field amplification} was close to equipartition, $B^2/8 \pi \approx 0.14 ~\rho
V_{\rm s}^2 $, i.e., $\epsilon_{\rm B} \approx 0.14$, orders of magnitude higher than expected if the circumstellar magnetic
field, of the order of a few mG, was just compressed, and strongly
argues for field amplification, similar to what was discussed in Sect. \ref{sec:_acc}. The energy density of relativistic electrons was, however, found to be much lower, $\epsilon_{\rm e} \sim 5\times 10^{-4}$.
Contrary to earlier, simplified models for SN 1993J\index{SN 1993J} 
based
on free-free absorption only \cite{Fransson1996,vanDyk1994}, the circumstellar
density  in these consistent models was found to have a  $\rho \propto r^{-2}$ profile.

 The high values of $B$ implied that synchrotron cooling was
important throughout most of the evolution for the electrons
responsible for the cm emission, and also for
the 21 cm emission before $\sim 100$ days.
 At early epochs, Coulomb losses\index{Coulomb losses} were important for
the low energy electrons. The injected electron spectrum was best
fit with $p_{\rm i} = 2.1$.

\begin{figure}[t]
\centering
\begin{center}
\includegraphics[scale=0.30,angle=-90,origin=c]{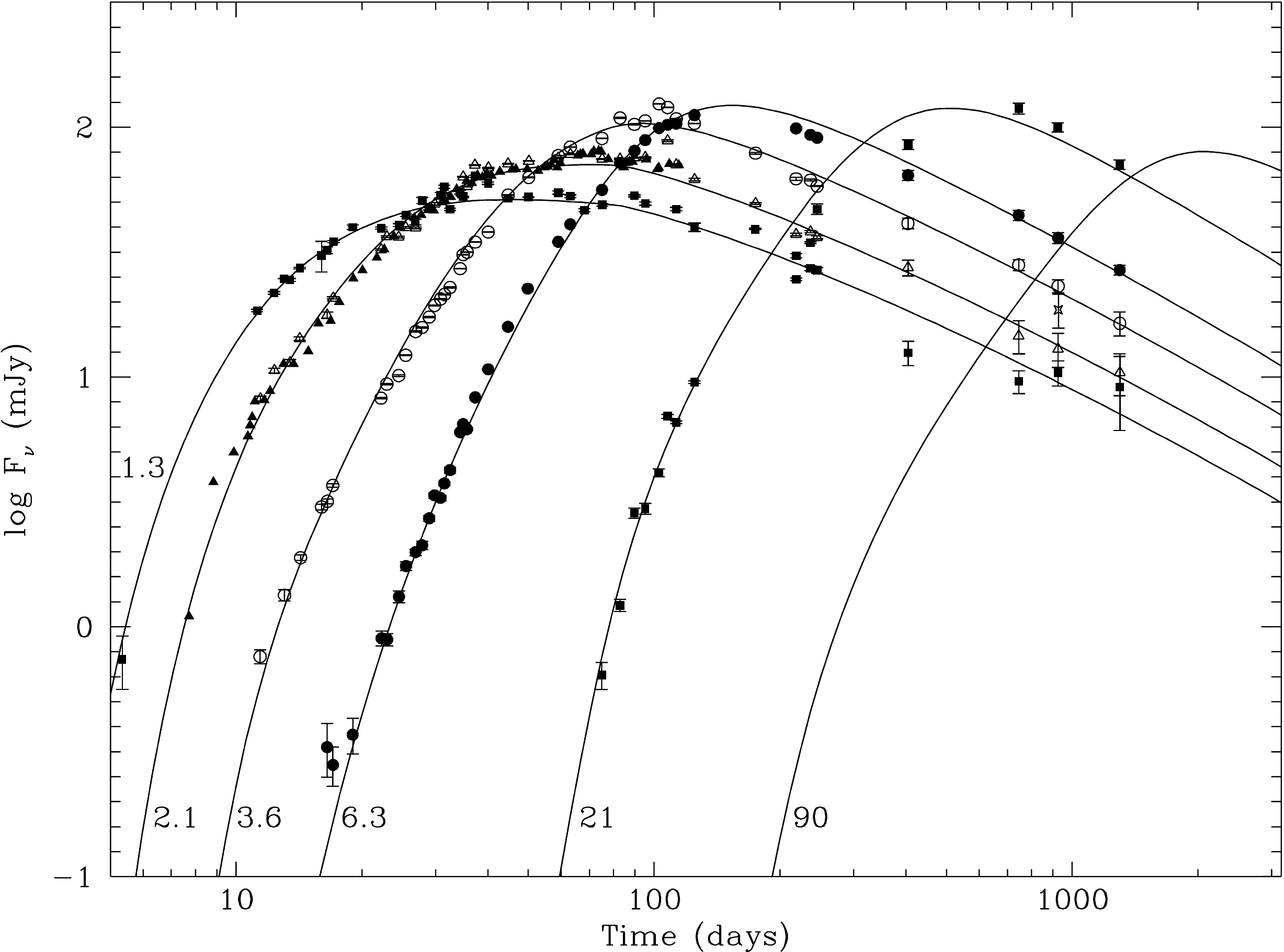}
\end{center}
\caption[]{Observed (dots) \cite{vanDyk1994} and model radio light curves (solid lines) of SN 1993J\index{SN 1993J}. With permission from \cite{Fransson1998}.}
\label{fig4a}
\end{figure}

In Fig. \ref{fig4a}, we show the excellent fit of the resulting light
curves (see also \cite{MartiVidal2011b}).
The form of the light curves can be understood if, for simplicity, we assume
equipartition, so that $B^2/8 \pi  \propto \rho V_{\rm
s}^2$. With $\rho \propto C_*~R_{\rm s}^{-2}$ and $V_{\rm s}
\propto R_{\rm s}/t$, we find that $B \propto C_*^{1/2} t^{-1}$. The
optically thick part is given by Eq. (\ref{eqss5}), so
\begin{equation} 
F_\nu(\nu) \propto ~R_{\rm s}^2~ \nu^{5/2} B^{-1/2} 
\propto C_*^{-1/4} \nu^{5/2}  t^{(5n-14)/2(n-2)},
\label{eq13c}
\end{equation}
since $R_{\rm s} \propto t^{(n-3)/(n-2)}$. For large $n$, we get
$F_\nu(\nu) \propto t^{5/2}$.  An additional curvature of the spectrum
is produced by free-free absorption in the wind, $\propto \exp(-\tau_{\rm ff})$, although this only
affects the spectrum at early epochs. The time dependence of the optically thin part is  given by Eq. (\ref{eq13r}). 

Although the discussion above assumes a spherically symmetric CSM there could well be irregularities and clumping\index{circumstellar clumps} in the CSM. 
Chugai \& Belous \cite{Chugai1999} propose a model in which the absorption
is by clumps.
The narrow line optical emission implies the presence of clumps,
but they are different from those required for the radio absorption.
The possible presence of clumps and irregularities introduces
uncertainties into models for the radio emission, although rough
estimates of the circumstellar density can still be obtained \cite{Bjornsson2013}.

SN 1993J  also has  the most extensive set of observations in the X-ray band with essentially all existing X-ray satellites. 
OSSE on board the Compton Observatory
 detected the SN in the hard 50-150 keV energy band on day 12 and
30 with an  X-ray luminosity of $\sim 5\times10^{40} \ergs$ on day 12
\cite{Leising1994}. However, by day 108 the 50-150 keV
emission had already faded below their detection limit. This is one of the very few hard X-ray observations of any SN. A fit to the spectrum gave a temperature of $82^{+61}_{-29}$ keV.

\begin{figure}[t]
\centering
\begin{center}
\includegraphics[scale=0.3,angle=0,origin=c]{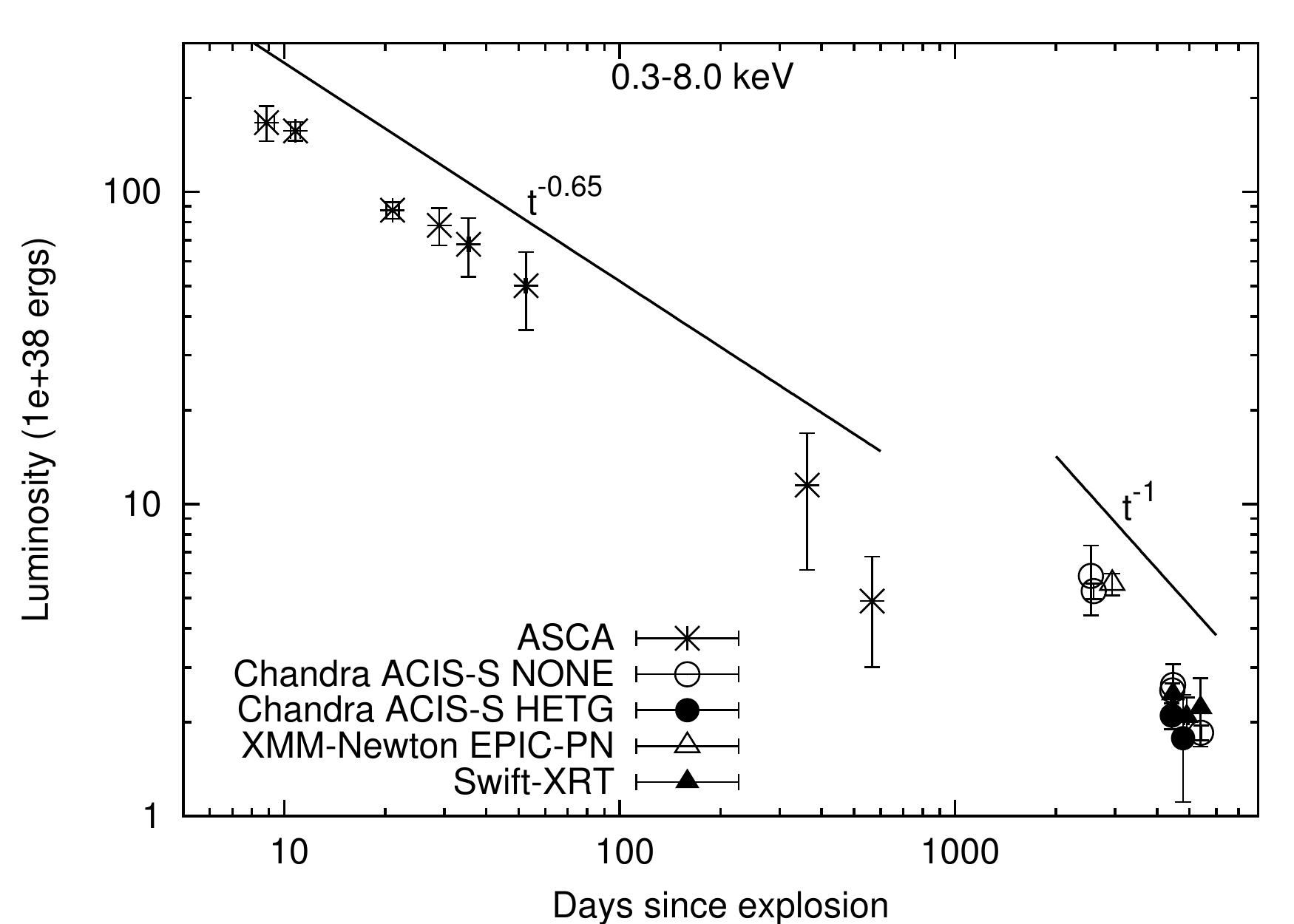}
\includegraphics[scale=0.3,angle=0,origin=c]{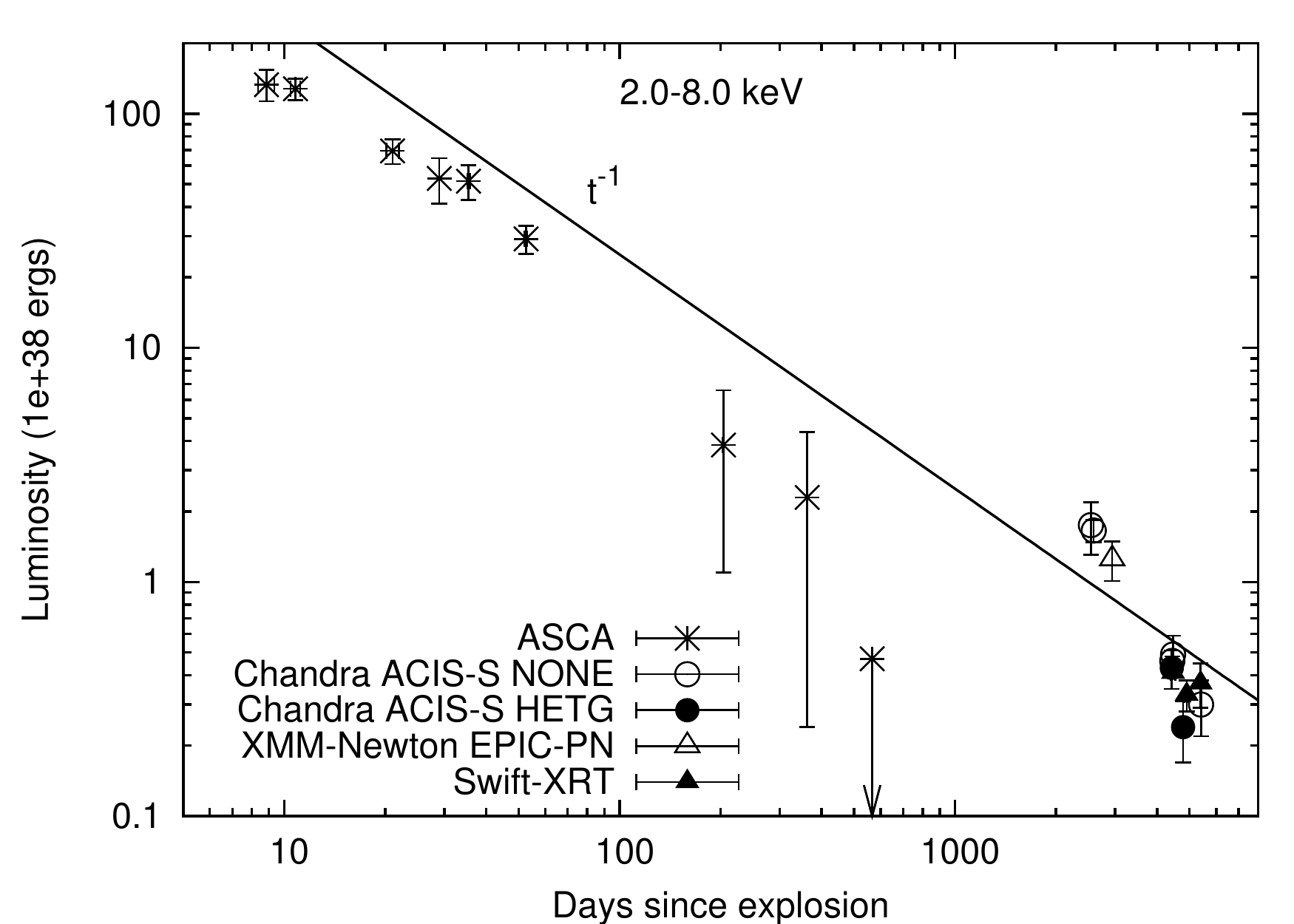}
\includegraphics[scale=0.3,angle=0,origin=c]{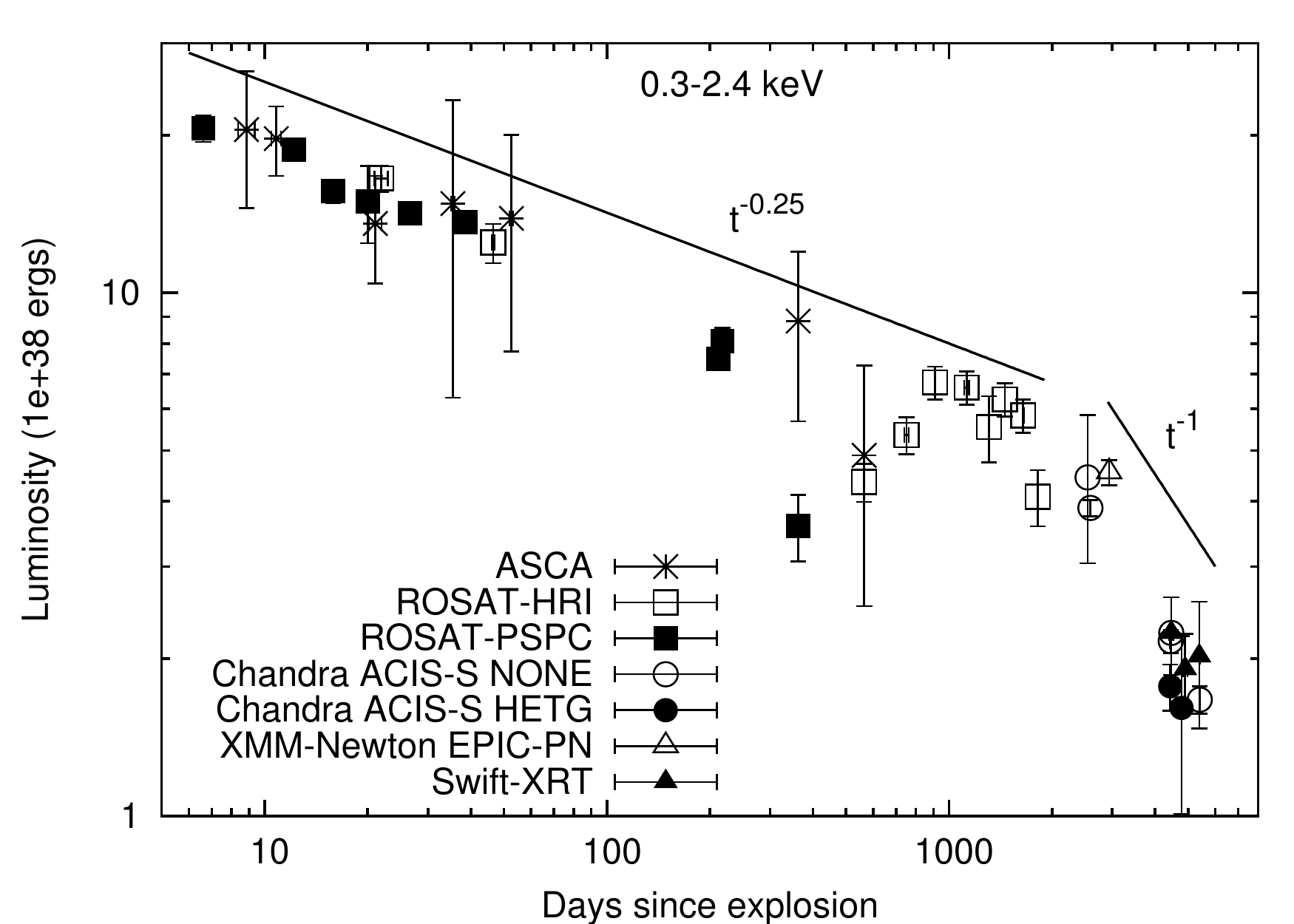}
\includegraphics[scale=0.3,angle=0,origin=c]{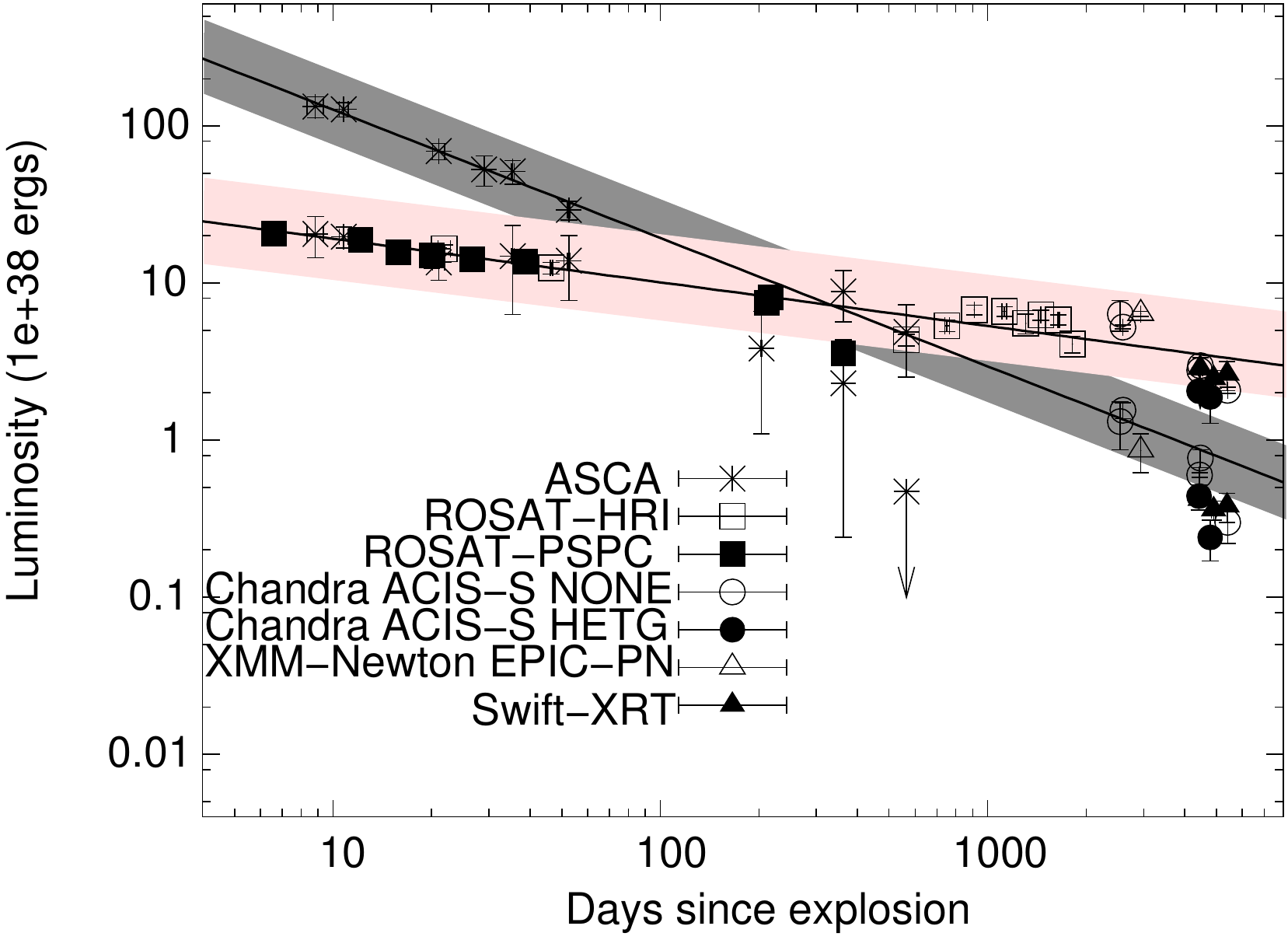}
\end{center}
\caption[]{Compilation of X-ray light cuves of SN 1993J in different energy bands. Note the slower decay rate in the soft 0.3 -- 2.4 keV band compared to the harder  2 -- 8 keV band. With permission from \cite{Chandra2009}.}
\label{fig93jx}
\end{figure}

ROSAT observed the SN in the 0.3-8 keV band from  day 6 up to day 1800, summarised in \cite{Zimmermann2003}. From the observations up to $\sim 50$ days only a lower limit to the temperature of $> 17$ keV could be obtained, consistent with the OSSE observation. However, when it was re-observed at $\sim 200$ days the spectrum had changed dramatically to a soft spectrum with $kT < 2$ keV. The ROSAT observations, together with additional ASCA, Chandra, Swift and XMM observations, have revealed a  continued softening of the spectrum, as illustrated by the slower luminosity decline in the 0.3 -- 2.4 keV band compared to the harder  2 -- 8 keV band, shown in Fig. \ref{fig93jx} where all observations below 10 keV are summarised \cite{Chandra2009}. 

\begin{figure}[t]
\begin{center}
\includegraphics[scale=0.22,angle=90,origin=c]{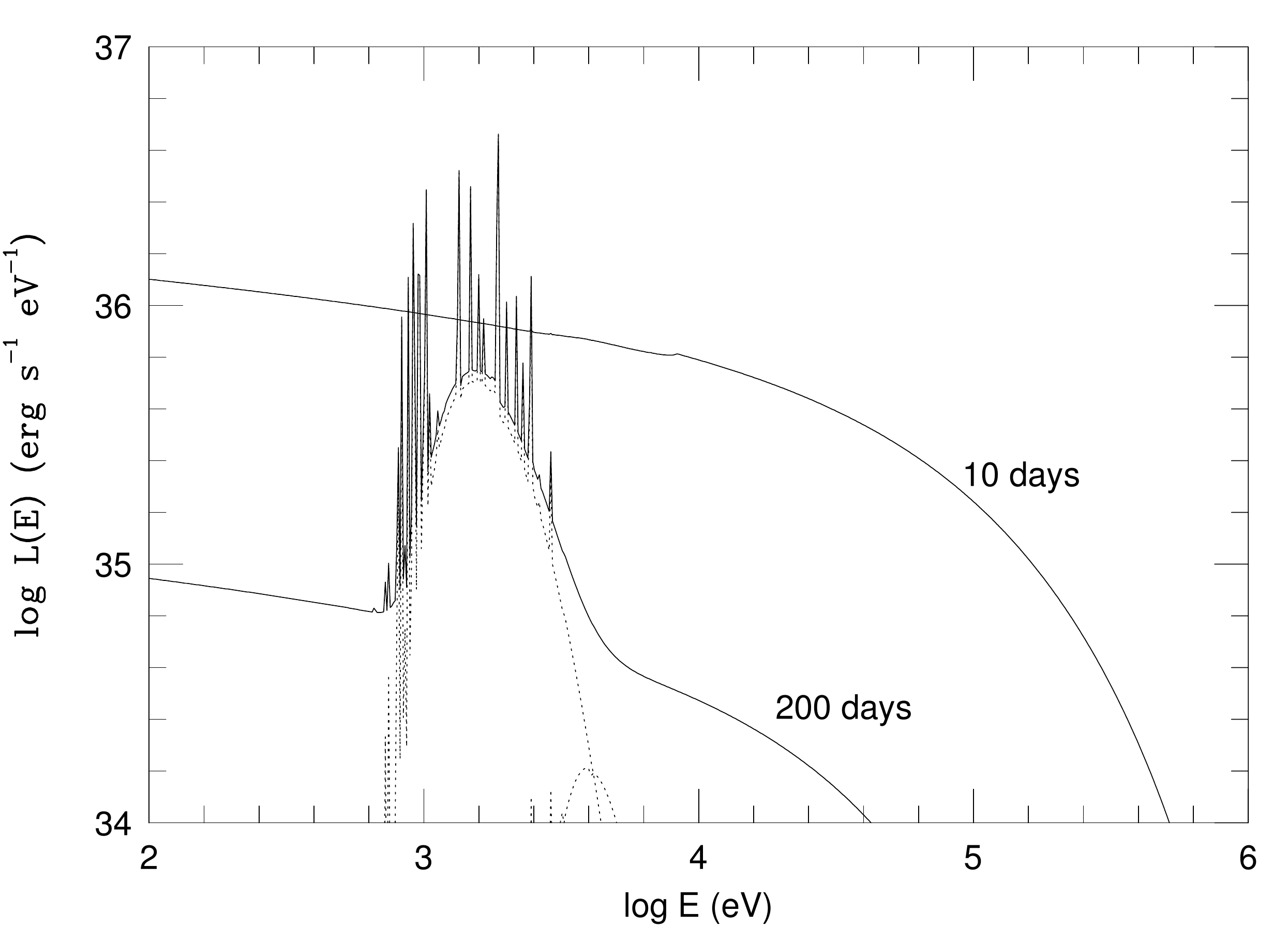}
\includegraphics[scale=0.22,angle=90,origin=c]{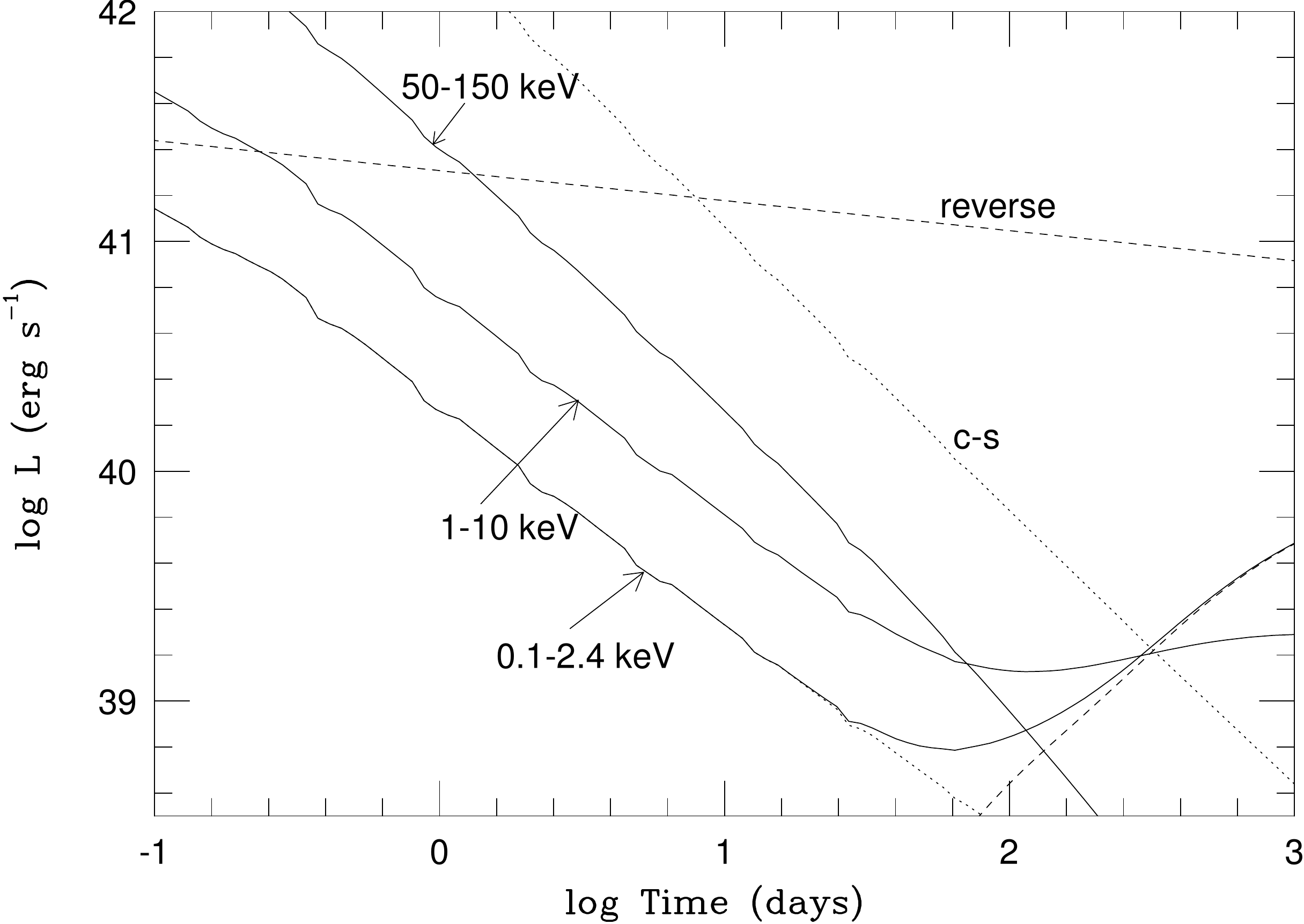}
\end{center}
\caption[]{left: Model X-ray spectra of SN 1993J at 10 days and at 200 days. At
10 days the free-free emission from the outer shock dominates, while
at 200 days the cool shell is transparent enough for the line
dominated spectrum from the reverse shock to dominate instead. Right:
The solid lines give the luminosity in the 0.1-2.4, 1-10, and 50-100
keV bands, corrected for absorption, as a function of time, while the
dotted lines give the total emitted luminosity from the reverse and
circumstellar shocks. With permission from \cite{Fransson1996}.}
\label{fig3}
\end{figure}

The transition from a hard to a soft X-ray spectrum can naturally be explained in the reverse - forward shock scenario discussed in Sect. \ref{sec_pscond}. At early epochs the column density of the cool, dense shell between the reverse and forward shocks is very large and X-rays from the reverse shock cannot penetrate the shell. Consequently, the emission is dominated by the hard flux from the forward shock\index{forward shock}.  At later epochs the soft spectrum from the reverse shock
penetrates the cool shell as the column density decreases, and the line dominated emission from the
cooling gas dominates. 
As an illustration, we show in Fig. \ref{fig3} the
calculated X-ray spectrum at 10 days and at 200 days for SN 1993J
\cite{Fransson1996}. At the first epoch the spectrum is dominated by the very
hard spectrum from the circumstellar shock, which reaches out to $\gsim
100$ keV.  The soft X-ray band is also dominated by the forward shock at this epoch. However, at 200 days the cool, dense shell\index{cool, dense shell} has become partly transparent and the soft spectrum from 
the line dominated emission from the
cooling gas behind the reverse shock\index{reverse shock} now dominates.  This evolution can also be seen in the light curves. 

Nomoto \& Suzuki  \cite{Nomoto1998}  modeled the X-ray light curve during the first $\sim 1000$ days with a similar scenario, but invoked clumping in the cool shell to explain the fact that soft X-rays were observed also at early epochs. Clumping is a natural consequence of the  Rayleigh-Taylor instabilities\index{Rayleigh-Taylor instability} expected in the decelerating, radiative shell \cite{Chevalier1995}.  The full evolution up to $\sim 15$ years after the explosion has been analyzed and modeled in \cite{Chandra2009}, who find that the reverse shock is likely to have been radiative up to $\sim 5$ years after explosion and adiabatic after that. This is consistent with the modelling in \cite{Nymark2009}, who modeled the 8 year XMM spectrum, requiring a mix of radiative and adiabatic shocks. As discussed in Sect. \ref{sec_pscond}, the time of the end of the radiative phase depends sensitively on both the composition and the structure of the ejecta. 

A clear signature of the cool, dense shell, and therefore  a radiative reverse shock, is the presence of broad, box-shaped line profiles, see in  H$\alpha$  and other lines after $\sim 1$ year, and present to at least 2500 days after explosion \cite{Matheson2000,Fransson2005}.

\subsection{Type IIn}
\label{sec_iin}
\index{Type IIn} Type IIn SNe probably originate from a wide range of progenitors.The main defining characteristic is the presence of narrow H and He lines originating in a dense CSM. The broad wings are a result of electron scattering\index{electron scattering} \cite{Chugai2001}.  Using these to estimate the expansion velocity of the ejecta can therefore be highly misleading. Instead, the line profile results from a random walk in frequency for each scattering.  
The thermal velocity of the electrons is
$V_{\rm therm} = 674 (T_{\rm e} /10^4  \ {\rm K})^{1/2} \kms$ and  the number of scatterings, $N \approx \tau_{\rm e}^2$. The FWHM of the line will therefore be
$\sim N^{1/2} V_{\rm therm} \approx 674 \ \tau_{\rm e} (T_{\rm e} /10^4 K)^{1/2} \ \kms$. For a FWHM $\sim 2000 \ \kms$, as observed in e.g., SN 2010jl \cite{Fransson2014}, one needs $ \tau_{\rm e} \gsim 3$ if $T_{\rm e} \sim 10^4$ K. 

The duration of this phase ranges from a few days, or weeks, like SN 1998S \cite{Fassia2001}  \index{SN 1998S} and SN 1995N \index{SN 1995N} \cite{Fransson2002}, to many years like SN 2010jl\index{SN 2010jl} \cite{Fransson2014}. The former have a fast decline of the optical luminosity, while the latter have broad light curves, lasting for a year or more. The integrated luminosity is consequently an order of magnitude higher for the SN 2010jl-like events. This suggests that these SNe may originate from two different classes of progenitors and it has been suggested that the SN 1998S-like SNe come from RSGs with enhanced mass loss rates, $\sim 10^{-4} \ \Msunyr$ \cite{Fransson2002}, like the super-wind of VY CMa, while the more long-lasting come from LBV progenitors, with episodic mass loss rates of $\sim 0.1 \  \Msunyr$. 
This is also consistent with their different host environments \cite{Taddia2015}, where the SN 1998S-like have metallicities similar to the Type IIP SNe, while the SN 2010jl-like have lower metallicities similar to the SN imposters. 
From a study of X-ray emission, Dwarkadas \& Gruszko \cite{Dwarkadas2012} found that many supernovae, especially
those of Type IIn, appear not to be expanding into a steady wind region.

\begin{figure}[t!]
\begin{center}
\includegraphics[width=8cm]{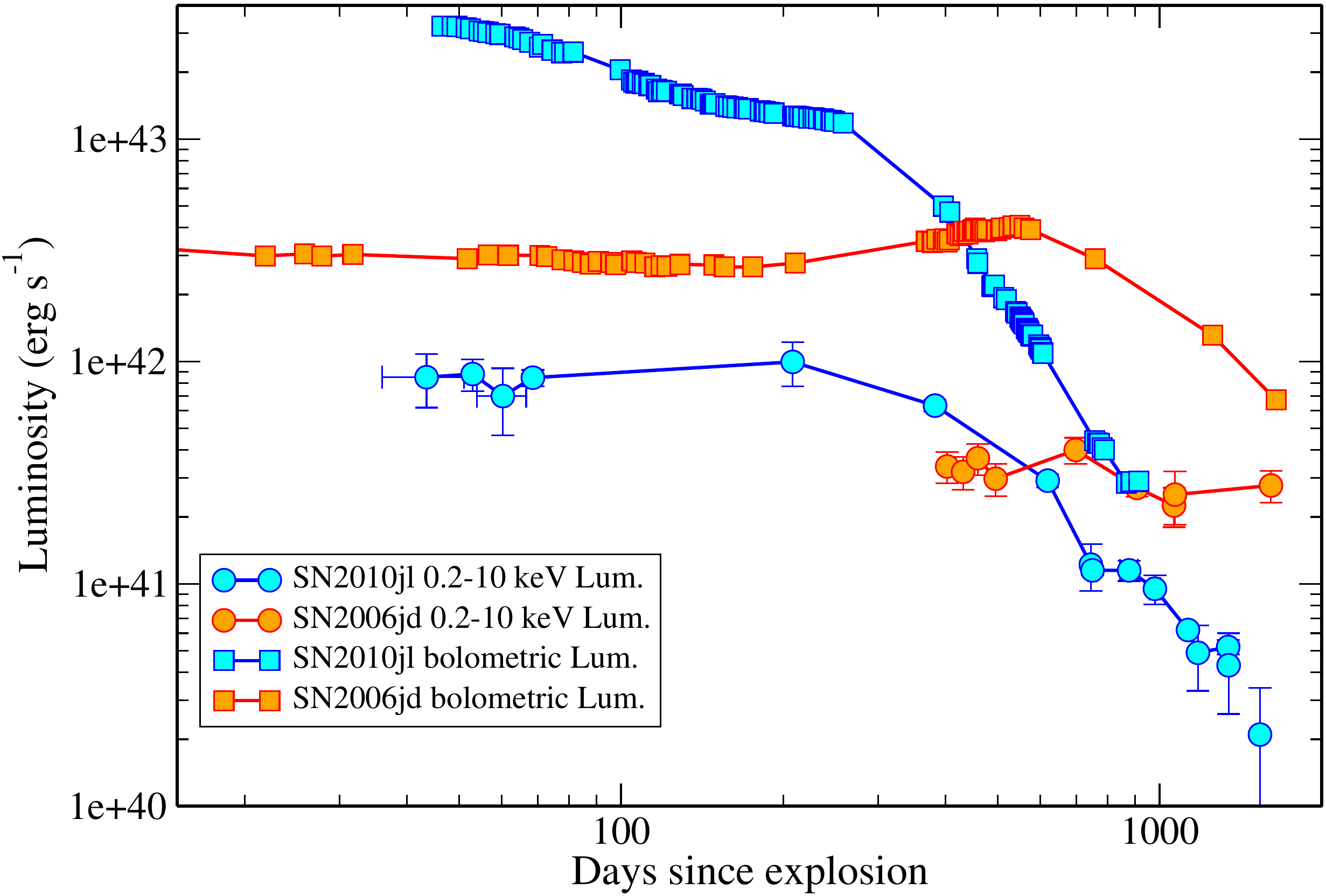}
\includegraphics[width=8cm]{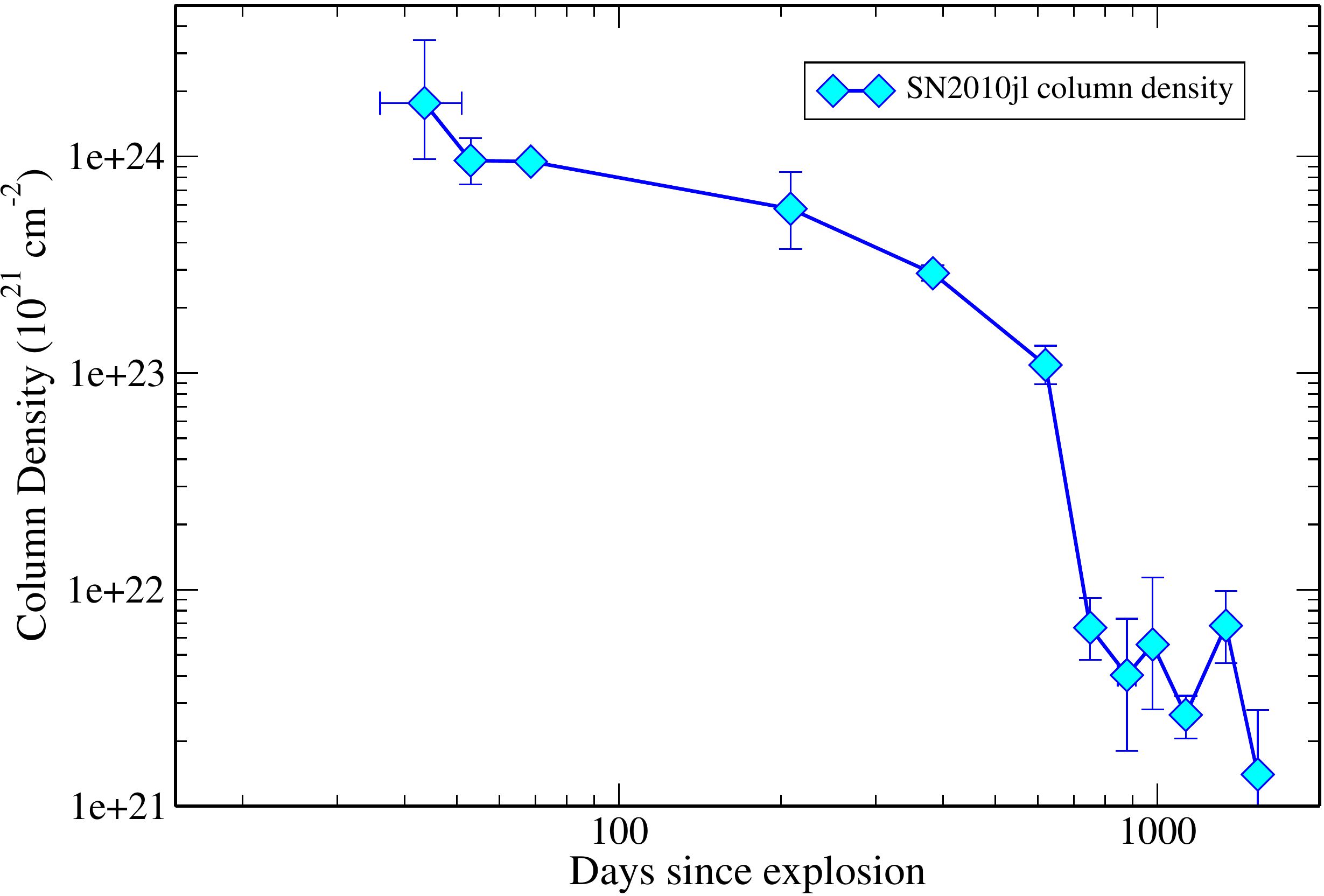}
\end{center}
\caption[]{Upper panel: Evolution of the 0.2-10 keV X-ray luminosity of the Type IIn SNe 2010jl and 2006jd (orange) together with the optical-NIR bolometric luminosity (blue). Lower panel: Evolution of the column density for SN 2010jl.  Note the decrease of the initially very large column density caused by the expansion of the ejecta. With permission from \cite{Chandra2015}.}  
\label{fig11}
\end{figure}

These long lasting, and therefore more energetic SNe, belong to the most  luminous SNe in the optical, radio and X-ray bands, including objects like SN 1988Z  \index{SN 1988Z}\cite{Aretxaga1999,Fabian1996,Williams2002}, and the well-observed SN 2010jl \cite{Fransson2014,Zhang2012,Ofek2014,Chandra2015}. The latter had an optical luminosity close to the superluminous Type II SNe. Fig. \ref{fig11} shows the bolometric light curve from the optical/NIR from \cite{Fransson2014}, together with the X-ray luminosity of SN 2010jl, as well as another luminous Type IIn, SN 2006jd  \cite{Stritzinger2012}. Importantly, it was also observed with NuStar, resulting in an X-ray temperature of $\sim 18$ keV \cite{Ofek2014}. Although very luminous, the initial X-ray luminosity of SN 2010jl is an order of magnitude lower compared to the optical luminosity \cite{Ofek2014,Chandra2015}. Later this ratio increases and after $\sim 1$ year the X-ray luminosity dominates. 

Equally interesting is the evolution of the X-ray column density, $N$, shown in the right panel of Fig. \ref{fig11}. At the first observation at $\sim 40$ days this was very high $N \sim 10^{24} \ \rm cm^{-2}$, but decreased with time. 
This may naturally be interpreted as a result of the decreasing column density as the ejecta are expanding. For a wind one expects a $N \propto t^{-1}$ decrease (Eq. \ref{eq11b}), while a shell may give a faster decrease.

Although very high, the X-ray column density was  lower than that needed in order to explain the electron scattering profiles, which were seen up to $\sim 1000$ days and require an optical depth of $\tau_{\rm e}\gsim 3$, corresponding to a column density of $\gsim 10^{25}  \ \rm cm^{-2}$. These two constraints are difficult to combine in a spherically symmetric model. 
In \cite{Fransson2014} it is instead proposed that, based on the hydrodynamical calculations by van Marle et al. \cite{vanMarle2010},  which preceded SN 2010jl, the SN ejecta are expanding in a dense  bipolar CSM, similar to that of Eta Car \cite{Smith2006}.  In this case, the density and column density are  considerably higher in the polar directions than in the equatorial direction, presumably due to mass loss from a fast rotating star. In this case von Zeipel's theorem implies that the strongest mass loss will occur from the poles. 

A SN exploding inside such a CSM would then have a shock velocity higher in the equatorial regions, $V_{\rm s} \propto  (\Mdot / u_{\rm w})^{-1/(n-2)}$ (Eqs. \ref{eq4} and \ref{eq4b}), and also deposit more of its energy in this direction. The shock would therefore first break through the shell in the equatorial direction, where the X-ray (and radio) emission would first be seen. It would then still be deep inside the CS shell at high latitudes, where most of the  emission is released as X-rays, which are then thermalized into optical radiation, resulting in a luminosity according to  Eq. (\ref{eqforlum}).

In radio, SN 2010jl\index{SN 2010jl} was first detected on day $\sim 700$ \cite{Chandra2015}, which is typical of these bright SNe, indicating that free-free absorption by a dense CSM absorbed the radio emission at earlier phases, consistent with a large mass loss rate parameter. 

High resolution optical spectra of SN 2010jl, as well as other IIns (e.g., \cite{Stritzinger2012}), showed narrow lines with velocity $ 100 - 1000 \ \kms$, which is in the range observed for the CSM from an LBV progenitor. From the state of ionisation, as well as from nebular diagnostics, the density of this component was inferred to correspond to a mass loss rate of $\sim 10^{-3} \ \Msunyr$, much lower than that needed to explain the large continuum flux. These two components resemble those also seen in Eta Car. 

Several of these bright Type IIn SNe have been followed for several decades. An interesting example is SN 1978K \index{SN 1978K}  \cite{Kuncarayakti2016}, where a high resolution spectrum from 2014 shows optical lines with a FWHM $\lsim 600 \ \kms$. This spectrum has hardly changed during 20 years, and shows that the SN is still interacting with the CSM. Kuncarayakti et al. \cite{Kuncarayakti2016} compare this spectrum to that of the ring of SN 1987A \cite{Groningsson2008} and find striking similarities, both with respect to line widths and degree of ionization. In both cases, the latter spans  a range from neutral, like H$\alpha$ and [O I] up to [Fe XIV]. In the case of SN 1987A we know that these lines arise as a result of radiative shocks propagating into dense clumps in the ring, and Kuncarayakti et al. speculate that a similar geometry may be present in SN 1978K, in spite of the symmetric lines. 

UV lines of C III], C IV, N III], N IV] and O III] are useful as abundance indicators and there are for several Type IIL and Type IIn, including SNe 1979C\index{SN 1979C},  1995N\index{SN 1995N}, 1998S\index{SN 1998S} and 2010jl\index{SN 2010jl}, strong evidence for CNO processing from lines in the circumstellar medium \cite{Fransson1984b,Fransson2005,Fransson2002,Fransson2014}. This provides independent evidence for strong mass loss from the progenitor. 

\index{SN 1986J} The Type IIn SN 1986J has been followed in both radio, optical and X-rays. The VLBI \index{VLBI} imaging is especially interesting since it gives a spatially resolved view of the interaction (e.g., \cite{Bietenholz2010b,PerezTorres2002}). These images show a highly non-spherical structure with a number of 'protrusions', indicating interaction with large scale clumps. The average expansion velocity measured between  1999 and 2008 was $5700 \pm1000 \kms$, considerably higher than in the optical at the same epochs \cite{Milisavljevic2008}, where the emission was probably mainly coming from the ionized oxygen emitting core of the SN. Up to $\sim 2002$ the  VLBI structure was dominated by one big blob off-center. At the same time as this decayed a new central blob appeared with increasing flux, showing a free-free absorbed spectrum with a decreasing optical depth. Bietenholz et al.  \cite{Bietenholz2010b} propose that this may either be the result of the emission from a compact object, explaining the central position, or as a result of the interaction of a large scale cloud which happens to be in the line of sight to the centre. An alternative interpretation is that the central emission is the result of circumstellar interaction with the inner boundary of a dense disk resulting from common envelope interaction of the progenitor \cite{Chevalier2012}. The continued evolution of the radio structure is of obvious interest. 

Chugai \cite{Chugai1993} has proposed that the X-ray emission from SN 1986J is the result of the forward shock front moving
into clumps\index{circumstellar clumps}, as opposed to the reverse shock emission.
One way to distinguish these cases is by the width of line emission;
emission from the reverse shock wave is expected to be broad.
It has not yet been possible to carry through with this test
\cite{Houck1998}.

\subsection{SN 1987A}

Recent developments in connection to both the ejecta and the ring interaction are reviewed in \cite{McCray2016}, including full references to the literature.
We therefore limit ourselves to some general remarks of interest for CS interaction. 

Because of its distance \index{SN 1987A} SN 1987A  is unique in that one can  resolve the CS environment around the SN. The most apparent structure of the CSM is the ring system, consisting of the equatorial ring at a distance of $\sim 6\times 10^{17}$  cm and the outer rings at $\sim 2 \times 10^{18}$ cm. 
 One can therefore study the different regions of the interaction in detail.  Although the CSM structure of SN 1987A probably is not typical to the majority of core collapse SNe, the fact that we can study it resolved, and also time dependent, means we can learn a great deal from a study of it in the different wavelength domains. 

\begin{figure}[t!]
\begin{center}
\includegraphics[width=10cm]{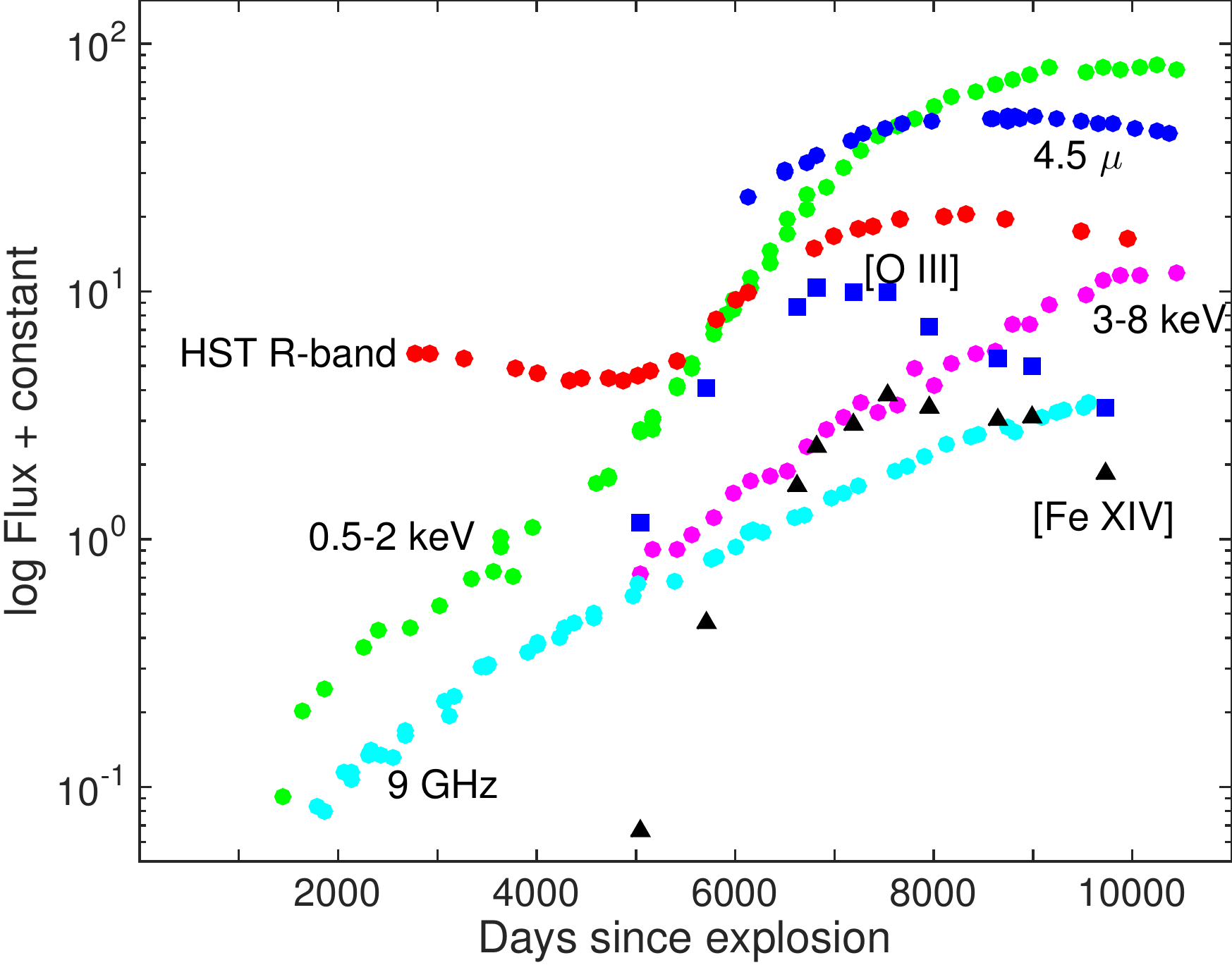}
\end{center}
\caption[]{Compilation of light curves in different wavelength bands and lines from the ring of SN 1987A. For clarity the different band and lines are arbitrarily scaled in flux. The HST R-band (red dots), [O III] 5007 \AA \ (blue squares)  and [Fe XIV] 5303 \AA \ (black triangles) points are  from \cite{Fransson2015}. The Chandra and ROSAT 0.5 - 2 keV (green dots) and 3 - 8 keV (magenta dots) points are  from \cite{Frank2016} and \cite{Haberl2006}, the IRAC 4.5 $\mu$m (blue dots) points from \cite{Arendt2016} and the ATCA 9 GHz (cyan dots) from  \cite{Ng2013} .}  
\label{fig_lc_87a}
\end{figure}

In Fig. \ref{fig_lc_87a} we show the evolution of the flux from the ring and the general CSM interaction in different wavelength ranges, which reflect the structure of the CSM. Initially SN 1987A was extremely faint in both radio and X-rays, consistent with the low mass loss rate, $\sim 10^{-7} \ \Msunyr$, and high wind velocity, $\sim 450 \kms$, associated with the B3Ia progenitor star \cite{Chevalier1995b}. At $\sim 1200$ days the radio flux, however, increased rapidly and has since day $\sim 2000$  increased almost exponentially \cite{Zanardo2010}. At about the same time as the radio increased also X-rays were observed and followed the evolution of the radio emission up to day $\sim 4500$. After the impact with the ring the soft X-rays increased even faster up to $\sim 7000$ days  \cite{Haberl2006}, after which it has levelled off  \cite{Frank2016}. The pre-impact evolution was explained as a result of the interaction of the nearly freely expanding ejecta with the photoionized H II region, formed by the progenitor, having a nearly constant density, $\sim 100 \ccm$  \cite{Chevalier1995b}. As the  decelerating ejecta started to interact with the clumps forming the CS ring, an increasing number of hotspots became visible in the optical, while at the same time  the soft X-ray flux increased at the same rate.

 At $\sim 7000$ days the optical flux reached a peak, and has since decreased, while new spots have appeared outside of the ring \cite{Fransson2015}. This evolution is seen most clearly for the different high ionization lines, like the [O III] 5007 \AA \ and [Fe XIV] 5303 \AA \ lines formed in the cooling gas behind the shocks (Fig. \ref{fig_lc_87a}).  The mid-IR shock heated dust emission \cite{Arendt2016} has also reached a peak, while the X-ray emission has leveled off \cite{Frank2016}. The hard X-rays and  the radio emission are continuing to rise \cite{Frank2016,Zanardo2010}. 
 
The basic scenario for this interaction, once the collision with the clumps in the ring  started,  is that of a forward shock propagating into the clumps\index{circumstellar clumps}, while the pressure behind the shock sends a reflected shock back into the ejecta which gives rise to a reverse shock\index{reverse shock}. Because the density of the clumps is $\gsim 10^4 \ccm$, while the ejecta density close to the shock is only $\sim 10^2 \ccm$ the velocity of the forward shock will be a factor $\sim (\rho_{\rm clump}/\rho_{\rm ej})^{1/2} \approx 10 $ lower than the forward shock (Eq. \ref{eq_vc}). From the expansion seen in the X-ray   and radio images the velocity of the  forward shock is $2000-4000 \kms$, giving a shock velocity into the clumps of $\sim 200 - 500 \kms$. 

 The difference between the evolution of the optical and mid-IR light curves on the one hand and the X-ray and radio on the other
  can be understood since the optical and dust emission are only coming from the shocked clumps in the ring; the X-ray and radio emission have a major contribution from lower density regions between the clumps as well as above and below the ring plane. The SN ejecta are now starting to probe the region outside the ring, which will provide insight into the mass loss history of the progenitor. 

The interaction of the SN ejecta and the CSM in SN 1987A is a nearly perfect shock laboratory, and are useful for understanding both the thermal and non-thermal processes discussed earlier in this review. From the radio \cite{Zanardo2010} and hard X-rays \cite{Frank2016} one can study the acceleration of relativistic particles at the reverse shock, in both time and space. The optical (e.g., \cite{Groningsson2008b}) and soft X-rays (e.g., \cite{Dewey2008}) in contrast give   information about the thermal processes from the  shocks propagating into the dense clumps in the ring, while the harder X-rays mainly come from non-radiative higher velocity shocks.

From the optical line widths of the hotspots one finds velocities between $200 - 700 \ \kms$. It is therefore clear that these lines are shock excited, in contrast to direct photoionization, which would give lines with FWHM $\sim 10 \kms$. For  $T_{\rm e} \lsim \ 2.6\EE7 \KK$ line emission dominates the cooling rate, so
$\Lambda(T_{\rm e}) \approx 2.4\EE{-23}~(T_{\rm e}/10^7 \ {\rm K})^{-0.48} \ \ergs cm^{3}$, and with the shock condition for $T_{\rm e}$, the cooling time behind the shocks propagating into the dense clumps becomes
\begin{equation}
t_{\rm cool} = 38 \left({V_{\rm s} \over 500 \ \kms}\right)^{3.4}
\left({n_{\rm e} \over 10^4 \ \ccm}\right)^{-1} \  \rm years.
\label{eq_cool}
\end{equation}
where $n_{\rm e}$ is the pre-shock electron density. The presence of lines with a velocity $\sim 700 \ \kms$ $\sim 15$ years after the impact therefore implies a pre-shock density of $\gsim 4 \times 10^4 \ccm$, in agreement with the highest densities determined from the un-shocked ring \cite{Lundqvist1996}. These shocks belong to the highest velocity radiative shocks we know of.

Shocks with higher velocity do not cool and are only seen in X-rays, with the exception of some coronal lines\index{coronal lines}, like the [Fe XIV] line shown in Fig. \ref{fig_lc_87a}. This applies to the outward propagating blast wave between the clumps, as well as below and above the ring plane, which now expands with a velocity of $\sim 1854 \kms$ \cite{Frank2016}.  It is also the case for the reflected and reverse shocks moving back into the ejecta. The X-ray emission is therefore the sum of all these different components. 

The maximum ejecta velocity seen from the broad H$\alpha$ component is $\sim 11,000 \kms$. This emission is coming from the region close to the reverse shock, where neutral H atoms are excited and ionized by the hot gas behind the reverse shock. From the radio and X-rays the expansion velocity in the observer frame of the reverse shock has been estimated to $2000-4000 \kms$. The  reverse shock velocity is therefore $\sim 8000 \kms$, and the emission is therefore expected mainly from hard X-rays. This is also the region where most of the acceleration of the relativistic electrons, responsible for the radio emission, takes place. 

\begin{figure}[t!]
\begin{center}
\includegraphics[width=5.5cm]{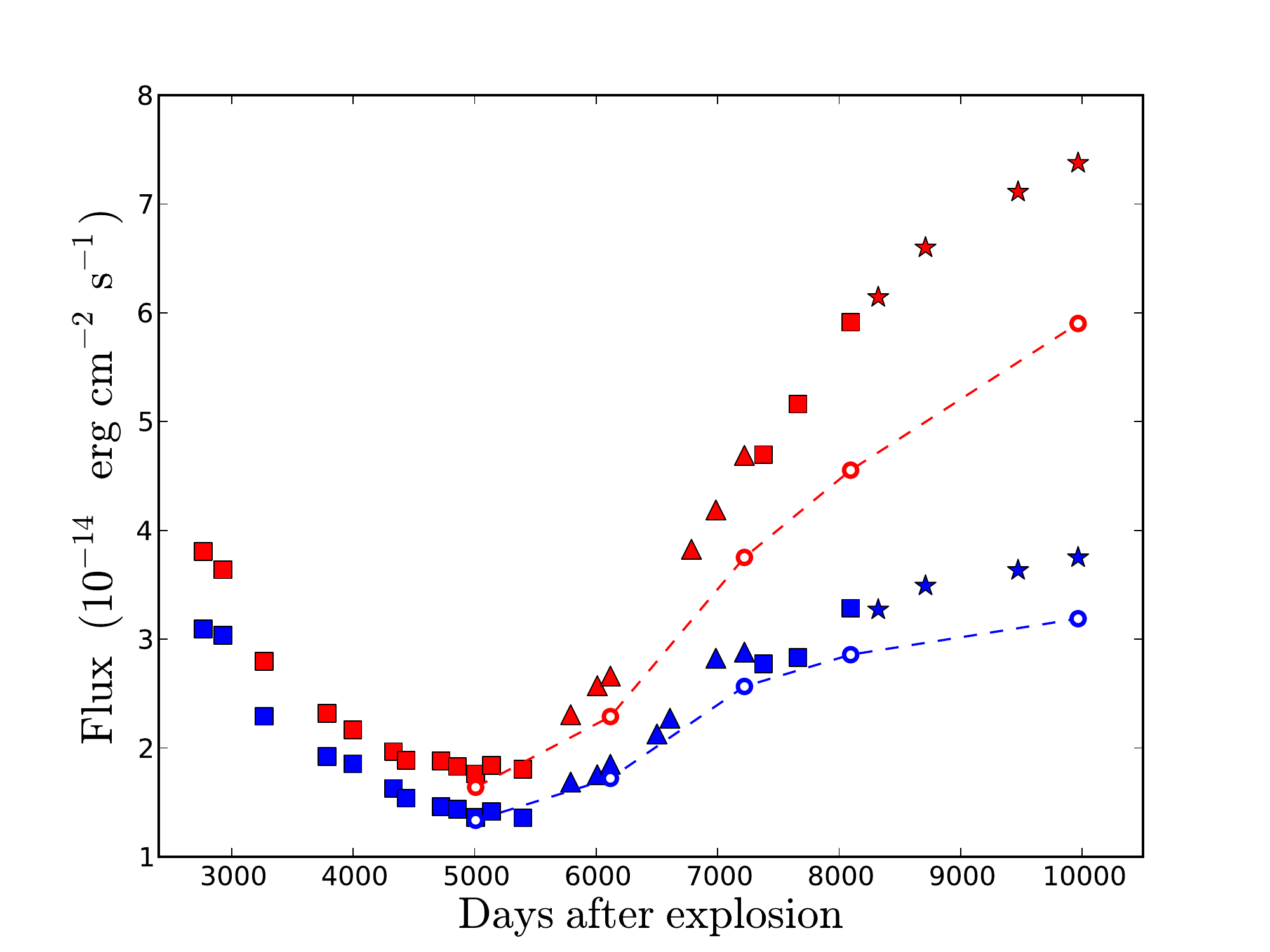}
\includegraphics[width=5.5cm]{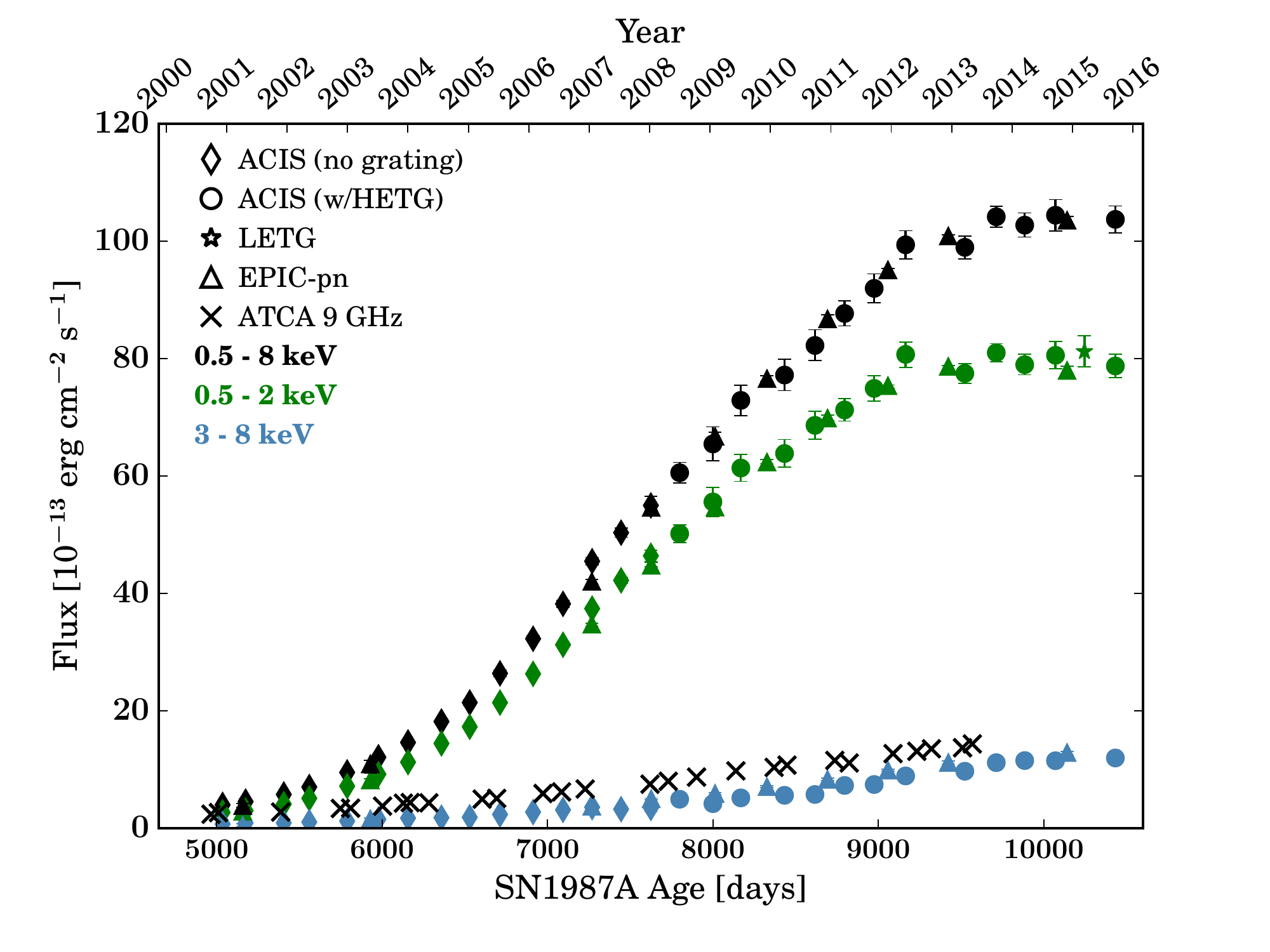}
\end{center}
\caption[]{Optical R (red) and B (blue) band light curves of the ejecta (left) (update of \cite{Larsson2011}) and X-rays (right)  of SN 1987A, with permission from \cite{Frank2016}.}  
\label{fig_lc_ring_ejecta_87a}
\end{figure}

SN 1987A provides one of the best examples of the thermalization of the X-rays from the circumstellar interaction  by the SN ejecta (Fig. \ref{fig_lc_ring_ejecta_87a}). Up to day $\sim 5000$  the light curve of the ejecta was decaying with a rate consistent with that of the radioactive decay. However, after that it has increased by a factor 2--4 \cite{Larsson2011}. 
At the same time the morphology of the ejecta changed from centrally peaked to a horseshoe like shape. 

This can easily be understood when we compare this light curve with the X-ray light curve in Fig. \ref{fig_lc_ring_ejecta_87a}. Both the increase of the optical flux and the change in morphology  can consistently be explained as a result of the X-ray deposition from the ring collision by the ejecta. The soft and hard X-rays can penetrate much of the H and He rich outer regions of the ejecta. At the boundary of the metal rich core the photoelectric opacity increase sharply and most of the X-rays are here deposited, giving rise to the limb brightened morphology \cite{Fransson2013}. We can therefore  directly study the effects of the CSM interaction on the ejecta with spatial resolution. 

The hydrodynamic interaction of the ejecta and the CSM of SN 1987A has been modeled by 3D hydrodynamic simulations by Potter et al. \cite{Potter2014} and by Orlando et al. \cite{Orlando2015}. The former authors had an emphasis on the radio evolution, while the latter had their emphasis on the X-ray evolution. Both groups assume a CSM structure similar to that by Chevalier \& Dwarkadas \cite{Chevalier1995b}. 
Because this structure was proposed to explain the early radio and X-ray evolution it is not surprising that they can explain the early evolution in these bands well. 

The second phase is dominated by the interaction with the ring, which dominates the soft X-rays as well as the optical emission. From their high resolution AMR FLASH simulations Orlando et al. model the fragmentation of the clumps in the ring in considerable detail. This was already done for individual clumps in 2D with high resolution by Borkowski et al. \cite{Borkowski1997}, although with a more simplified model for the ejecta and CSM.  The main result of this is that the clumps are crushed by the radiative shock and shredded by the Richtmyer-Meshkov and Rayleigh-Taylor instabilities behind it.

In the final phase, beginning at $\sim 32$ years, the blast wave has propagated past the ring, while the reverse shock is propagating deep into the envelope of the SN.  The X-ray emission is dominated by the ejecta and the morphology is characterized by shocked ejecta clumps. It is now becoming a more classical remnant. The fact that the optical, X-ray, mid-IR and radio fluxes are decreasing or leveling off, as discussed above, shows that we are now in this transition phase. 

The radio evolution is complex, showing a varying spectral index, increasing rapidly from $\alpha\approx 0.75$ on day 1200 (the impact at the termination shock of the BSG wind) to  $\alpha \approx 1.0$  on day 2400 after which it decreased linearly to $\alpha \approx 0.7$ on day 8000 \cite{Zanardo2010}. Another interesting aspect is the east -- west asymmetry of the radio emission. In contrast to the X-rays, the radio intensity is strongest on the eastern side, which is also where the expansion velocity measured from the radio images is considerably faster ($\sim 6000 \ \kms$) than on the western side ($\sim 1900 \ \kms$). From the relation $B \propto V_{\rm s}^{1.5}$, predicted by the Bell instability \cite{Bell2004}, Potter et al. \cite{Potter2014} argue that the higher intensity on the eastern side is a direct consequence of the stronger magnetic field there, which in turn is a consequence of the higher expansion velocity in this direction. The latter may be a result of an anisotropic explosion geometry in the ejecta. 

There are several things we can learn from these observations, relevant to the observations of more distant SNe. The most important is perhaps that the CSM around the SN progenitor can be very complex, reflecting the evolution of the progenitor star. The simple spherically symmetric structure most often assumed in more distant unresolved SNe, may  sometimes be a poor approximation. This applies to both the CSM and the SN ejecta, which in the case of SN 1987A are both  highly asymmetric. In addition, SN 1987A shows that even if no CS interaction is seen at early times, it may become important after several years, depending on structures at large distances from the SN. In this respect it is interesting to compare with the late optical, radio and X-ray turn-on for the Type Ib/c SNe (Sect. \ref{sec_ibc})

There are also other aspects of SN 1987A  which are seen in other SNe.  E.g.,  the  coronal lines which are seen in Type IIn and Ibn SNe are also observed in spectra from the SN 1987A ring \cite{Groningsson2006}. In SN 1987A these are clearly formed in the post-shock gas. For more distant SNe high resolution spectra are, however, needed to distinguish between shock excitation and photoionization by X-rays of unshocked gas. If shocked emission, one can then from Eq. (\ref{eq_cool})  estimate the pre-shock density.

For the X-rays seen from shell interaction in Type Ib/c SNe (Sect. \ref{sec_ibc}) or from a clumpy CSM one may have a similar situation to that of SN 1987A with several shock components contributing to the flux as from the blast wave and reverse shock in SN 1987A.  

\subsection{Type Ib/c}
\label{sec_ibc}

A review of radio and X-ray observations of \index{Type Ib/c} Type Ibc SNe is presented in \cite{Chevalier2007}.  In general, these SNe have weaker radio and X-ray emission (Fig. \ref{figxraylc}) and a more gradual power law  turn-on of their light curves compared to Type II SNe. The weaker emission is most likely a result of the higher wind velocities, $\gsim 1000 \ \kms$, of the progenitor stars, thought to be Wolf-Rayet stars. The CSM density is therefore a factor of $\gsim 100$ lower compared to a red supergiant progenitor of the same mass loss rate. 

The power law turn-on is typical  for synchrotron-self-absorption, which well fits the light curves and spectra \cite{Chevalier2006}. From the optically thin emission it is found that the spectral index in the radio in general is steeper than for other types of SNe, with $\alpha \approx 1$. Because of the low CSM density synchrotron cooling is not important at radio frequencies. At early times Compton cooling\index{Compton cooling} may be important as long as the bolometric luminosity is $\gsim 2\times 10^{42} \ \ergs$ \cite{Bjornsson2004}, but becomes unimportant as the SN fades. The steep spectral index therefore implies a steep electron injection spectrum with $p \approx 3$, possibly a result of a cosmic ray dominated shock. 

Because of the high shock velocity and low CSM density both the forward and reverse shocks are likely to be adiabatic, unless the reverse shock is propagating into a heavy element dominated core. 

\begin{figure}[t!]
\begin{center}
\includegraphics[width=8cm]{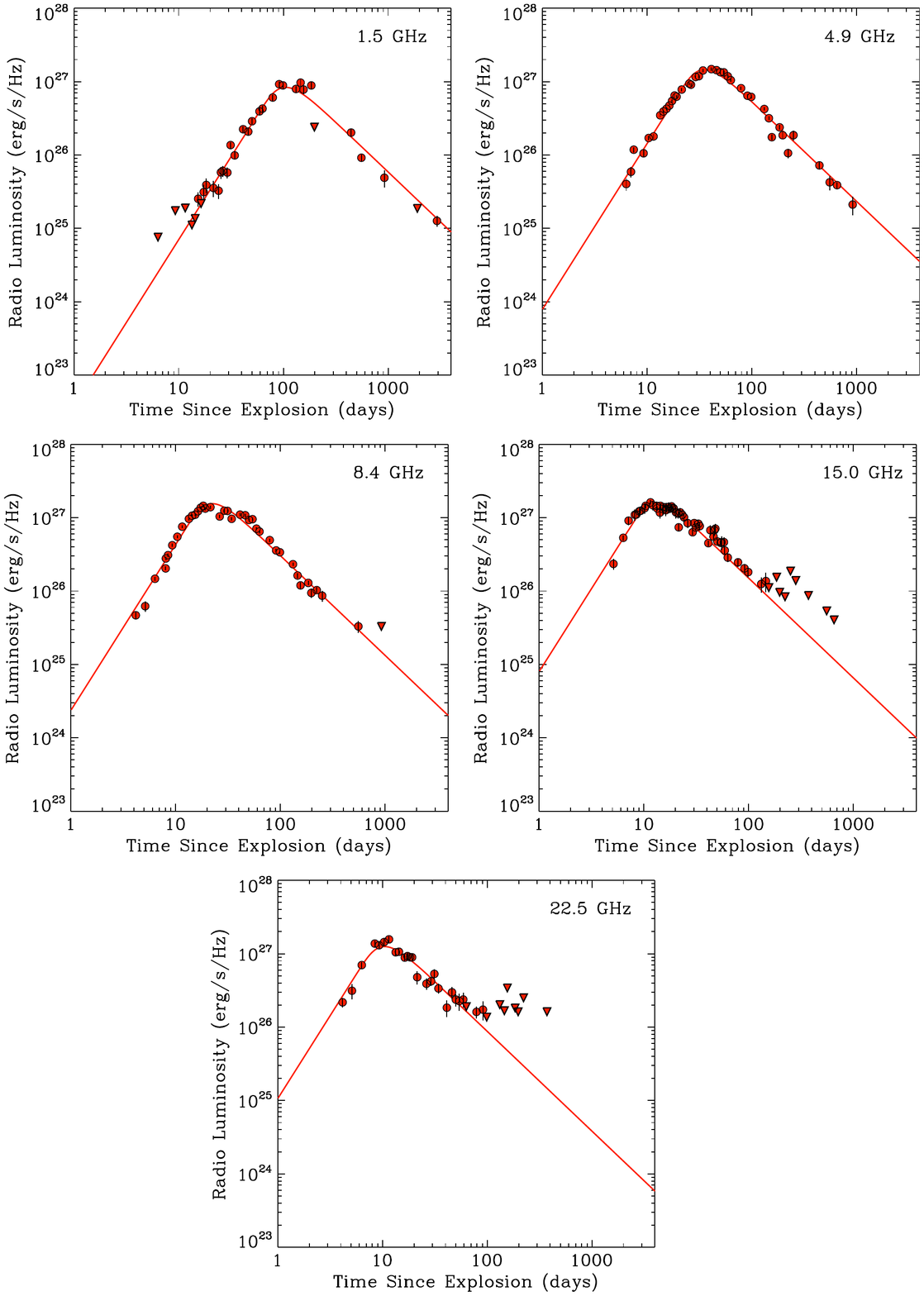}
\end{center}
\caption[]{Radio light curves of SN 1994I at different frequencies. Note the power law turn-on characteristic of synchrotron-self-absorption. The upside-down triangles mark upper limits. With permission from \cite{Alexander2015}.
}  
\label{fig11_94i}
\end{figure}

The best observed Type Ib/c in radio is SN 1994I observed by Weiler et al. with the VLA \cite{Weiler2011}. Originally classified as a Type Ic, the He I $\lambda $10830 line was  clearly detected in an early NIR spectrum with a maximum velocity $29,900 \kms$ \cite{Filippenko1995}. 
A review of radio and X-ray observations of Type Ibc SNe is presented in \cite{Chevalier2007}. 
In  Fig. \ref{fig11_94i} we show the radio light curves at different wavelengths from an analysis by \cite{Alexander2015}, showing the gradual turn-on, and  a steep decline after the peak, common to other Type Ib/c SNe.   

Alexander et al. \cite{Alexander2015} found that these light curves could be well fitted  with a SSA model, without any free-free absorption (solid line in Fig.  \ref{fig11_94i}). The mass loss was estimated to $\Mdot \approx 10^{-5} \ \Msunyr$ for a wind velocity of $1000 \ \kms$, but depends on uncertain assumptions for $\epsilon_{\rm e}$ and $\epsilon_{\rm B}$. The derived
expansion velocity is less sensitive to these (Eq. \ref{eqss6}) and was found to be  $\sim 35,000 \kms$, somewhat larger than that obtained from the optical, and typical of other Type Ib/c SNe (Fig. \ref{fig4d}). 

X-rays from SN 1994I\index{SN 1994I} were detected  at an age of 6 - 7 years  \cite{Immler2002}, although the identification has been challenged by \cite{VanDyk2016}. This very late detection is difficult to explain either by thermal shock emission, inverse Compton or a single power-law  synchrotron spectrum, extending from radio to X-rays. In \cite{Chevalier2006b} an explanation was suggested based on the broken power-law spectrum of the electrons in a cosmic ray dominated shock wave (Sect. \ref{sec:_acc}), where the low energy electrons have a steeper spectrum than those at high energy, giving a high X-ray to radio ratio. 

Although most Ib/c SNe are faint in radio and X-rays at early epochs, there are cases where strong emission in these bands has been observed to become bright on time scales of years. Examples of this are SN 2001em\index{SN 2001em} as well as the recent SN 2014C\index{SN 2014C} (\cite{Chugai2006} and \cite{Margutti2016,Milisavljevic2015} and references therein). Both of these exhibited abrupt increases of the radio and X-ray flux at an age of $\sim 700$ days and $\sim 150$ days, respectively (Fig. \ref{fig14c}).  At about the same time as the radio increase a strong, narrow H$\alpha$ line with FWHM $\sim 1800 \ \kms$ was seen in SN 2001em. 

In \cite{Chugai2006}, the X-ray and radio evolution of SN 2001em was modeled by  a He core with mass $\sim 2.5 \  \Msun$ interacting with a massive H rich shell. The H$\alpha$ line was explained as a result of the accelerated, shocked shell, while the X-rays, assumed to have $kT_{\rm e} \sim 80$ keV, and radio came from the much faster reverse shock with a velocity of $\sim 5500 \ \kms$. To yield the required velocity ratio of the shocked shell and reverse shock together with the observed X-ray luminosity a shell with mass $\sim 2-3 \ \Msun$ at $\sim 6 \times 10^{16}$ cm was required. Assuming a pre-shock velocity of $\sim 20 \ \kms$ this shell may have been ejected $\sim (1-2) \times10^3$ years before the explosion, and with a mass loss rate $\sim (2-10)\times 10^{-3}  \ \Msunyr$. {

  The analysis of SN 2014C yielded a similar set of parameters \cite{Margutti2016}. However, NuSTAR observations of SN 2014C gave a well-determined temperature of  $\sim 18$ keV, much lower than that assumed for SN 2001em It is therefore likely that the temperature of SN 2001em was considerably lower than estimated above, and may possibly have come from the forward shock, illustrating the importance of hard X-ray observations. The thermal X-ray spectrum of SN 2014C also showed strong emission at  6.7-6.9 keV from He-like and H-like Fe K transitions, clearly demonstrating the thermal nature of the spectrum. 
\begin{figure}[t!]
\begin{center}
\includegraphics[width=8cm]{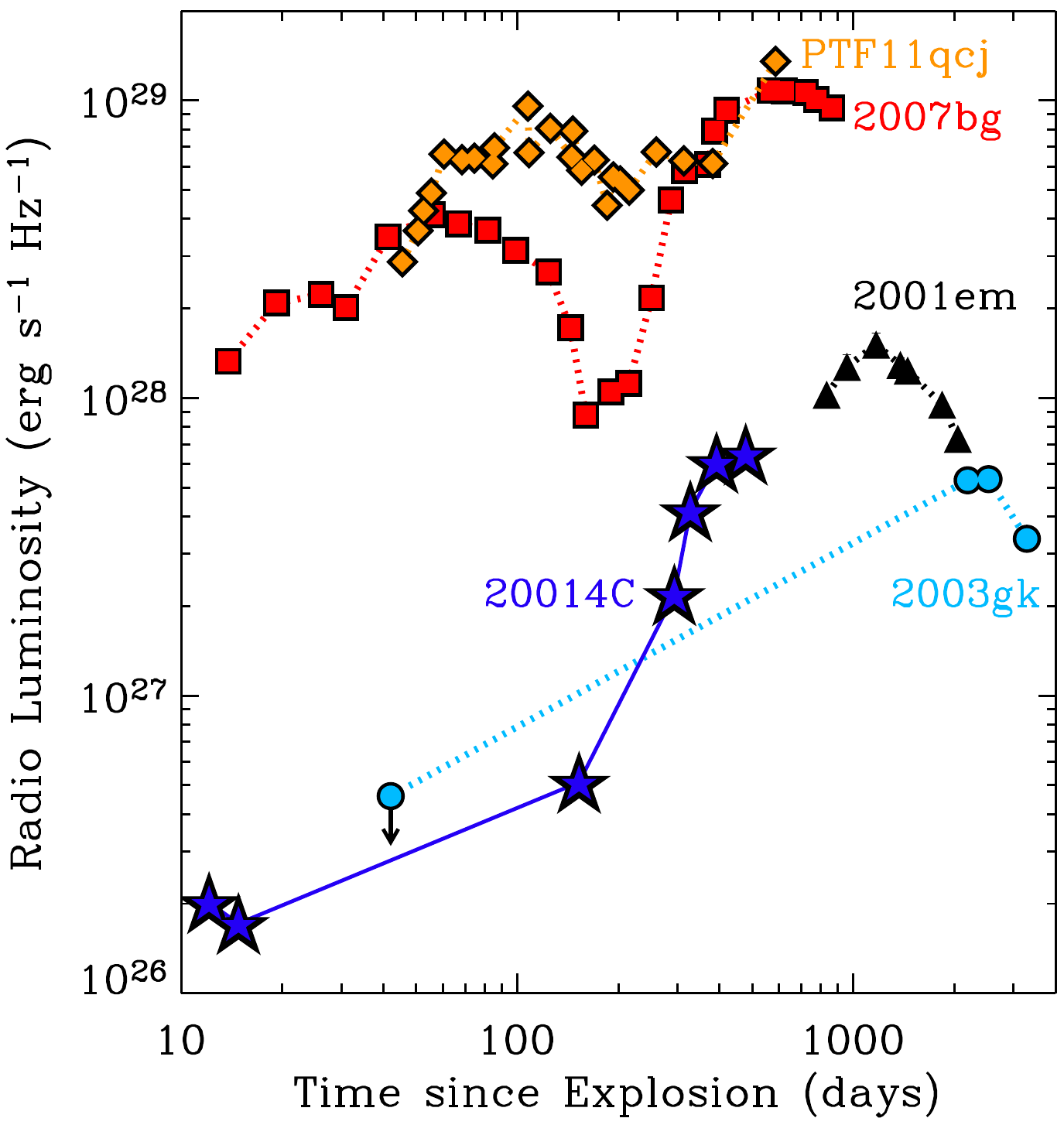}
\end{center}
\caption[]{Radio light curves at 7.1 - 8.5 GHz of the type Ib/c SNe 2001em, 2003gk, 2007bg, PTF11qcj and 2014C. With permission from \cite{Margutti2016}.
}  
\label{fig14c}
\end{figure}

Several other Type Ic SNe show variability at radio wavelengths on time scales of weeks to months \cite{Wellons2012}.

\subsection{Type Ibn}
While optical lines from the CSM are rare for \index{Type Ibn} Type Ib/c SNe, there  exist a number of interesting cases where this has been observed, which are classified as Ibn, in analogy with the IIn. For a summary, see \cite{Pastorello2016} and references therein. 

The best studied object of this class is SN 2006jc. Early spectra displayed He I lines with a width extending to $\sim 5000 \kms$, which is low for a Ib SN, but similar to Type IIn lines  \cite{Foley2007} . The bluer He I lines had P-Cygni absorptions extending to $\sim 2000 \ \kms$ with the minimum at $\sim 1000 \ \kms$, typical of WR winds. An outburst of the progenitor was observed $\sim 2$ years before explosion \cite{Pastorello2007}. The last spectra at $\sim 175$ days showed narrow lines with FWHM $\sim 1800 \ \kms$, including H$\alpha$.

SN 2006jc was detected with Swift and Chandra with a 0.5 - 10 keV luminosity of $(1-4) \times 10^{39} \ergs$, peaking at $\sim 110$ days after which it decayed \cite{Immler2008}. The Chandra spectrum displayed a very hard spectrum $dN/dE\propto E^{-0.24}$ \cite{Ofek2013}. No radio emission was detected. 
On day $\sim 75$ a strong red continuum excess appeared. At the same time the red wing of the initially symmetric He I lines faded. Both these aspects are indicative of dust formation \cite{Smith2008}, which was consistent with the observation of a strong NIR and mid-IR continuum from Spitzer \cite{Mattila2008}. 

Chugai \cite{Chugai2009} has modeled the optical and X-ray light curves, assuming a dense shell with a mass of $0.02-0.05 \ \Msun$ within $\sim 2 \times 10^{15}$ cm, ejected one year before explosion.    
The dust formation is assumed to take place in the cool, dense shell behind the reverse shock, as was earlier proposed by Mattila et al., \cite{Mattila2008}  based on Spitzer observations. 

\subsection{Relativistic expanding supernovae}

There are a number of SNe Ic, SNe 1998bw, 2003lw, 2006aj, and 2009bb, that are inferred to have  mildly relativistic expansion based
on their high radio luminosity at an early time and a synchrotron self-absorption model.
It is possible to estimate the minimum energy in the synchrotron emitting gas from the synchrotron luminosity and it is surprisingly
high for ejecta accelerated in a normal supernova.
In three of these cases (SNe 1998bw, 2003lw, 2006aj), there is a low luminosity gamma-ray burst (GRB) associated with the supernova, opening the possibility
that the energetic emission is associated with a collimated flow from a central engine.
However, Nakar \& Sari \cite{Nakar2012} found that  the basic properties of these events could be explained by a relativistic shock breakout model, but
the required supernova energy is very large.
In a revised model for SN 2006aj (GRB 060218), Nakar  \cite{Nakar2015} suggested that a normal GRB jet was slowed and deposited its
energy in an extended envelope around the progenitor star.
The deposition of the jet energy, about $10^{51}$ ergs, into a small amount of mass gives an energetic shock breakout event, as observed.
The radio emission from SN 2006aj can then be explained by the interaction of the supernova ejecta with a circumstellar wind
\cite{Barniol2015}.

\subsection{Type Ia}

\index{Type Ia}Except for objects with very dense interaction like SN 2006gy, essentially all types of core collapse supernovae  have shown radio and
X-ray emission from circumstellar interaction.
On the other hand, Type Ia supernovae have not been detected as radio and X-ray sources.
In some cases, this may be due to absorption by a very dense  circumstellar medium.
The Type IIn optical features in supernovae like SN 2002ic imply that a dense medium is present.
It is possible in these cases that radio and X-ray emission is attenuated by absorption processes. 
However, the great majority of SNe Ia show no sign of circumstellar interaction at optical wavelengths
and the low radio and X-ray emission is attributed to a low circumstellar density.
The limits set on the surrounding density are  more constraining than for other methods used for normal SNe Ia.

The occurrence of relatively nearby SNe Ia, e.g., SN 2011fe in M101 \cite
{Chomiuk2016,Horesh2012} and SN 2014J in M82 \cite{Margutti2014,PerezTorres2014}, has allowed especially strong limits 
on the radio and X-ray luminosities.
Limits on the radio luminosity of individual events are in the range $10^{23}-10^{24}$ ergs s$^{-1}$ Hz$^{-1}$.
As can be seen in Fig. \ref{fig4d}, the only core collapse supernova with a luminosity in this range is the early luminosity of SN 1987A. The X-ray limits of $\sim (4-7) \times 10^{36} \ \ergs$ \cite{Margutti2014} are also lower than any core collapse SNe, with the exception of SN 1987A. 

\begin{figure}[t!]
\begin{center}
\includegraphics[width=8cm]{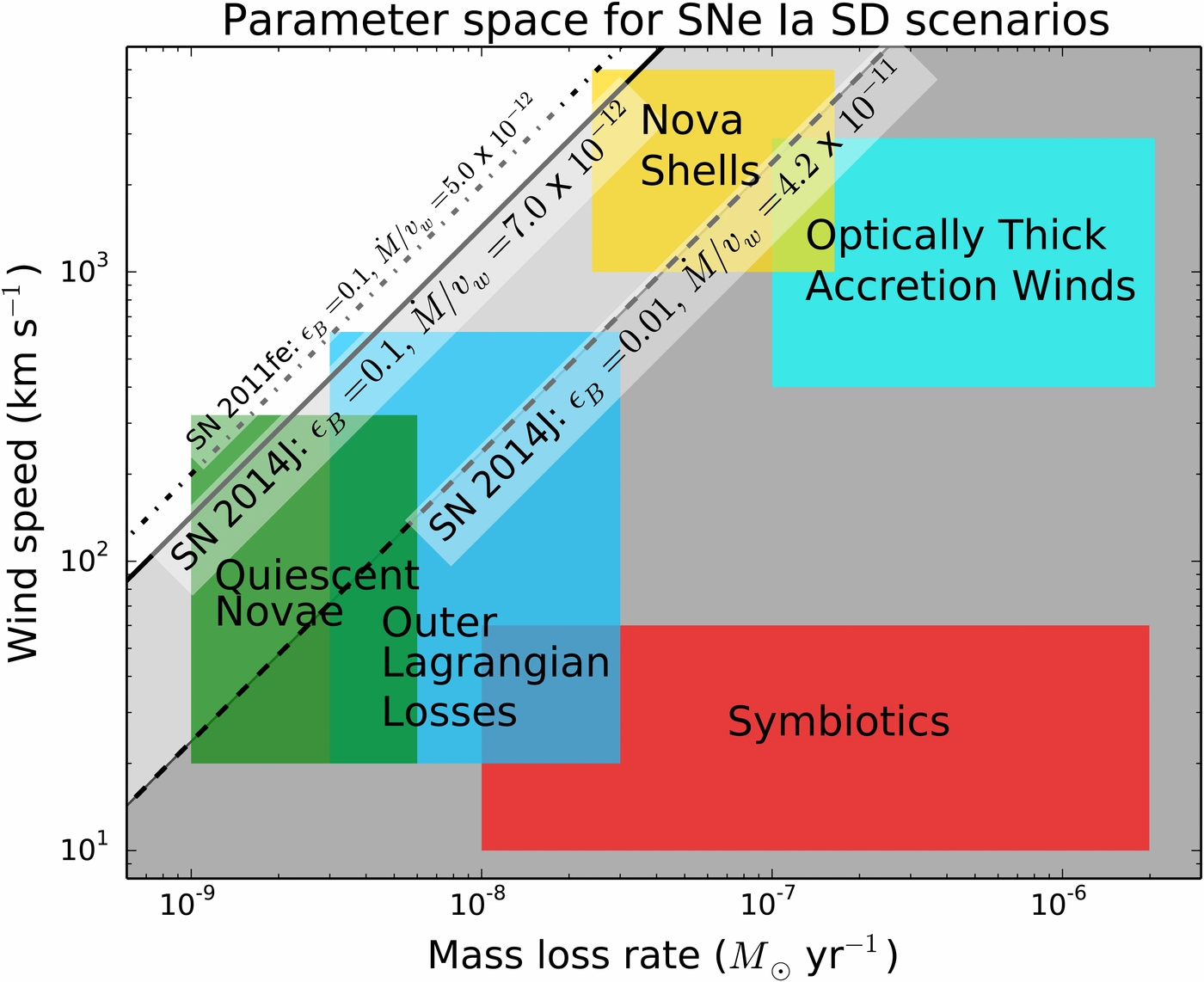}
\end{center}
\caption[]{Constraints on Type Ia single degenerate progenitors from radio observations of SN 2014J and SN 2011fe. Progenitors falling into the gray shaded areas should have been detected by the observations, and therefore are ruled out. With permission from \cite{PerezTorres2014}.
}  
\label{fig_ia_radio}
\end{figure}
There are uncertainties in setting limits on the mass loss density because of the lack of a complete theory for particle acceleration and magnetic field amplification, i.e., values of $\epsilon_{\rm e}$ and $\epsilon_{\rm B}$.
With plausible assumptions on the efficiency of production of particles and fields, some proposed progenitors of SNe Ia can be
ruled out, including ones where the companion star is a red giant (Fig. \ref{fig_ia_radio}).
Interaction with the interstellar medium easily produces emission below the observed limits, providing support for a double degenerate progenitor scenario. These limits, however, only apply to the individual objects with good limits, and there may be other channels which apply to other objects.

\section{Conclusions}
\label{sec_6}

During the last four decades, radio and X-ray observations have  given us a new tool to study the CSM of different types of SNe. This has in turn yielded a completely new view of the evolution of the SN progenitors and the evolution during the last phases before the explosion. Mass loss has become the most important factor for understanding the different types of core collapse SNe. 

Although there has been considerable  progress in many respects, there remains much to be done. Here we list a few of the areas that need further study: 
\begin{itemize}
\item A better understanding of the generation and efficiencies of magnetic field amplification and relativistic particle acceleration. These are  crucial for deriving accurate mass loss rates from the radio observations. 
\item The importance of the geometry.  Observations of supergiants in our Galaxy and of CS interaction give evidence for non-spherical CSM structure. SN 1987A is an obvious example, but there are many other indications that deviations from spherical geometry, as well as shell structures, are important.
\item CSM interaction on timescales of years. We have been able to follow a few supernovae on long timescales, in some cases  revealing the presence of shells and other structure. A more systematic monitoring of older supernovae would give important insight to the mass loss history preceding the final stage. 
\item The understanding of the mass loss mechanisms in the final stages before  explosion. Although there are many proposals, there is no consensus as to what causes the strong mass loss in the final stages as observed in particular cases.
\item  Effects of binary evolution. Common envelope evolution and the related mass loss are clearly important. This process is expected to lead to an asymmetric distribution of the gas, possibly as a toroidal structure in the orbital plane. The structure of the ring of SN 1987A is an important lesson.
\end{itemize}

\begin{acknowledgement}
RAC's research was partly supported by NASA grant NNX12AF90G, while that of CF by the Swedish Research Council.
\end{acknowledgement}

{\bf Cross-References}
\begin{itemize}
\item The Physics of Supernova 1987A
\item Shock breakout theory.
\item The Supernova - Supernova Remnant Connection
\item Interacting Supernovae: Types IIn and Ibn
\item Interacting Supernovae and the Influence on Spectra and Light Curves
\end{itemize}


\printindex
\end{document}